\documentclass[a4paper,11pt]{article}
\usepackage{jheppub}

\usepackage{adjustbox} 

\usepackage{amsmath}
\usepackage{amsfonts}
\usepackage{amssymb}
\usepackage{mathrsfs}
\usepackage{graphicx}
\usepackage{color}
\usepackage[utf8]{inputenc}
\usepackage{amsfonts}
\usepackage{booktabs}
\usepackage{multirow}
\usepackage{makecell}
\usepackage{siunitx}
\usepackage{ragged2e}
\usepackage{rotating}
\usepackage{float}
\usepackage[dvipsnames]{xcolor}
\definecolor{darkred}{rgb}{0.5,0,0}
\definecolor{darkblue}{rgb}{0,0,0.5}
\definecolor{firebrick}{rgb}{0.75,0.125,0.125}
\definecolor{darkgreen}{rgb}{0,0.5,0}
\hypersetup{urlcolor=darkblue,
	    citecolor=darkgreen,
	    linkcolor=firebrick}

\usepackage{orcidlink}
\usepackage{lipsum}
\usepackage[normalem]{ulem}
\usepackage[shortlabels]{enumitem}
\usepackage{fontawesome5}

\long\def\exclude#1{}

\newcommand{\ie}{{\it i.e.}}

\newcommand{\eg}{{\it e.g.}}

\newcommand{\cf}{{\it cf.}}

\newcommand{\eq}{Eq.}

\newcommand{\fig}{Fig.}

\newcommand{\Refe}{Ref.}
\newcommand{\Refes}{Refs.}
\newcommand{\equ}[1]{\eq~(\ref{equ:#1})}
\newcommand{\figu}[1]{\fig~\ref{fig:#1}}
\newcommand{\orcid}[1]{\href{https://orcid.org/#1}{\includegraphics[width=10pt]{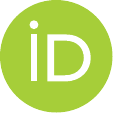}}}
\begin{document}

\title{No Flavor Anisotropy in the High-Energy Neutrino Sky Upholds Lorentz Invariance}

\author[a]{Bernanda Telalovic
\orcid{0000-0002-1406-502X}}

\author[a]{Mauricio Bustamante
\orcid{0000-0001-6923-0865}}

\affiliation[a]{Niels Bohr International Academy, Niels Bohr Institute,\\University of Copenhagen, DK-2100 Copenhagen, Denmark}

\date{March 19, 2025}

\abstract{
Discovering Lorentz-invariance violation (LIV) would upend the foundations of modern physics. Because LIV effects grow with energy, high-energy astrophysical neutrinos provide the most sensitive tests of Lorentz invariance in the neutrino sector. We examine an understudied yet phenomenologically rich LIV signature: compass asymmetries, where neutrinos of different flavors propagate preferentially along different directions. Using the directional flavor composition of high-energy astrophysical neutrinos, \ie, the abundances of $\nu_e$, $\nu_\mu$, and $\nu_\tau$ across the sky, we find no evidence of LIV-induced flavor anisotropy in 7.5 years of IceCube High-Energy Starting Events. Thus, we place upper limits on the values of hundreds of LIV parameters with operator dimensions 2--8, tightening existing limits by orders of magnitude and bounding hundreds of parameters for the first time~\href{https://github.com/BernieTelalovic/LIV_constraints_from_HESE_flavour_isotropy}{\faGithubSquare}.
}

\maketitle

\section{Introduction}

\begin{figure*}[t!]
 \centering
 \includegraphics[trim={0.25cm 1.25cm 0.25cm 1.25cm}, clip, width=\textwidth]{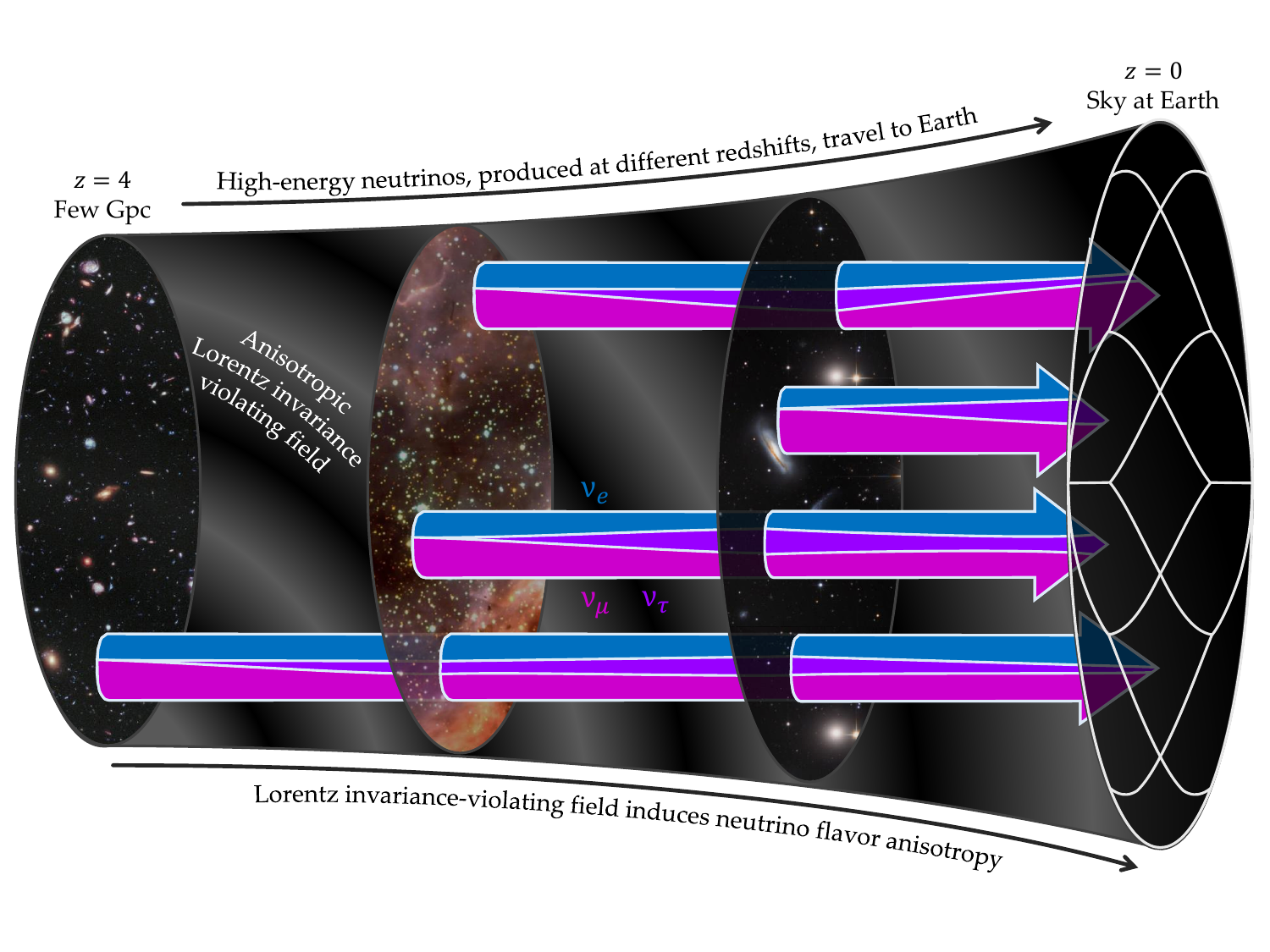}
 \caption{\textbf{\textit{Effect of anisotropic Lorentz-invariance violation on the flavor composition of high-energy astrophysical neutrinos.}} Neutrinos are emitted by astrophysical sources located at different distances from Earth, of up to a few Gpc, distributed in redshift, $z$.  They are emitted with different proportions of the different flavors, $\nu_e$, $\nu_\mu$, and $\nu_\tau$, as determined by the neutrino production mechanism.  In this figure, as an example, they are emitted with a 1:2:0 ratio between $\nu_e$, $\nu_\mu$, and $\nu_\tau$; in producing our results, we explore all possibilities.  En route to Earth,  the neutrinos interact with a pervading Lorentz-invariance-violating field that, together with standard flavor-transition processes, modifies the proportions of the different neutrino flavors that reach Earth.  If the field is anisotropic, this modification is different along different directions in the sky.  \textbf{\textit{We use the measurement of the ``directional flavor composition'', extracted from 7.5 years of IceCube HESE data, to constrain hundreds of possible forms of anisotropic Lorentz-invariance violation.}}  See the main text for details.}
 \label{fig:sketch}
\end{figure*}

Does Nature have preferred directions?  If yes, then not all inertial frames of reference would be equivalent, and Lorentz invariance would not be a fundamental symmetry of Nature---in contrast to what the Standard Model assumes.  Proposed theories of quantum gravity posit that such a violation of Lorentz invariance might occur in particle processes, its presence obscured by a high energy scale, possibly as high as the Planck scale~\cite{Ellis:2011ek, Tasson:2014dfa, Carlip:2022pyh, Basile:2024oms}.  Thus, Lorentz-invariance violation (LIV), if it exists, would manifest more prominently, even if still subtly, the higher the particle energies involved~\cite{Addazi:2021xuf, AlvesBatista:2023wqm}.

Motivated by this, we look for signs of LIV in high-energy astrophysical neutrinos.  These neutrinos are incisive probes of fundamental physics, including, but not limited to, LIV~\cite{Anchordoqui:2013dnh, Ahlers:2018mkf, Ackermann:2019cxh, Arguelles:2019rbn, Ackermann:2022rqc, MammenAbraham:2022xoc, Arguelles:2022tki, AlvesBatista:2023wqm}. Because they have some the highest known neutrino energies---from tens of TeV to a few PeV---they can probe new-physics effects that are ordinarily suppressed by high energy scales.  Because they travel cosmological-scale distances virtually unscathed, they can accrue LIV effects during their long travel time to Earth, making them more easily detectable.  (These same features also make high-energy astrophysical neutrinos powerful probes of astrophysics~\cite{Anchordoqui:2013dnh, Ackermann:2019ows, AlvesBatista:2019tlv, AlvesBatista:2021eeu, Ackermann:2022rqc}.)  

Previous works have pointed out ways to test LIV with high-energy astrophysical neutrinos~\cite{Barenboim:2003jm, Christian:2004xb, Hooper:2005jp, Bazo:2009en, Ando:2009ts, Bhattacharya:2009tx, Bustamante:2010nq, Mehta:2011qb, Arguelles:2015dca, Bustamante:2015waa, Rasmussen:2017ert, Moura:2022dev, PerezdelosHeros:2022izj, Carmona:2023mzs, AlvesBatista:2023wqm} and used their detection to constrain various forms of it~\cite{Diaz:2013wia, Wang:2016lne, Wei:2016ygk, IceCube:2017qyp, Lai:2017bbl, Wang:2020tej, IceCube:2021tdn, PerezdelosHeros:2022izj, Bustamante:2024fbj, KM3NeT:2025mfl, Satunin:2025uui, Amelino-Camelia:2025lqn, Yang:2025kfr}.  (Lower-energy astrophysical neutrinos, terrestrial neutrinos, and relic neutrinos also test LIV~\cite{Kostelecky:2004hg, Katori:2006mz, IceCube:2010fyu, Kostelecky:2011gq, Guo:2012mv, DoubleChooz:2012eiq, Super-Kamiokande:2014exs, T2K:2017ega, SNO:2018mge, Mishra:2023tdj, Torri:2024jwc}; see \Refe~\cite{Kostelecky:2008ts} for a summary.)  
We focus on a largely understudied, but phenomenologically rich manifestation of LIV: flavor-dependent \textit{compass asymmetries}~\cite{Diaz:2009qk, Diaz:2011ia, Klop:2017dim, Lin:2025aym}.  These are persistent anisotropies in the distribution of arrival directions at Earth of neutrinos of different flavors, $\nu_e$, $\nu_\mu$, and $\nu_\tau$.  We adopt the formalism from the Standard Model Extension (SME)~\cite{Kostelecky:2003xn, Kostelecky:2003cr, Kostelecky:2003fs, Kostelecky:2011gq}, an effective field theory that contains LIV operators of different dimensions, regulated by coefficients whose values are a priori unknown, to be determined by experiment.

Figure~\ref{fig:sketch} sketches the effect of LIV compass asymmetries on high-energy astrophysical neutrinos.  Neutrinos are emitted by astrophysical sources located at different distances away from Earth, distributed in redshift.  They are emitted with an initial flavor composition---\ie, the ratio of each neutrino flavor to the total---that is determined by their production mechanism.  In \figu{sketch},  the ratio between emitted $\nu_e$, $\nu_\mu$, and $\nu_\tau$ is $1:2:0$; in our main results, we explore all possibilities.  En route to Earth, neutrinos \textit{oscillate}, \ie, the proportions of neutrinos of different flavors evolve~\cite{Super-Kamiokande:1998kpq, SNO:2001kpb}.  Our standard expectation---devoid of LIV---is that, upon reaching Earth, the flavor composition of their flux is the same from every direction in the sky.  In \figu{sketch}, the expectation is of roughly equal proportion of each flavor at Earth~\cite{Bustamante:2015waa}, but different production mechanisms lead to alternatives.

Figure~\ref{fig:sketch} illustrates how this expectation changes when neutrinos propagate through a pervasive, anisotropic LIV field.  If neutrinos of different flavors couple to this field differently---via LIV operators that depend on the neutrino direction---this would modify neutrino oscillations by making them direction-dependent.  As a consequence, there would be different preferred directions of propagation for neutrinos of different flavors.  Figure~\ref{fig:sketch} shows that this would manifest as \textit{flavor anisotropies}, \ie, as a directional dependence across the sky of the flavor composition of the neutrinos upon reaching Earth.

Different forms of the neutrino coupling to the LIV field would induce different forms of flavor anisotropy and have a different dependence with neutrino energy.  We search for flavor anisotropy in present-day IceCube data, specifically, in the public 7.5-year sample of High-Energy Starting Events (HESE)~\cite{IceCube:2020wum, IC75yrHESEPublicDataRelease}, by adopting the methods and results introduced by \Refe~\cite{Telalovic:2023tcb}, which account for particle-physics and astrophysical uncertainties.  Finding no statistically significant evidence for it, we place upper limits on hundreds of LIV coefficients with operator dimensions from 2 to 8.  \textbf{\textit{We tighten existing limits on LIV parameters by orders of magnitude, due to the high neutrino energies, and bound hundreds of LIV parameters for the first time ever.}}

The rest of this paper is organized as follows.  Section~\ref{sec:nu_oscillations} introduces neutrino oscillations and the effect of LIV on them.  Section~\ref{sec:astro_nu} introduces high-energy astrophysical neutrinos, their flavor composition, and their detection.  Section~\ref{sec:liv_strategies} presents and contrasts and our two strategies to explore the LIV parameter space.  Section~\ref{sec:stat_methods} introduces the statistical methods that we use to extract limits on LIV parameters.  Section~\ref{sec:results} shows our resulting limits and compares them to existing ones.  Section~\ref{sec:summary} summarizes and concludes.  Appendix~\ref{sec:constraint_tables} contains tables with our limits; they are also available digitally on GitHub~\href{https://github.com/BernieTelalovic/LIV_constraints_from_HESE_flavour_isotropy}{\faGithubSquare}.


\section{Neutrino oscillations and Lorentz-invariance violation}
\label{sec:nu_oscillations}


\subsection{Standard oscillations}
\label{sec:nu_oscillations-std}

In the Standard Model augmented to include neutrino masses, neutrinos interact as flavor states, $\nu_e$, $\nu_\mu$, and $\nu_\tau$, but propagate as propagation states, each a superposition of flavor states.  The mixing of flavor and propagation states---possibly augmented by new physics---and its evolution during their propagation induces neutrino oscillations.  Below, we give an overview of this.

Under standard oscillations, the propagation states in vacuum are neutrino mass eigenstates, $\nu_1$, $\nu_2$, $\nu_3$, with masses $m_1$, $m_2$, and $m_3$.  They are connected to the flavor states via the Pontecorvo-Maki-Nakagawa-Sakata (PMNS) mixing matrix, $U_\text{PMNS}$, \ie, $\nu_\alpha = \sum_i (U_\text{PMNS})_{\alpha i}^\ast \nu_i$ ($\alpha = e, \mu, \tau$, $i = 1, 2, 3$).  Following convention~\cite{ParticleDataGroup:2024cfk}, we parametrize the matrix via three angles, $\theta_{12}$, $\theta_{23}$, and $\theta_{13}$, and one CP-violation phase, $\delta_\text{CP}$. 

The high-energy astrophysical neutrinos that we focus on in our analysis travel from their sources to Earth practically in vacuum.  [The exception is when they reach the Earth and propagate inside it to reach the detector, during which they undergo interactions with matter.  These interactions are included in the detector response we use (Sec.~\ref{sec:astro_nu-detection}).]
The Hamiltonian that describes the propagation of neutrinos of energy $E$ in vacuum, written in the flavor basis, is
\begin{equation}
 \label{equ:hamiltonian_vac}
 H_{\rm vac}
 =
 \frac{1}{2E} 
 U_{\rm PMNS}M^2 U^\dagger_{\rm PMNS} \;,
\end{equation}
where $M^2 = \text{diag}(0, \Delta m^2_{21}, \Delta m^2_{31})$ is the neutrino mass matrix and $\Delta m^2_{ij} \equiv m_i^2-m_j^2$.  The values of $\theta_{12}$, $\theta_{23}$, $\theta_{13}$, $\delta_\text{CP}$, $\Delta m^2_{21}$, and $\Delta m^2_{31}$ are measured in neutrino oscillation experiments; later, we use these measurements in our analysis (Table~\ref{tab:fit_params}).  A neutrino created as $\nu_\alpha$, after propagating a distance $L$, is in a state $\nu_\alpha(L, E) = e^{-iH_\text{vac} L} \nu_\alpha$, where we have assumed that neutrinos are relativistic so that their propagation time $t \approx L$.  The probability of detecting it as $\nu_\beta$ is $\vert 
\nu_\beta^\dagger \nu_\alpha(L, E) \vert^2$ ($\alpha, \beta = e, \mu, \tau$), which is an oscillatory function of $L/E$~\cite{ParticleDataGroup:2024cfk}. [In our calculations, however, we use its average value (Sec.~\ref{sec:nu_oscillations-prob}).]

The presence of additional, flavor-dependent neutrino interactions, like those coming from Lorentz-invariance violation, modifies the probability.  Below, we show how.


\subsection{Lorentz-invariance violation in neutrinos}
\label{sec:nu_oscillations-liv}

The Standard Model is thought to be a low-energy effective field theory of a more fundamental description of Nature.  Proposed extensions of the Standard Model, including theories of quantum gravity, posit departures from it that become significant at particle energies that could be as high as the Planck energy scale of $E_{\rm Pl} \approx 10^{19}$~GeV, but possibly smaller and within reach of the highest particles currently observed.  Most egregiously, these departures include the violation of Lorentz invariance~\cite{Ellis:2011ek, Tasson:2014dfa, Addazi:2021xuf, Carlip:2022pyh, AlvesBatista:2023wqm, Basile:2024oms}, a pillar of the Standard Model, and the possible associated violation of CPT invariance.

The Standard Model Extension (SME) is an effective field theory that augments the Standard Model and general relativity by introducing all possible operators that could induce LIV.  In the neutrino sector, it contains terms that induce CPT-preserving (or ``CPT-even'') and CPT-violating (or ``CPT-odd'') LIV, and neutrino-anti-neutrino mixing.  However, because we cannot distinguish between neutrinos and anti-neutrinos using the detected high-energy events that we use in our analysis (Sec.~\ref{sec:astro_nu-detection}), we ignore the latter and restrict ourselves to CPT-even and CPT-odd operators in neutrino--neutrino and anti-neutrino--anti-neutrino mixing.  

Following Eq.~(44) in \Refe~\cite{Kostelecky:2011gq}, we describe the effect of LIV on neutrinos via a Hamiltonian, written in the flavor basis, made up of effective LIV operators, \ie,
\begin{equation}
 \label{equ:hamiltonian_liv}
 H_\text{LIV}
 =
 \frac{1}{E}(\hat{a}_\text{eff}-\hat{c}_\text{eff}) \;,
\end{equation}
where $\hat{a}_\text{eff}$ and $\hat{c}_\text{eff}$ are towers of CPT-odd and CPT-even operators, respectively, of varying operator dimensions, represented by $3\times 3$ complex matrices with components $\hat{a}_{\rm eff}^{\alpha\beta}$ and $\hat{c}_{\rm eff}^{\alpha\beta}$. 
Inside each tower, the operator of dimension $d$ has a dependence on neutrino momentum magnitude, $\vert p \vert \approx E$ for relativistic neutrinos, of $\vert E \vert^{d-2}$.

We expand each effective operator as a power series in neutrino energy and a spherical harmonic series that captures its angular dependence, \ie,
\begin{eqnarray}
 \label{equ:sph_exp_a}
 \hat{a}_{\rm eff}^{\alpha\beta} 
 &=&
 \sum_{d=3~\text{(odd)}}^\infty E^{d-2} 
 \sum_{\ell=0}^{d-1}
 \sum_{m=-\ell}^{\ell} 
 Y_\ell^m(\hat{\boldsymbol{p}})
 ({a}_{\rm eff}^{(d)})_{\ell m}^{\alpha\beta}
 \;,
 \\
 \label{equ:sph_exp_c}
 \hat{c}_{\rm eff}^{\alpha\beta} 
 &=&
 \sum_{d=2~\text{(even)}}^\infty E^{d-2} 
 \sum_{\ell=0}^{d-1}
 \sum_{m=-\ell}^{\ell} 
 Y_\ell^m(\hat{\boldsymbol{p}})
 ({c}_{\rm eff}^{(d)})_{\ell m}^{\alpha\beta}
 \;,
\end{eqnarray}
where $Y_{\ell}^m$ are complex spherical harmonic functions of the neutrino propagation direction, given by the unit vector in the direction of the neutrino three-momentum, $\hat{\boldsymbol{p}}$ (expressed in the Sun-centered reference frame described below).  The coefficients  
$(a^{(d)}_\text{eff})_{\ell m}^{\alpha\beta}$ and $(c^{(d)}_\text{eff})_{\ell m}^{\alpha\beta}$ are complex scalar constants, independent of the neutrino energy and direction, that can be arranged into matrices.  CPT-odd operators have odd-valued $d$; CPT-even operators, even-valued $d$.  In matrix form, from Eqs.~(\ref{equ:sph_exp_a}) and (\ref{equ:sph_exp_c}), the LIV Hamiltonian is
\begin{equation}
 \label{equ:h_liv_d}
 H_\textrm{LIV}
 =
 \sum_{d=2}^\infty
 H_\textrm{LIV}^{(d)} \;,
\end{equation}
where, for operator dimension $d$,
 \begin{equation}
  \label{equ:h_liv_matrix}
  H_\textrm{LIV}^{(d)}
  \equiv
  E^{d-3}
  \sum_{\ell=0}^{d-1}
  \sum_{m=-\ell}^{\ell} 
  Y_\ell^m(\hat{\boldsymbol{p}})
  \times
  \left\{
   \begin{array}{ll}
    \begin{pmatrix}
      (a_{\rm eff}^{(d)})_{\ell m}^{ee}      & 
      (a_{\rm eff}^{(d)})_{\ell m}^{e\mu}    & 
      (a_{\rm eff}^{(d)})_{\ell m}^{e\tau}   \\
      (a_{\rm eff}^{(d)})_{\ell m}^{\mu e}   & 
      (a_{\rm eff}^{(d)})_{\ell m}^{\mu\mu}  & 
      (a_{\rm eff}^{(d)})_{\ell m}^{\mu\tau} \\
      (a_{\rm eff}^{(d)})_{\ell m}^{\tau e}  & 
      (a_{\rm eff}^{(d)})_{\ell m}^{\tau\mu} & 
      (a_{\rm eff}^{(d)})_{\ell m}^{\tau\tau} 
    \end{pmatrix} &
    \textrm{if $d$ is odd (CPT-odd),}        \\[2.em]
    -
    \begin{pmatrix}
      (c_{\rm eff}^{(d)})_{\ell m}^{ee}      &
      (c_{\rm eff}^{(d)})_{\ell m}^{e\mu}    & 
      (c_{\rm eff}^{(d)})_{\ell m}^{e\tau}   \\
      (c_{\rm eff}^{(d)})_{\ell m}^{\mu e}   &
      (c_{\rm eff}^{(d)})_{\ell m}^{\mu\mu}  &
      (c_{\rm eff}^{(d)})_{\ell m}^{\mu\tau} \\
      (c_{\rm eff}^{(d)})_{\ell m}^{\tau e}  &
      (c_{\rm eff}^{(d)})_{\ell m}^{\tau\mu} &
      (c_{\rm eff}^{(d)})_{\ell m}^{\tau\tau}
    \end{pmatrix}  &
    \textrm{if $d$ is even (CPT-even)}   
   \end{array}
  \right.
  \;.
 \end{equation}
Coefficients with $\ell = 0$ are called \textit{isotropic}, since $Y_0^0$ is a constant; those with $\ell > 0$ are called \textit{anisotropic}.

Our goal is to constrain the values of the coefficients $(a_{\rm eff}^{(d)})_{\ell m}^{\alpha\beta}$ and $(c_{\rm eff}^{(d)})_{\ell m}^{\alpha\beta}$.  While their number grows quickly with rising $d$, the Hermiticity of the operators $\hat{a}_{\rm eff}$ and $\hat{c}_{\rm eff}$ enforces that $(a_{\rm eff}^{(d)})^{\alpha\beta}_{\ell m} = (-1)^{m} (a_{\rm eff}^{(d)})^{\beta\alpha}_{\ell (-m)}$, and equivalently for $(c_{\rm eff}^{(d)})^{\alpha\beta}_{\ell m}$, which means we need only constrain coefficients with $m \geq 0$.  As a result, the number of free LIV coefficients that we need to constrain at each value of $d > 2$ is $9 d^2$.  We consider operators up to $d = 8$, placing the total number of complex coefficients that we explore at 1062, and we treat their real and imaginary parts separately, resulting in 1818 free LIV parameters, overall.  (From the Hermiticity condition, the coefficients with $\alpha = \beta$ and $m = 0$ are real-valued.)  In Sec.~\ref{sec:liv_strategies}, we introduce our strategy to constrain them.

(For the $d = 2$ CPT-even case, we only constrain 27 out of the 36 $(c_{\rm eff}^{(d)})^{\alpha\beta}_{\ell m}$ coefficients, specifically, the anisotropic ones.  The remaining 9 isotropic coefficients induce direction-independent changes to the neutrino flavor-transition probabilities that have the same $1/E$ energy dependence as standard oscillations, and cannot, therefore, be constrained independently of them~\cite{Kostelecky:2011gq}.)

Because $H_\text{LIV}$ depends on the neutrino direction, the values of the LIV coefficients depend on our choice of the observer frame where they are measured. In the literature, the conventional reference frame where these coefficients are defined is the Sun-centered reference frame~\cite{Kostelecky:2002hh, Kostelecky:2011gq}.  For the rest of this paper, we consider that the LIV coefficients are always written in this frame.  Reference~\cite{Telalovic:2023tcb} contains a detailed explanation of how we express the LIV coefficients in this frame.


\subsection{Sensitivity to the LIV parameters}
\label{sec:nu_oscillations-liv_sensitivity}

The total Hamiltonian includes the contribution from standard [\equ{hamiltonian_vac}] and LIV-induced mixing [\equ{hamiltonian_liv}], \ie,
\begin{equation}
 \label{equ:hamiltonian_total}
 H_\text{tot} = H_\text{vac} + H_\text{LIV} \;.
\end{equation}
The relative contribution of $H_\text{LIV}$ grows with energy for $d > 2$.  Considering a single nonzero CPT-odd LIV parameter with specific operator dimension---which is our default strategy (Sec.~\ref{sec:liv_strategies-single_parameter})---the vacuum and LIV contributions are comparable when the LIV parameter has a value of approximately
\begin{equation}
 \label{equ:liv_coeff_sensitivity}
 ({a}_{\rm eff}^{(d)})_{\ell m}^{\alpha\beta}
 \approx
 10^{-23} 
 \left( \frac{\Delta m_{ij}^2}{10^{-3}~{\rm eV}^2} \right)
 \left( \frac{E}{{\rm GeV}} \right)^{2-d}
 {\rm GeV}^{4-d} \;,
\end{equation}
for odd-valued $d$, and neglecting the dependence on neutrino direction in this estimate.  (For a CPT-even LIV parameter, $({c}_{\rm eff}^{(d)})_{\ell m}^{\alpha\beta}$, the expression is the same but with even-valued $d$.)

Equation~(\ref{equ:liv_coeff_sensitivity}) determines the range of values of the LIV parameters to which our analysis is sensitive.  We consider astrophysical neutrinos with energies in excess of 10~TeV, with a steeply falling power-law energy spectrum (Sec.~\ref{sec:astro_nu-flux}), so that low-energy neutrinos are more abundant and, therefore, drive the sensitivity sensitive.  For 10-TeV neutrinos, \equ{liv_coeff_sensitivity} estimates that we are sensitive to LIV parameters with values larger than about $10^{-15-4d}~{\rm GeV}^{4-d}$.  This is, indeed, roughly the size of the upper limits on the LIV parameters that our analysis places, from $d = 2$ to 8 (see Tables \ref{tab:constraint_tables_d2}--\ref{tab:constraint_tables_d8}).  The actual limits we place are worse than our estimated sensitivity due to the uncertainties in the model parameters involved in the statistical analysis 

(The exact value of the minimum astrophysical neutrino energy assumed in the analysis impacts the range of values of LIV parameters to which our analysis is sensitive.  We choose 10~TeV following the choice that was made in \Refe~\cite{Telalovic:2023tcb} to extract the directional flavor composition from IceCube data, on which we base our analysis.  Our choice yields conservative upper limits on the LIV parameters.  We elaborate on this in Sec.~\ref{sec:results-comparison_previous_limits}.)


\subsection{Neutrino oscillation probability with LIV}
\label{sec:nu_oscillations-prob}

In the presence of LIV, the new propagation states, $\nu^\prime_i$, are the eigenstates of $H_\text{tot}$, \equ{hamiltonian_total}.  The transformation between them and the flavor states is given by a new mixing matrix, $U$, that diagonalizes  $H_\text{tot}$, \ie, $\nu_\alpha = \sum_i U_{\alpha i}^\ast \nu^\prime_i$.  Unlike the PMNS matrix, the matrix $U$ depends on the neutrino energy and direction, on all of the standard oscillation parameters, $\boldsymbol{\theta}_{\rm std} \equiv \{ \theta_{12}, \theta_{23}, \theta_{13}, \delta_\text{CP}, \Delta m_{21}^2, \Delta m_{31}^2 \}$, and on the collection of effective LIV coefficients from \equ{h_liv_matrix}, $\boldsymbol{\Lambda}_{\rm LIV} \equiv \{ ({a}_{\rm eff}^{(d)})_{\ell m}^{\alpha\beta}, ({c}_{\rm eff}^{(d)})_{\ell m}^{\alpha\beta} \}_{d, \ell, m, \alpha, \beta}$.  For a fixed neutrino energy, the values of these parameters determine the relative contribution of the standard and LIV terms to $H_\text{tot}$. 

In analogy to the standard-oscillation case (Sec.~\ref{sec:nu_oscillations-std}), the state of a neutrino created as a $\nu_\alpha$, after propagating a distance $L$, is $\nu_\alpha(L,E) = e^{-i H_\text{tot} L} \nu_\alpha$ and the $\nu_\alpha \to \nu_\beta$ transition probability is $\vert \nu_\beta^\dagger \nu_\alpha(L, E) \vert^2$.  Unlike the standard case, however, this is no longer necessarily a periodic function of $L$ nor can it be, in general, written in a simple closed form, but rather must be evaluated numerically.  However, for the high-energy astrophysical neutrinos that are our focus, we simplify the computation as follows.

Under standard oscillations in vacuum, \ie, when $H_\text{tot} \approx H_\text{vac}$, the neutrino oscillation probability is a superposition of three periodic functions, each with oscillation length of approximately 
\begin{eqnarray}    
 L_{\text{osc}, ij}^\text{vac}
 &\equiv&
 4 \pi \frac{E}{\Delta m_{ij}^2} \\
 &\approx&
 8 \cdot 10^{-56}~{\rm Mpc}
 \left( \frac{E}{\rm GeV} \right)
 \left( \frac{{\rm eV}^2}{\Delta m_{ij}^2} \right) \;.
\end{eqnarray}
For neutrinos with energies $E \gtrsim 1$~TeV, and using the known values of $\Delta m_{ij}^2$~\cite{Esteban:2020cvm}, these lengths are tiny compared to the typical extragalactic distances traveled by high-energy astrophysical neutrinos, \ie, $L_{\text{osc}, ij}^\text{vac} \ll L \gtrsim 100$~Mpc.  As a result, on the scale of the distances traveled by the neutrinos, oscillations are rapid.  Coupled to this, high-energy neutrino telescopes, like IceCube, have limited energy resolution (Sec.~\ref{sec:astro_nu-detection}), insufficient to resolve rapid oscillations.  Further, when studying oscillations in the diffuse flux of high-energy astrophysical neutrinos (Sec.~\ref{sec:astro_nu-flux}), the rapid oscillations are smeared out by the distribution of distances to the unresolved neutrino sources that, combined, make up the flux.  Because of the above arguments, in practice we are sensitive only to the average $\nu_\alpha \to \nu_\beta$ flavor-transition probability~\cite{Learned:1994wg}, $\sum_i \vert (U_\text{PMNS})_{\alpha i} \vert^2 \vert (U_\text{PMNS})_{\beta i} \vert^2$, which, unlike the original oscillatory probability, no longer depends on $E$ or $\Delta m_{ij}^2$.

Under dominant LIV-induced mixing, \ie, when $H_\text{tot} \approx H_\text{LIV}$, the ``oscillation'' length (a misnomer, since flavor transitions under LIV are not periodic) due to a single dominant CPT-odd coefficient is approximately
\begin{eqnarray}
 \label{equ:osc_length_liv}
 L_{\text{osc}, \alpha\beta\ell m}^{\text{LIV}, (d)}
 &\equiv&
 2 \pi \frac{E^{3-d}}{({a}_{\rm eff}^{(d)})_{\ell m}^{\alpha\beta}} \\
 \nonumber
 &\approx& 
 4 \cdot 10^{-38}~{\rm Mpc} 
 \left(\frac{E}{{\rm GeV}}\right)^{3-d}
 \left(\frac{{\rm GeV}^{4-d}}{({a}_{\rm eff}^{(d)})_{\ell m}^{\alpha\beta}} \right) \; ,
\end{eqnarray}
ignoring the angular dependence from the spherical harmonics.  For a single dominant CPT-even coefficient, the expression is similar, with $({a}_{\rm eff}^{(d)})_{\ell m}^{\alpha\beta} \to ({c}_{\rm eff}^{(d)})_{\ell m}^{\alpha\beta}$.  Hence, the total oscillation length is approximately 
\begin{equation}
 L_{\text{osc},ij\alpha\beta\ell m}^{(d)}
 \equiv
 \left( 
 \frac{1}{L_{\text{osc}, ij}^\text{vac}}
 +
 \frac{1}{L_{\text{osc}, \alpha\beta\ell m}^{\text{LIV}, (d)}}
 \right)^{-1} \;,
\end{equation}
which accounts for the relative contributions of the standard and LIV terms, depending on the energy and the size of the LIV coefficients.  
As long as the total oscillation length is tiny compared to the typical baseline, \ie, $L_{\text{osc},ij\alpha\beta\ell m}^{(d)} \ll L \sim 100$~Mpc, we are still justified in using the average oscillation probability.  In our analysis, this condition is always satisfied within the range of values of the LIV coefficients that we explore (Table~\ref{tab:fit_params}).

\begin{figure}[t!]
 \centering
 \includegraphics[width=0.75\textwidth]{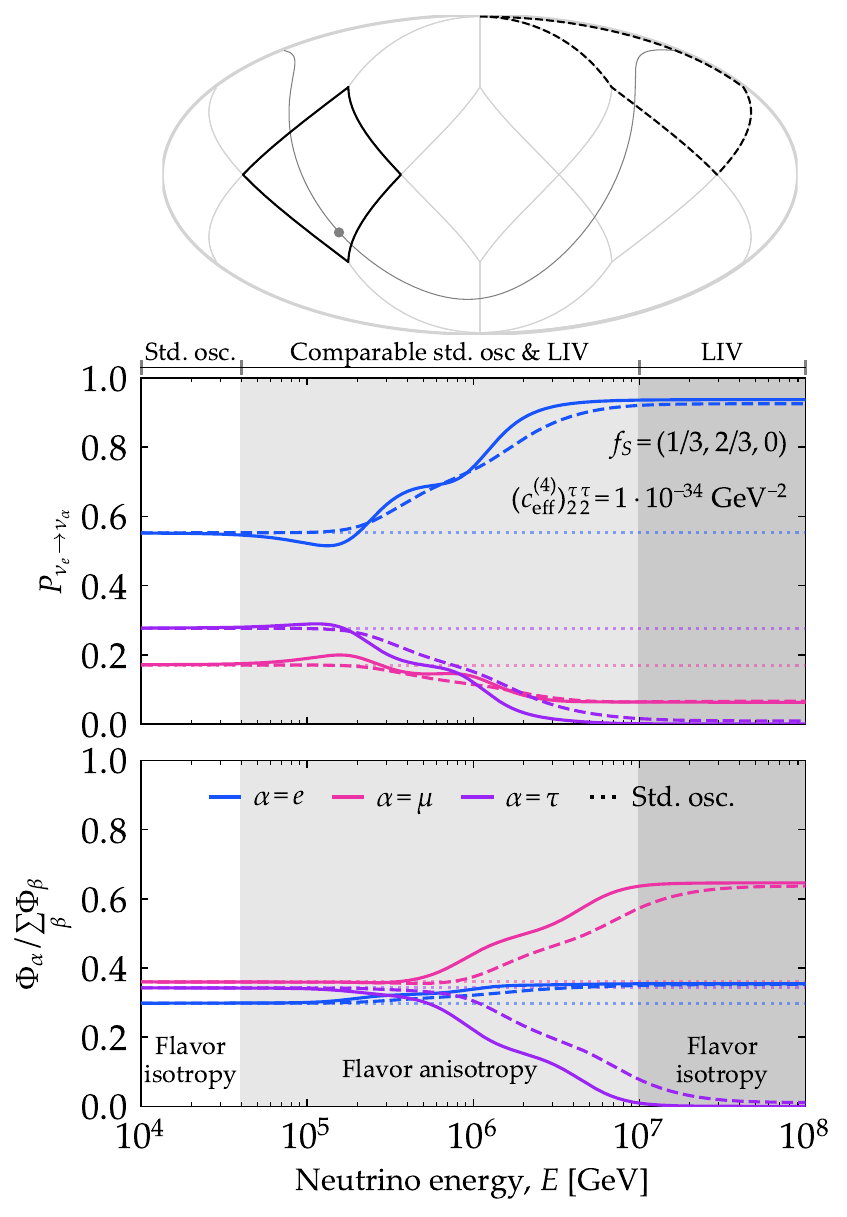}
 \caption{\textbf{\textit{Energy dependence of the LIV effects on high-energy astrophysical neutrinos.}}  \textit{Top:} Tessellated sky map highlighting the two pixels for which we show results in this figure. Solid/dashed lines in the bottom two pannels correspond to the values inside the solid/dashed pixel, respectively. \textit{Center:} Flavor-transition probability, \equ{probability}, averaged inside each of the two pixels, computed for a fixed illustrative value of the sole nonzero LIV parameter $(c_{\rm eff}^{(4)})_{22}^{\tau\tau}$.  \textit{Bottom:} High-energy neutrino flavor composition at Earth [\textit{before} the energy averaging in \equ{flavor_ratio}], also averaged inside each of the two pixels, computed with the same LIV parameter.  For this plot, we assume the flavor composition at the sources to be $\left( \frac{1}{3}, \frac{2}{3}, 0 \right)_{\rm S}$.  \textbf{\textit{Our limits on the LIV parameters come from the regimes where LIV and standard oscillations are comparable and where LIV dominates.}}  (However, we use energy-averaged flavor ratios, \equ{flavor_ratio}, to place limits.) See Secs.~\ref{sec:nu_oscillations-prob} and \ref{sec:astro_nu-flavor} for details.}\label{fig:probability_vs_energy}
\end{figure}

Thus, in our analysis we consider both the standard and LIV contributions to $H_\text{tot}$ and use the average $\nu_\alpha \to \nu_\beta$ flavor-transition probability, computed as
\begin{equation}
 \label{equ:probability}
 P_{\nu_\alpha\to\nu_\beta}
 (E, \hat{\boldsymbol{p}}, \boldsymbol{\theta}_{\rm std}, \boldsymbol{\Lambda}_{\rm LIV})
 =
 \sum_{i=1}^3 
 \vert U_{\beta i} \vert^2
 \vert U_{\alpha i} \vert^2 \;,
\end{equation}
where $U \equiv U(E, \hat{\boldsymbol{p}}, \boldsymbol{\theta}_{\rm std}, \boldsymbol{\Lambda}_{\rm LIV})$.  If $H_\text{LIV} = 0$, the flavor-transition probability becomes energy- and direction-independent, and $U = U_\text{PMNS}$, recovering standard oscillations.  The interpretation of \equ{probability} is that---as long as the total oscillation length is small---by the time the high-energy astrophysical neutrino reach Earth, the wave packets of the $\nu_i^\prime$ propagation states, which travel at different speeds, have separated enough not to interfere with one another anymore.  As a result, the flavor-transition probability depends only on the flavor content of $\nu_1^\prime$, $\nu_2^\prime$, and $\nu_3^\prime$.  However, unlike standard oscillations, the LIV contribution to the Hamiltonian, depends, in general, on the neutrino energy and direction.  

Figure~\ref{fig:probability_vs_energy} shows the evolution with energy of the $P_{\nu_e \to \nu_\alpha}$ probabilities, computed under LIV with a single nonzero LIV parameter set to a fixed value.  The probabilities are computed for two example directions in the sky.  Specifically, we show the probabilities averaged inside two example pixels of the sky tessellation~\cite{Telalovic:2023tcb} that we use in our analysis (Sec.~\ref{sec:stat_methods-inferring_flavor_composition}); the results for both are similar.  Here and below, ``tessellation'' refers to our division of the sky into $N_{\rm pix} = 12$ equal-area pixels using the HEALPix scheme~\cite{Gorski:2004by, healpix_url}. We adopt this coarse pixelation to account for the limited angular resolution of cascade events.  This is the same scheme used in \Refe~\cite{Telalovic:2023tcb} to search for flavor anisotropies; we revisit it in Sec.~\ref{sec:stat_methods-inferring_flavor_composition}.  We identify three energy regimes, which we describe below.  Other parameters show similar behavior, with the transition energies between regimes and the energies where specific features appear depending on the choice of parameter and its value.

At low energies---below a few tens of TeV for the choice of LIV parameter in \figu{probability_vs_energy}---standard oscillations dominate and, therefore, the probabilities are independent of energy.  At high energies---above about $10^7$~GeV---LIV dominates.  Because we consider a single nonzero LIV parameter in $H_{\rm LIV}$ [\equ{h_liv_matrix}], the total Hamiltonian becomes a matrix with a single, large component that saturates the LIV effects, so that increasing the energy further has no appreciable impact on the probabilities.  

Finally, at intermediate energies, standard oscillations and LIV are comparable in size.  There, the probabilities are energy-dependent.  Their interplay leads to cancellations at different energies that show up as dips and bumps in the probabilities in \figu{probability_vs_energy}.  This is also the energy regime where the directional flavor composition is anisotropic, which is one of the observables from which we extract our LIV constraints  (Sec.~\ref{sec:liv_strategies-sensitivity_origin}).  Later (Sec.~\ref{sec:liv_strategies-single_parameter}), we explain how we place constraints.


\subsection{Sidereal vs.~compass anisotropies}

In a beam of neutrinos placed on the surface of the Earth, the rotation of the Earth on its axis and around the Sun may induce \textit{sidereal} LIV-induced anisotropies, induced by the change over time of the direction of the beam relative to the direction of a background Lorentz-invariant tensor~\cite{Kostelecky:2003xn}.  Such effects have been searched for, \eg, in LSND~\cite{LSND:2005oop}, MINOS~\cite{MINOS:2008fnv, MINOS:2010kat}, MiniBooNE~\cite{MiniBooNE:2011pix}, and T2K~\cite{T2K:2017ega}, using accelerator neutrinos with GeV-scale energies, and in Double Chooz~\cite{DoubleChooz:2012eiq}, using reactor neutrinos with MeV-scale energies.

Because our analysis uses, not a beam of neutrinos fixed on Earth, but a flux of astrophysical neutrinos that arrives at Earth from all directions at all times during its orbit, it is sensitive instead to \textit{compass} LIV-induced anisotropies that are persistent in time, decoupled from the motion of Earth~\cite{Kostelecky:2003xn}.  (See, however, the definition of the Sun-centered reference frame that we adopt~\cite{Telalovic:2023tcb}.)

Because in our analysis we use experimental data made by neutrinos of all flavors (Sec.~\ref{sec:astro_nu-detection}) and treat each flavor separately in our analysis (Sec.~\ref{sec:astro_nu-flavor}), we are able to probe compass anisotropies that are flavor-dependent.  In Sec.~\ref{sec:astro_nu}, we show how they manifest in the high-energy neutrino sky.


\section{High-energy astrophysical neutrinos}
\label{sec:astro_nu}


\subsection{The diffuse high-energy neutrino flux}
\label{sec:astro_nu-flux}

In our analysis, we look for signs of LIV in the flavor composition of the diffuse flux of high-energy astrophysical neutrinos, \ie, the sum of the contributions from all sources of high-energy neutrinos. 

We compute the diffuse flux at Earth by assuming it is due to a  population of identical, nondescript extragalactic neutrino sources distributed in redshift.  Each source in it emits neutrinos with the same spectrum, which we assume to be a power law $\propto E^{-\gamma}$, where $E$ is the neutrino energy and the value of $\gamma$, the spectral index, is fixed by fits to observations.  A power-law spectrum is nominally expected for neutrino production via proton-proton interactions~\cite{Kelner:2006tc}, but in some circumstances can also approximately describe production via proton-photon interactions~\cite{Kelner:2008ke}, which typically yields bump-like neutrino spectra; see, \eg, \Refe~\cite{Fiorillo:2022rft}.  (While a power-law spectrum describes well the IceCube HESE data sample that we use~\cite{IceCube:2020wum, IC75yrHESEPublicDataRelease}, there is nascent evidence for deviations from it~\cite{Naab:2023xcz} that we, however, do not consider here.)

Yet, present-day observations are insufficient to establish what is the dominant neutrino production mechanism, whether there is more than one neutrino source population, and whether different sources emit neutrinos with different luminosities and energy spectra.  Rather than modeling  poorly constrained alternatives in detail, we opt to follow common practice and assume a single source population made up of identical neutrino sources whose internal workings we do not model in detail.  A growing number of detected high-energy neutrinos has started to offer discrimination power between competing neutrino spectra~\cite{Fiorillo:2022rft, IceCube:2023qpn}, and future versions of our analysis could be made flexible enough to account for them.

The resulting diffuse flux of $\nu_\alpha + \bar{\nu}_\alpha$ is the contribution from sources located at all redshifts [see, \eg, Eq.~(B4) in \Refe~\cite{Bustamante:2016ciw}], \ie,
\begin{equation}
\begin{split}
 \label{equ:diffuse_flux}
 \Phi_\alpha
 (E, \hat{\boldsymbol{p}}, \boldsymbol{\theta}_{\rm std}, \gamma, f_{\beta, {\rm S}}, \boldsymbol{\Lambda}_{\rm LIV})
 =
 &\frac{\Phi_0}{E^2} \int_0^{z_{\text{max}}}
 dz~
 \frac{\rho_{\rm src}(z)}{h(z)(1+z)^2} 
 [E(1+z)]^{2-\gamma} \\
 &\sum_{\beta = e, \mu, \tau} 
 P_{\nu_\beta\to\nu_\alpha}
 (E(1+z), \hat{\boldsymbol{p}}, \boldsymbol{\theta}_{\rm std}, \boldsymbol{\Lambda}_{\rm LIV}) 
 f_{\beta, \text{S}}
 \;, 
 \end{split}
\end{equation}
where $z_{\rm max} = 4$ (see why below), $\Phi_0$ is the all-flavor, sky-averaged flux normalization---whose value, in our analysis, is determined by fits to observations---$\rho_\text{src}$ is the number density of sources, $h(z) \equiv H(z)/H_0$, $H(z) \equiv H_0\sqrt{\Omega_m(1+z)^3+\Omega_\Lambda}$ is the Hubble parameter, $\Omega_m$ and $\Omega_\Lambda$ are the adimensional matter and vacuum energy densities, respectively, and $H_0$ is the Hubble constant.  The flavor composition at the sources is $(f_{e, {\rm S}}, f_{\mu, {\rm S}}, f_{\tau, {\rm S}})$, where $f_{\alpha, {\rm S}}$ is the ratio of $\nu_\alpha + \bar{\nu}_\alpha$ to the total flux emitted; we elaborate on this in Sec~\ref{sec:astro_nu-nuisance_fS}.   In what follows, we refer to $\nu_\alpha + \bar{\nu}_\alpha$ simply by $\nu_\alpha$, unless otherwise specified, since the interactions of high-energy $\nu_\alpha$ and $\bar{\nu}_\alpha$ in IceCube generate nearly identically looking events.  [The one exception is interactions with electrons, which are experienced only by $\bar{\nu}_e$ (Sec.~\ref{sec:astro_nu-detection}).]  

To produce our main results, we assume that $\rho_\text{src}$ follows the star formation rate (SFR), which we parametrize following \Refe~\cite{Yuksel:2008cu}, \ie,
\begin{equation}
 \label{equ:sfr}
 \rho_\text{src}(z)
 =
 \rho_0
 \left[
 (1+z)^{a\eta} 
 + \left(\frac{1+z}{B}\right)^{b\eta} 
 + \left(\frac{1+z}{C}\right)^{c\eta}
 \right]^{1/\eta}\; ,
\end{equation}
where we fix the constants $a$, $b$, $c$, $B$, $C$, and $\eta$ to their best-fit values from \Refe~\cite{Yuksel:2008cu}.  The SFR peaks at $z = 1$, and falls by a factor of $10^{-5}$ at our assumed value of the maximum redshift in \equ{diffuse_flux}, $z_\text{max} = 4$.  Contributions from higher redshifts are negligible.  In our computation of the neutrino flux, the local SFR is absorbed into the value of the flux normalization, $\Phi_0$ in \equ{diffuse_flux}, so we set $\rho_0 = 1$ without loss of generality.  Later (Sec.~\ref{sec:astro_nu-nuisance_redshift}), we show the effect of adopting a different choice for $\rho_\text{src}$.

The sum on the right-hand side of \equ{diffuse_flux} accounts for the flavor transitions of neutrinos en route to Earth.  In it, $P_{\nu_\beta \to \nu_\alpha}$ is the average $\nu_\beta \to \nu_\alpha$ flavor-transition probability, \equ{probability}---including possible LIV effects---and $f_{\beta, {\rm S}}$ is the fraction of $\nu_{\beta}$ emitted by the sources, which depends on the neutrino production mechanism; we elaborate on it in Sec.~\ref{sec:astro_nu-nuisance_fS}.  Interaction with matter in the sources likely does not modify the flavor composition~\cite{Mena:2006eq, Razzaque:2009kq, Sahu:2010ap, Varela:2014mma, Xiao:2015gea} (but see \Refe~\cite{Dev:2023znd}), and there is interaction with matter en route to Earth due to the negligible column densities traversed by the neutrinos.

The standard calculation of the diffuse flux in \equ{diffuse_flux}, \ie, computed without including LIV in it, yields isotropic fluxes for the individual neutrino flavors and for the all-flavor flux, $\sum_\alpha \Phi_\alpha$.  Our calculation does not capture the sub-dominant contribution to the diffuse flux from high-energy neutrinos from the Galactic Plane, which is about 10\% in the 10--100~TeV range~\cite{IceCube:2023ame, Bustamante:2023iyn}, because publicly available information about it is presently inadequate to model this contribution in detail.  While this is not a serious concern when performing our analysis on present-day IceCube data---which is limited in size---it should be accounted for in future revised versions of our analysis once more data become available.

Below, we show that including anisotropic LIV in \equ{diffuse_flux} makes the diffuse fluxes of individual neutrino flavors anisotropic---which we use to constrain LIV---while preserving the isotropy of the all-flavor flux.


\subsection{Neutrino flavor composition at Earth}
\label{sec:astro_nu-flavor}

Even without the presence of LIV, the diffuse flux of high-energy neutrinos might not be fully isotropic, as evidenced by the detection of neutrinos from the Galactic Plane~\cite{IceCube:2023ame}.  Even after subtracting the Galactic component from the total diffuse flux, however, the remaining extragalactic flux might retain small-scale anisotropy, reflecting the angular distribution of extragalactic neutrino sources in the sky.  Thus, blindly attributing to LIV any anisotropy observed in the (all-flavor) diffuse neutrino flux could be misleading.

We circumvent this risk by adopting a different strategy, focusing instead on the flavor composition of the diffuse flux at Earth, $f_{\alpha, \oplus}$, \ie, the proportion of $\nu_\alpha$ in it (we define it precisely below).  This is because, even if the all-flavor diffuse flux were to be anisotropic due to the angular distribution of the sources, it is less likely that the neutrino production mechanism inside sources varies significantly across the sky, \ie, that the flavor composition at the sources, $f_{\beta, {\rm S}}$, is anisotropic.  Thus, any anisotropy seen in the flavor composition at Earth could more safely be attributed to the effects of LIV.

Figure~\ref{fig:probability_vs_energy} shows the evolution with energy of the flavor ratios, $\Phi_\alpha / \sum_\beta \Phi_\beta$,  where the diffuse flux, $\Phi_\alpha \equiv \Phi_\alpha(E, \hat{\boldsymbol{p}}, \boldsymbol{\theta}_{\rm std}, \gamma, \boldsymbol{\Lambda}_{\rm LIV})$, is computed using \equ{diffuse_flux}.  In the case of \figu{probability_vs_energy}, there is a single nonzero LIV parameter set at a fixed value and, like in the flavor-transition probabilities shown in the same figure, the flavor ratios are averaged inside two example pixels of our sky tessellation.  The behavior of the flavor ratios reflects that of the probabilities, exhibiting the same transitions from standard-oscillation and LIV dominance.  It is from the intermediate energies, where standard oscillations and LIV effects are comparable, that we expect flavor anisotropy.  (Figure~\ref{fig:probability_vs_energy} shows this only for two example pixels. 
 Later, in \figu{energy_resonance}, we show how the amount of flavor anisotropy evolves with energy.)

Since there is presently no sensitivity to infer the energy dependence of the flavor composition~\cite{Liu:2023flr}, in our analysis we use instead the energy-averaged flavor ratios at Earth.  For $\nu_\alpha + \bar{\nu}_\alpha$ of $\nu_\alpha$, this is 
\begin{equation}
 \label{equ:flavor_ratio}
 f_{\alpha,\oplus}(\hat{\boldsymbol{p}}, \boldsymbol{\theta}_{\rm std}, \gamma, f_{\beta, {\rm S}}, \boldsymbol{\Lambda}_{\rm LIV}) 
 =
 \frac{\int dE \, \Phi_\alpha}
 {\int dE \, \sum_\beta \Phi_\beta} \; .
\end{equation}
We integrate over a broad energy range of $10^4$--$10^8$~GeV that encloses that of the IceCube HESE events~\cite{IceCube:2020wum}.  The flavor ratios are independent of the all-flavor flux normalization, $\Phi_0$ in \equ{diffuse_flux}.  They do depend on the standard oscillation parameters---whose values are known relatively well (Table~\ref{tab:fit_params})---and, most sensitively, on the far more weakly constrained values of the LIV parameters and the flavor composition at the sources; we elaborate on this below.  

\begin{figure*}[t!]
 \centering
 \includegraphics[trim={0.75cm 0.50cm 0.5cm 0cm}, clip, width=\textwidth]{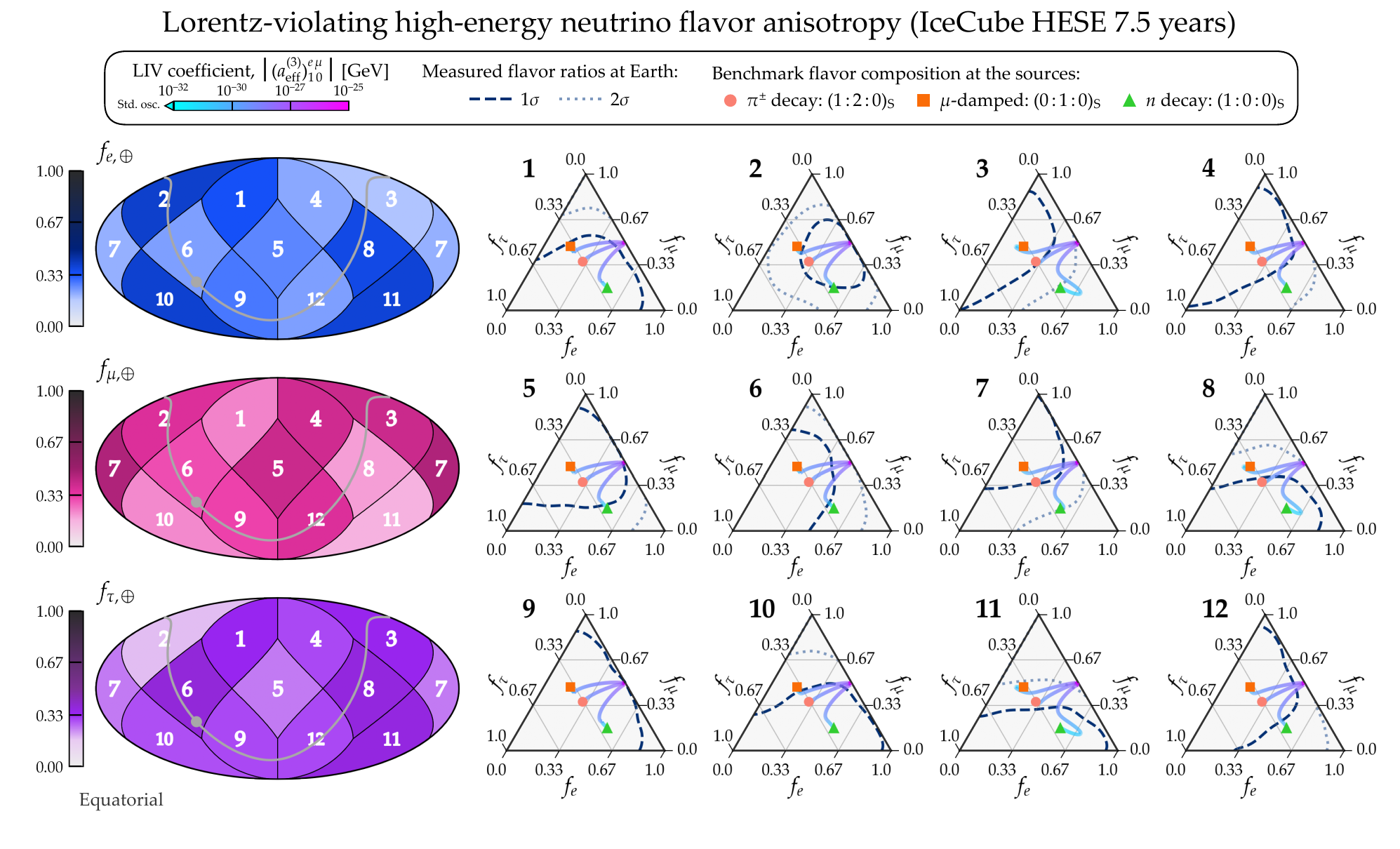}
 \caption{\textbf{\textit{Directional flavor composition of high-energy astrophysical neutrinos at Earth, as predicted under Lorentz-invariance violation.}}  In this plot, we compute the energy-averaged flavor composition at Earth, \equ{flavor_ratio}, by varying a single, illustrative LIV parameter, $(a^{(3)}_{\rm eff})_{10}^{e\mu}$, inside each pixel of our sky tessellation (Sec.~\ref{sec:stat_methods-inferring_flavor_composition}).  We show results obtained under our three benchmark choices of flavor composition at the sources (Sec.~\ref{sec:astro_nu-nuisance_fS}).  We also show the measured flavor composition in each pixel, extracted from the public 7.5-year IceCube HESE sample~\cite{IceCube:2020wum, IC75yrHESEPublicDataRelease} in \Refe~\cite{Telalovic:2023tcb}.  \textbf{\textit{Our constraints on the LIV parameters come from contrasting our predictions for the flavor composition at Earth \textit{vs.}~the measurement from IceCube data.}}  See Secs.~\ref{sec:astro_nu-flavor}, Sec.~\ref{sec:stat_methods}, and Sec.~\ref{sec:results}, respectively, for details about the prediction of flavor composition at Earth, the measurement of the directional flavor composition from IceCube data, and the constraints we set on the LIV parameters.} 
 \label{fig:all_triangles}
\end{figure*}

Figure~\ref{fig:all_triangles} shows the energy-averaged flavor composition at Earth, \equ{flavor_ratio}, inside each pixel of our sky tessellation, as predicted by varying a single illustrative nonzero anisotropic LIV parameter, $(a_{\rm eff}^{(3)})_{10}^{e\mu}$.  (Inside each pixel, the flavor ratios are averaged in direction; we defer to Sec.~\ref{sec:stat_methods-constraining_liv_parameters} for details.)  We show predictions made under each of our three benchmark choices of flavor composition at the sources---pion decay, muon-damped, and neutron decay---which we introduce in Sec.~\ref{sec:astro_nu-nuisance_fS}.  

Figure~\ref{fig:all_triangles} illustrates three behaviors that are common to all the LIV parameters that we study.  First, when the LIV parameter is small, the flavor composition reverts to the expectation from standard oscillations.  Second, as the LIV parameter grows, the features in the trajectory that the flavor composition traces inside each pixel---the wiggles---reflect the same interplay between standard oscillations and LIV shown in \figu{probability_vs_energy} (shown there for a different choice of LIV parameter).  Because the LIV parameter varied in \figu{all_triangles} introduces anisotropic effects, these features are different in the different pixels.

Third, when the LIV parameter is large, dominant LIV makes the flavor composition at Earth the same, approximating $\left( \frac{1}{2}, \frac{1}{2}, 0 \right)_\oplus$, regardless of what is the flavor composition at the sources.  This can be understood from the shape of the LIV Hamiltonian, \equ{h_liv_matrix}: since only the $(a_{\rm eff}^{(3)})_{10}^{e\mu}$ component (and, by Hermiticity, also $(a_{\rm eff}^{(3)})_{10}^{\mu e}$) is nonzero, when LIV is dominant only transitions between $\nu_e$ and $\nu_\mu$ can occur---and they are as large as possible, yielding maximum mixing between them.  And, because our benchmark choices of flavor composition at the sources have null $\nu_\tau$ content (\ie, $f_{\tau, {\rm S}} = 0$), it remains null also at Earth.  Different LIV parameters predict different flavor composition at Earth under LIV dominance, but this same underlying explanation holds. 

(When varying multiple nonzero LIV parameters jointly, as in Sec.~\ref{sec:liv_strategies-multiple_parameters}, the above observations still hold broadly, except that, depending on which parameters are varied, there could be transitions between three flavors.)

It is from the comparison between the predicted flavor ratios at Earth \textit{vs}.~the flavor composition inferred from IceCube data in each pixel that we extract our constraints on the LIV parameters.  In \figu{all_triangles}, these inferred compositions are shown as contours representing the Bayesian posterior probability distributions (68\% and 95\% credible regions) derived from the HESE event topology in each pixel (detailed in Sec.~\ref{sec:stat_methods-inferring_flavor_composition}).  We elaborate on how the posteriors are computed later, in Sec.~\ref{sec:stat_methods}.


\subsection{The neutrino flavor composition at the sources}
\label{sec:astro_nu-nuisance_fS}

The flavor composition at the sources, $(f_{e, {\rm S}}, f_{\mu, {\rm S}}, f_{\tau, {\rm S}})$ in \equ{diffuse_flux}, is determined by the neutrino production mechanism, which, in turn, depends on the physical conditions present in the production regions.  Because the identity of the bulk of the neutrino sources is unknown, the flavor composition at the sources is only weakly constrained by current IceCube observations~\cite{Bustamante:2019sdb, Song:2020nfh}.  It is one of the largest sources of uncertainty in our analysis, as the LIV effects on the predicted flavor composition at Earth depend strongly on it.  Later (Sec.~\ref{sec:results}), we show that this results in a degeneracy between the inferred values of the flavor composition at the sources and the LIV parameters in our statistical analysis.

\smallskip

\textbf{\textit{Scenarios of flavor composition.---}}In lieu of modeling in detail different models of neutrino production, we consider three benchmark scenarios considered often in the literature---pion decay, muon-damped, and neutron decay (\eg, \Refes~\cite{Barenboim:2003jm, Pakvasa:2007dc, Lipari:2007su, Arguelles:2015dca, Bustamante:2015waa, Bustamante:2019sdb})---and one general, \textit{flavor-agnostic} scenario (\eg, \Refes~\cite{Bustamante:2015waa, Rasmussen:2017ert, Song:2020nfh}) where we marginalize over all likely possibilities and under which we produce our main results later.
\begin{itemize}
 \item
  \textbf{Pion decay:} Inside our nondescript astrophysical neutrino sources, high-energy protons with energies of tens of PeV interact with surrounding matter and photons and produce charged pions (and less quantities of other mesons). The pions then decay to TeV--PeV neutrinos  via $\pi^+ \to \mu^+ + \nu_\mu$ and $\mu^+ \to e^+ + \bar{\nu}_\mu + \nu_e$.  If this is the primary mechanism of neutrino production, the expected flavor composition at the sources would be $\left( \frac{1}{3}, \frac{2}{3}, 0 \right)_{\rm S}$. 
 \item
  \textbf{Muon-damped:} If the astrophysical sources harbor strong magnetic fields, the above intermediate muons might cool significantly via synchrotron radiation before they decay \cite{Kashti:2005qa}.  In this case, the neutrinos from muon decay would be produced at lower energies than the neutrino from pion decay, so that the high-energy flavor composition would be $(0,1,0)_{\rm S}$.  (We ignore the possibility that intermediate mesons are re-accelerated~\cite{Winter:2014tta, Kawanaka:2015qza}.)
 \item
  \textbf{Neutron decay:} The beta-decay of high-energy neutrons and neutron-rich isotopes into $\bar{\nu}_e$ would yield a flavor composition of $(1,0,0)_{\rm S}$.  However, this scenario is unlikely to be the dominant one because the neutrinos from beta decay receive only a small fraction of the parent neutron energy---on account of the proton and neutron masses being very similar---making their production reliant on significantly higher-energy neutrons, which are scarce. 
 \item
  \textbf{Flavor-agnostic:} We consider a generic flavor composition at the sources of the form $(f_{e, {\rm S}}, 1-f_{e, {\rm S}}, 0)$, parametrized only by a single flavor ratio, $f_{e, {\rm S}}$, whose value we let float freely between 0 and 1 in our statistical analysis below.  Upon doing so, the above benchmark scenarios are allowed when $f_{e, {\rm S}}$ takes on specific values: $f_{e, {\rm S}} = 1/3$ for pion decay, 0 for muon-damped, and 1 for neutron decay.  We fix $f_{\tau, {\rm S}} = 0$ because $\nu_\tau$ emission requires the production of charmed mesons that are made in negligible quantities (see, \eg, \Refe~\cite{Farzan:2021gbx}).
\end{itemize}
Figure~\ref{fig:all_triangles} shows the flavor composition at Earth, \equ{flavor_ratio}, that is expected from the pion decay, muon-damped, and neutron decay benchmarks, computed under standard oscillations and under LIV.  References~\cite{Bustamante:2015waa, Song:2020nfh} show the flavor composition at Earth expected under the flavor-agnostic scenario (see also \Refe~\cite{Arguelles:2015dca}).

In each of the above cases, we make three assumptions, on the energy dependence, the number of source populations, and the isotropy of the neutrino emission, all of which are physically sound and common in the literature.

\smallskip

\textbf{\textit{Energy dependence.---}}Different production mechanisms and energy-loss processes may be present at different energies, making the flavor composition at the sources vary with energy~\cite{Kashti:2005qa}, including possibly transitioning from pion decay, at low energy, to muon-damped, at high energy.  However, \Refe~\cite{Liu:2023flr} showed that there is presently no sensitivity to test this possibility, and that it might be challenging to test it in the future.  Thus, we consider the neutrino flavor composition at the sources to be constant in energy, as is commonly done in the literature; see \eg, \Refe~\cite{Bustamante:2015waa}.  Any energy dependence that the flavor composition at the Earth might have (prior to our averaging it over energy) is due solely to the effects of LIV.

\smallskip

\textbf{\textit{A single source population.---}}Similarly, it is possible that different populations of sources emit neutrinos with different flavor composition.  The flavor composition measured at Earth might be due to a single population or to multiple ones---in effect, the measurement is of the population-averaged flavor composition.  This encompasses both averaging over the variation in the properties of the sources within their population and averaging over potentially multiple source populations.  However, there is presently no sensitivity in IceCube data to test this, though there might be in the future~\cite{Song:2020nfh}.  Thus, in our analysis, we compute the diffuse flux as being due to a single population of extragalactic sources that emit neutrinos with a common flavor composition.

\smallskip

\textbf{\textit{Flavor-isotropic neutrino emission.---}}Regardless of what is the flavor composition at the sources, we assume that it is uniform across the sky, \ie, that the flavor composition with which neutrinos are emitted is isotropic.  In other words, there is no special direction in the sky along which neutrinos are emitted preferentially with a certain flavor composition.  Under our assumption of the diffuse flux being due to a single source population, this means that even if the positions of the sources were not distributed isotropically across the sky, the flavor composition with which they emit neutrinos is, since the neutrino production mechanism is the same in all of sources.  (Our assumption of flavor isotropy would hold even if there were multiple source populations, as long as the angular distribution of sources in all populations is isotropic.)  Any anisotropy that the flavor composition at the Earth might have is due solely to the effects of LIV.

\smallskip

\begin{figure}[t!]
 \centering
 \includegraphics[width = 0.75\textwidth]{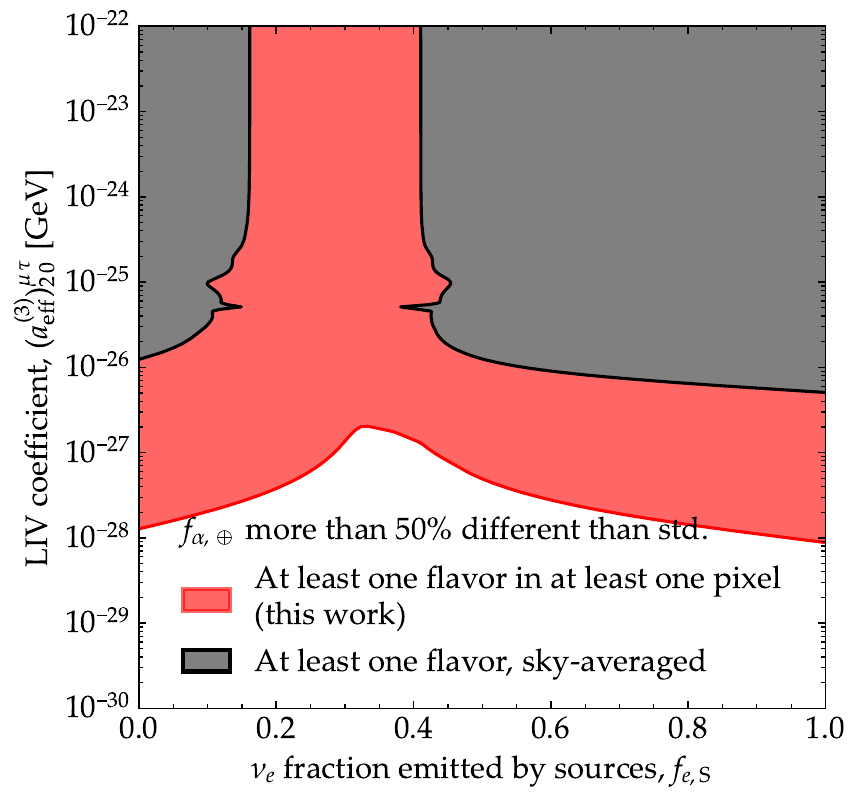}
 \caption{\textbf{\textit{Impact of the neutrino flavor composition at the sources on the LIV effects.}}  For this plot, we show results for an illustrative LIV parameter, $(a_{\rm eff}^{(3)})_{20}^{\mu\tau}$. The astrophysical neutrino sources responsible for the diffuse flux emit high-energy neutrinos with flavor composition $(f_{e, {\rm S}}, 1-f_{e, {\rm S}}, 0)$.  For each choice of values of $f_{e, {\rm S}}$ and $(a_{\rm eff}^{(3)})_{20}^{\mu\tau}$, we compute the expected neutrino flavor composition at Earth, \equ{flavor_ratio}, and find where it is different from the standard-oscillation expectation by more than 50\%, either in at least one pixel our sky tessellation---representing how we place constraints on LIV parameters---or averaged across the sky.  \textbf{\textit{We rely on spotting differences in the flavor composition in different directions of the sky to boost the sensitivity to anisotropic LIV effects.}} See Sec.~\ref{sec:astro_nu-nuisance_fS} for details.}
 \label{fig:constrainability_with_fes}
\end{figure}

\textbf{\textit{Milder impact of the flavor composition at the sources.---}}Figure \ref{fig:constrainability_with_fes} illustrates how significant the choice of flavor composition at the sources is on the effect that LIV has on the flavor composition at Earth.  Assuming a flavor composition of $(f_{e, {\rm S}}, 1-f_{e, {\rm S}}, 0)$, we vary the value of $f_{e, {\rm S}}$ and of an example anisotropic LIV parameter, $(a_{\rm eff}^{(3)})_{20}^{\mu\tau}$ (other LIV parameters show similar behavior).  For each choice of their values, we compute the predicted energy-averaged flavor composition at Earth, \equ{flavor_ratio}, and contrast it against the expectation from standard oscillations, identifying when the deviations from it are larger than 50\% for at least one flavor.  Doing this reveals how powerful using the directional flavor distribution is when constraining anisotropic LIV effects---as expected.

Figure \ref{fig:constrainability_with_fes} shows that, for the anisotropic LIV parameter $(a_{\rm eff}^{(3)})_{20}^{\mu\tau}$, no large deviations (\ie, more than 50\% relative to the standard-oscillation expectation) occur in the sky-averaged flavor composition when $0.2 \lesssim f_{e, {\rm S}} \lesssim 0.4$, even if the LIV parameter is large.  The specific range of $f_{e, {\rm S}}$ where the sensitivity to LIV is lost  varies for different choice of LIV parameter; see, \eg, \Refe~\cite{IceCube:2021tdn}.  However, even when LIV-induced deviations in the sky-averaged flavor composition are small, deviations due to anisotropic LIV parameters can be large along specific directions.  The underlying reason is that, while the spherical harmonics average to zero across the sky, they evaluate to nonzero values in specific directions.  

Indeed, \figu{constrainability_with_fes} shows that the window of $f_{e, {\rm S}}$ where sensitivity to LIV is lost using the sky-averaged flavor composition is reclaimed using the directional flavor composition.  This behavior illustrates where the sensitivity to anisotropic LIV parameters stems from in our work.  In our statistical methods (Sec.~\ref{sec:stat_methods}), we look inside each of the pixels of our sky tessellation (see, \eg, \figu{all_triangles}) for appreciable deviations of the flavor composition relative to the standard-oscillation expectations.  In addition to patching the loss of sensitivity that we had when using the sky-average flavor composition, this makes our analysis more sensitive to smaller values of the LIV parameter, since now it must be only large enough to affect appreciably individual pixels, not only the entire sky.

Therefore, because the impact of $f_{e, {\rm S}}$ on the directional flavor composition at Earth under anisotropic LIV is relatively mild, we are justified in making our default results the ones obtained under the flavor-agnostic scenario (see above), where we marginalize over $0 \leq f_{e, {\rm S}} \leq 1$.

\subsection{Influence of the neutrino energy spectrum}
\label{sec:astro_nu-nuisance_energy}

Section~\ref{sec:nu_oscillations-prob} showed that, under standard oscillations, the average flavor-transition probability, \equ{probability}, is independent of neutrino energy.  As a consequence, so is the flavor composition at Earth, $\Phi_\alpha / \sum_\beta \Phi_\beta$, with the diffuse flux given by \equ{diffuse_flux}.   [This is also because we assume the flavor composition at the sources to be independent of energy (Sec.~\ref{sec:astro_nu-nuisance_fS}).]  

Under LIV, the flavor-transition probability becomes energy-dependent and so does the flavor composition at Earth prior to averaging it over energy (see \figu{probability_vs_energy}).  When computing the energy-averaged flavor composition at Earth, $f_{\alpha, \oplus}$ in \equ{flavor_ratio}, its energy dependence of the latter is determined not only by that of the LIV operator, via the probability, but also by the energy spectrum with which the sources emit neutrinos.  In our computation of the diffuse flux,  \equ{diffuse_flux}, the latter is a power law $\propto E^{-\gamma}$, with $\gamma > 0$.  This makes lower-energy neutrinos more abundant than higher-energy ones, their relative abundance set by the value of $\gamma$, which we vary in our statistical analysis later.  

The steeper the spectrum, \ie, the higher the value of $\gamma$, the more abundant lower-energy neutrinos are \textit{vs.}~higher-energy ones.  This relative abundance matters because, for LIV operators with dimension $d > 2$, the LIV effects grow with energy $\propto E^{d-3}$ relative to the standard-oscillation contribution, which is $\propto 1/E$. As a consequence, the LIV effects in the diffuse neutrino flux at Earth are more prominent when there are relatively more high-energy neutrinos in it, \ie, when $\gamma$ is lower.  Because we use energy-averaged flavor ratios, these effects are not as stark, but they have residual influence.  

\begin{figure}
 \centering
 \includegraphics[width=0.5\linewidth]{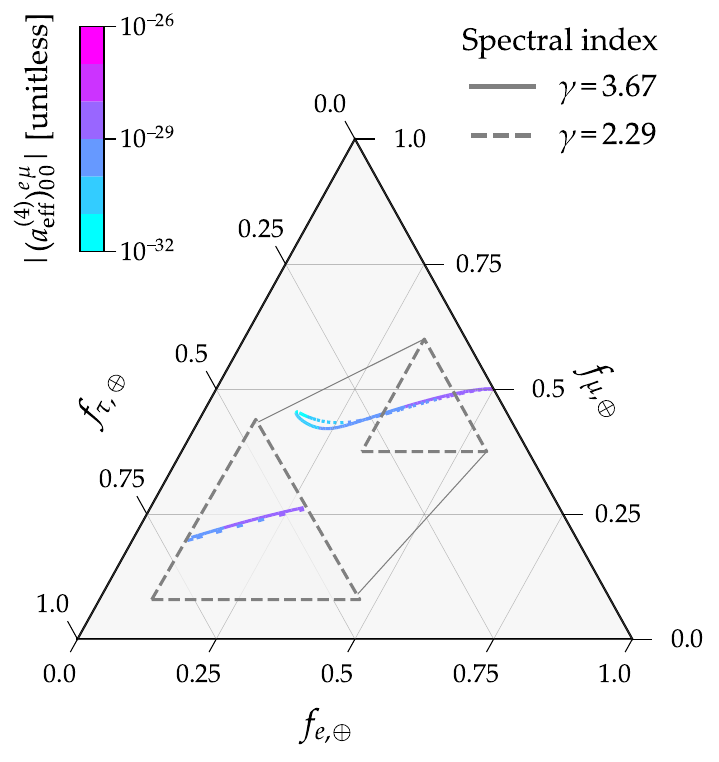}~
 \includegraphics[width=0.5\linewidth]{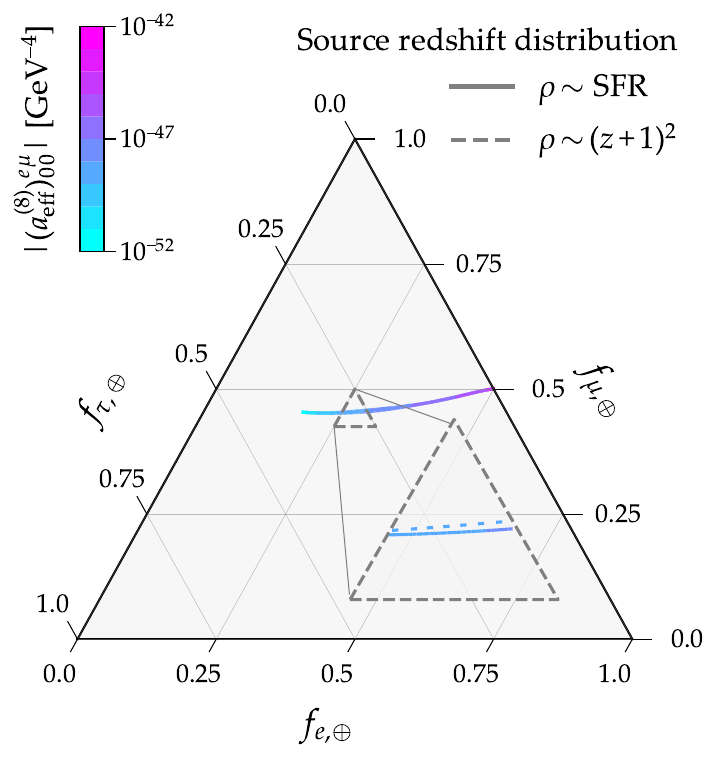}
 \caption{\textbf{\textit{Influence of the spectral index (left) and source redshift distribution (right) on the energy-averaged flavor composition at Earth under LIV.}} We compute the energy-averaged flavor composition at Earth, \equ{flavor_ratio}. Left, assuming a power-law neutrino energy spectrum with two different values of the spectral index, $\gamma = 3.67$ and 2.29, and varying the single nonzero isotropic LIV parameter, $(a_{\rm eff}^{(4)})_{00}^{e\mu}$.  Because, in this case, LIV is isotropic, the flavor composition in this plot is the same everywhere in the sky.  For comparison, from the public 7.5-year IceCube HESE sample, $\gamma = 2.89 \pm 0.23$~\cite{IceCube:2020wum}.  Right, assuming that the number density of sources follows either the star-formation rate (SFR), which peaks at redshift $z = 1$, or is $\propto (1+z)^2$, which peaks at $z = z_{\rm max} = 4$, the maximum redshift in the computation of the diffuse neutrino flux, \equ{diffuse_flux}.  We vary the value of the single nonzero isotropic LIV parameter, $(a_{\rm eff}^{(8)})_{00}^{e\mu}$.\textbf{\textit{The uncertainties in the values of the spectral index and source redshift distribution have negligible effects on the flavor composition at Earth and, therefore, on the constraints we place on LIV parameters.}}}
 \label{fig:LIV_vs_gamma}
\end{figure}

Figure \ref{fig:LIV_vs_gamma} illustrates this via two different predictions of the energy-averaged flavor composition at Earth, computed for two different values of the spectral index, as we vary the value of one illustrative isotropic LIV parameter, $(a_{\rm eff}^{(4)})_{00}^{e\mu}$.  The differences in the predictions are negligible compared to the uncertainty with which the flavor composition is measured by IceCube; see, \eg, \Refe~\cite{IceCube:2020fpi}.  The differences are also negligible for other LIV parameters, including anisotropic ones.  Nevertheless, we vary $\gamma$ as one of the free parameters in our statistical analysis. 

\subsection{Influence of the source redshift distribution}
\label{sec:astro_nu-nuisance_redshift}

Although the identity of the astrophysical sources responsible for the diffuse high-energy neutrino flux remains unknown, it is reasonable to assume that they are predominantly extragalactic sources located at cosmological-scale distances away from us, from tens of Mpc up to a few Gpc.  

Because of the cosmological expansion, a neutrino emitted with energy $E$ by a source located at redshift $z$ reaches the Earth with energy $E/(1+z)$.  Collectively, this effect stretches the neutrino energy spectrum.  The distribution of sources in redshift determines how large this effect is.  In our work, we compute the diffuse neutrino flux at Earth, \equ{diffuse_flux}, assuming that the number density of sources follows the SFR (see Sec.~\ref{sec:astro_nu-flux}), which peaks at $z \approx 1$, and integrate the contributions of sources up to $z_{\rm max} = 4$.  

While it is reasonable to expect that neutrinos are produced at redshifts where there is a large number of candidate sources---which is what the SFR traces---there is little to no actual experimental evidence for the redshift distribution of neutrino sources.  At best, studies between the incoming direction of IceCube neutrinos and the distribution of large-scale structure reveal a weak correlation with structures that lie at $z = 1$~\cite{Ouellette:2024ggl}; see also \Refe~\cite{IceCube:2022ham, Zhou:2024kzp}.  However, it turns out that the uncertainty in the source redshift distribution is not really a limiting factor in our work.

Figure \ref{fig:LIV_vs_gamma} illustrates this via two different predictions of the energy-averaged flavor composition at Earth, computed for two different source redshift distributions---SFR and a simpler one $\propto (1+z)^2$---as we vary the value of one illustrative isotropic LIV parameter, $(a_{\rm eff}^{(8)})_{00}^{e\mu}$.  While the SFR peaks at $z = 1$, the alternative peaks at $z = z_{\rm max} = 4$ (where the SFR has dropped by a factor of about $10^{-5}$ relative to its peak value).  In spite of this large contrast, the differences in the flavor composition in \figu{LIV_vs_gamma} between the two redshift distributions are negligible compared to the uncertainty with which the flavor composition is measured by IceCube.  The differences are equally negligible for other LIV parameters, including anisotropic ones.

This is because, as long as the sources lie at cosmological-scale distances, regardless of their precise redshift distribution, the ``oscillation'' length associated to LIV, \equ{osc_length_liv}, is tiny compared to the distanced traveled by the neutrinos to Earth.  As a result, the LIV effects on the flavor composition saturate quickly after neutrino emission, regardless of the  redshift distribution.  This justifies why, in our work, we keep the source redshift evolution fixed to the SFR.


\subsection{Neutrino detection at IceCube}
\label{sec:astro_nu-detection}

IceCube is the largest neutrino telescope in operation.  It consists of about 1~km$^3$ of clear Antarctic ice, instrumented by thousands of photomultipliers (PMTs) buried at depths beyond 1.5~km.  IceCube is a Cherenkov detector: the PMTs collect the Cherenkov light emitted by the fast-moving charged particles produced when high-energy neutrinos collide with the ice.  From the amount of light deposited in the ice, and from its spatial and temporal profiles, it is possible to infer the energy, direction, and flavor of the neutrino that interacted.

\smallskip

\textbf{\textit{HESE events.---}}In our work, we use IceCube High-Energy Starting Events (HESE).  These are events where the neutrino interacts inside the instrumented detector volume.  The outermost layer of PMTs acts as a self-veto that mitigates the otherwise overwhelming background of atmospheric neutrinos~\cite{Schonert:2008is, Gaisser:2014bja}.  As a result, samples of detected HESE events have a high content of astrophysical neutrinos~\cite{Beacom:2004jb, IceCube:2013low}.  Further, HESE events are made by neutrinos of all flavors, which is why they are used to infer the flavor composition of the diffuse high-energy neutrino flux---sometimes complemented by through-going muons to tighten the measurement of the $\nu_\mu$ content.  In high-energy neutrino telescopes, events made by neutrinos and anti-neutrinos of the same flavor are currently indistinguishable from one another.

There are three types of HESE events: cascades, tracks, and double cascades.  Cascades are made primarily by the charged-current interaction of $\nu_e$ and $\nu_\tau$, and also by the neutral-current interaction of neutrinos of all flavors.  Their light profiles are roughly spherical, expanding radially outwards from the neutrino interaction vertex, which gives them poor pointing resolution. 

Tracks are made by the charged-current interaction of $\nu_\mu$, which produces a final-state energetic muon whose propagation range exceeds 1~km.  As it propagates, it leaves a long track of Cherenkov light in its wake that is easily identifiable.  Because of this, the direction of tracks can be reconstructed precisely, to less than $1^\circ$ at the highest energies.  There is also a sub-dominant contribution of tracks made by the 17\% of charged-current $\nu_\tau$ interactions that create a tau that decays into a muon.

Double cascades~\cite{Learned:1994wg, Beacom:2003nh, Bugaev:2003sw} are made by the charged-current interaction of $\nu_\tau$ where two causally connected cascades are identified.  The first cascade is from the $\nu_\tau$ interaction with the ice.  This interaction creates a final-state tau whose decay, some distance away, creates a second cascade.

HESE events have a resolution on their energy, $E_{\rm evt}$, of about 10\% in $\log_{10}(E_{\rm evt}/{\rm GeV})$~\cite{IceCube:2013dkx}.   Because a large part of  the shower energy is deposited in the detector, HESE events offer good resolution to reconstruct the energy of the parent neutrino.  The angular resolution varies depending on the type of event.  In the 7.5-year IceCube HESE sample that we use, cascades, tracks, and double cascades have a median resolution in zenith angle of about $6.3^\circ$, $1.5^\circ$, and $5.0^\circ$, respectively~\cite{IceCube:2020wum}.  However, there is a large spread in the angular resolution, especially for cascades, some of which can be as bad as tens of degrees.  The angular uncertainty of HESE events limits the size of the LIV-induced features that we can look for in the skymap of flavor composition~\cite{Telalovic:2023tcb}.

\smallskip

\textbf{\textit{Measuring flavor composition.---}}Because showers are made by neutrinos of all flavors---though most likely by $\nu_e$ or $\nu_\tau$---and tracks are made by $\nu_\mu$ and, sometimes, $\nu_\tau$, it is not possible to indisputably infer the flavor of any single detected event (other than double cascades).  

Instead, what is measured is the flavor composition of the  diffuse flux of astrophysical neutrinos, \ie, $f_{e, \oplus}$, $f_{\mu, \oplus}$, and $f_{\tau, \oplus}$.  The ratios are inferred from comparing the relative numbers of cascades, tracks, and double cascades in a sample of detected HESE events~\cite{Beacom:2003nh}.  This is the strategy adopted by analyses performed within~\cite{IceCube:2015rro, IceCube:2015gsk, IceCube:2020fpi} and without~\cite{Mena:2014sja, Palomares-Ruiz:2015mka, Vincent:2016nut, Liu:2023flr} the IceCube Collaboration, including ours.  The main difference between our analysis and previous ones is that ours allows for the flavor composition to have directional dependence.

In addition to the flavor degeneracies outlined above, events of different types are occasionally mis-identified, \eg, a bright track segment could be mis-identified as a cascade and the two cascades of a double cascade could be mis-identified as a single one if they are too close.

To account for the above complications, \Refe~\cite{Telalovic:2023tcb}, on whose results we base ours, used the Monte Carlo sample of simulated IceCube HESE events provided by the Collaboration~\cite{IC75yrHESEPublicDataRelease} to compute expected event rates.  By using the Monte Carlo sample, we adopt a nuanced description of the detector response.  Further, we account for the propagation of neutrinos inside Earth, which attenuates the flux that reaches the detector in an energy-, direction-, and flavor-dependent manner.  We contrast our predictions against the 7.5-year HESE sample in order to extract the directional flavor composition.  Section~\ref{sec:stat_methods-inferring_flavor_composition} outlines the method used by \Refe~\cite{Telalovic:2023tcb} to extract the directional flavor composition.


\section{LIV exploration strategies}
\label{sec:liv_strategies}


\subsection{Where does our sensitivity come from?}
\label{sec:liv_strategies-sensitivity_origin}

The origin of the constraint placed on an LIV parameter depends on the degree, $\ell$, of the parameter. Parameters with smaller values of $\ell$ introduce large-angle features in the flavor-composition skymaps (though the features trickle down to smaller sizes, too) and so can be captured even by the coarse sky tessellation that we use (Sec.~\ref{sec:stat_methods-inferring_flavor_composition}).  In this case, our constraints on the LIV parameter comes predominantly from the anisotropy that it induces in the flavor-composition skymap.

For larger values of $\ell$, approximately $\ell > 3$, the LIV-induced features become too small to be resolved by our coarse sky tessellation.  This depends not only on the value of $\ell$, but also on the size of the LIV parameter: large values of one LIV parameter suppress anisotropies in the flavor-composition skymap, regardless of the value of $\ell$ (see discussion below).  The resulting skymap is effectively isotropic, though does not necessarily have the same expectation as that from standard oscillations.  In this case, our constraints on LIV parameters come from contrasting the nearly isotropic flavor-composition skymap under LIV \textit{vs.}~the isotropic skymap from standard oscillations (the comparison is still performed by comparing pixel-by-pixel). 

The separation above is merely pedagogical: when constraining an LIV parameter, we do not pick between the two methods above.  Instead, our statistical procedure (Sec.~\ref{sec:stat_methods}) tacitly incorporates both methods, pulling what sensitivity is available from each of them.


\subsection{The high-dimensional LIV parameter space}
\label{sec:liv_strategies-high_dim}

Although there are many free LIV parameters that affect neutrinos in the SME (Sec.~\ref{sec:nu_oscillations-liv}), experimental constraints on their values are  typically placed on a single parameter at a time, attributing any predicted departure from the standard-oscillation prediction to the one ``live'' LIV parameter.  The reason for this simplification is practical: it is computationally unfeasible to explore the high-dimensional parameter space that is spanned by allowing all LIV parameters to simultaneously float freely in a fit to observations.  Considering a single nonzero LIV parameter at a time makes computing constraints feasible, but does so at the cost of foregoing accounting for potential degeneracies and correlations between multiple LIV parameters---which exist, as we show below---and, as a result, of potentially overestimating the statistical significance of constraints placed on single parameters.  

[This scenario is reminiscent of non-standard neutrino interactions (NSI), where traditionally constraints were placed on single free parameters and, only recently, as computational power has grown, on multiple parameters simultaneously; see, \eg, \Refe~\cite{Coloma:2023ixt}.  However, the number of free NSI parameters is in the tens, while the number of free LIV parameters is in the thousands.]

Constraining a single LIV parameter at a time betrays the underlying complexity of the effects of LIV on neutrino flavor anisotropy.  This manifests most egregiously in the two following ways:
\begin{itemize}
  \item
   \textbf{Blind spots:} Single-parameter constraints disregard the possibility of running into ``blind spots,'' \ie, regions of the full LIV parameter space where different LIV parameters have opposite effects that cancel out, producing no net departure from the standard-oscillation prediction even for large value of the LIV parameters.  
 \item
  \textbf{Fake flavor isotropy:} Single-parameter constraints disregard the possibility that combinations of multiple nonzero LIV parameters that would individually produce anisotropic flavor-composition skymaps may produce skymaps that are approximately isotropic at large angular scales, though different from the standard-oscillation prediction.
\end{itemize}

With this in mind, below we describe two strategies to explore the LIV parameter space.  The first strategy (``single-parameter'') considers a single nonzero LIV parameter a time and is, therefore, subject to the above caveats and limitations.  We produce our main results under this strategy, including the tables of parameter constraints in Appendix~\ref{sec:constraint_tables}.  The second strategy (``diagonalizable models'') considers multiple nonzero LIV parameters simultaneously.  We introduce it to illustrate the importance of the caveats raised above.

\begin{figure*}[t!]
 \centering
 \includegraphics[trim={0.2cm 0cm 0.5cm 2cm}, clip, width=\textwidth]{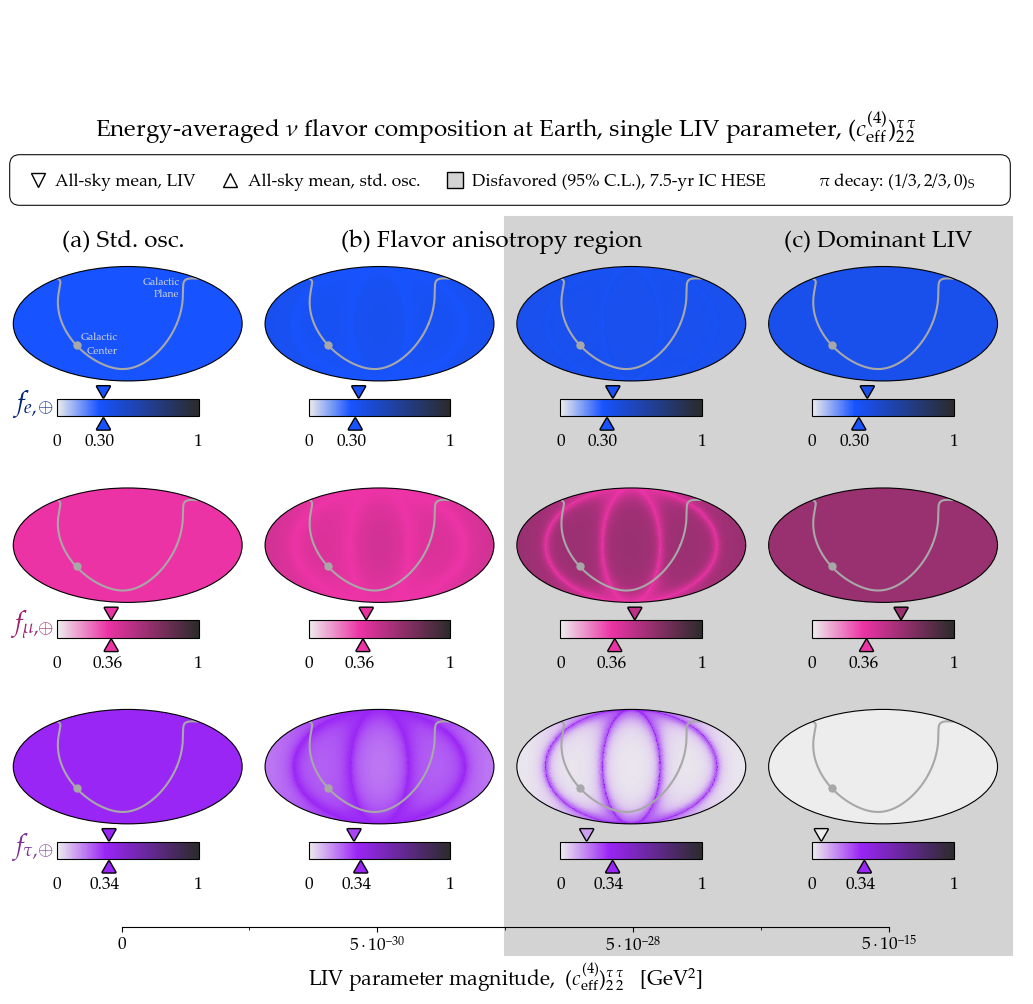}
 \caption{\textbf{\textit{Effect on the flavor-composition skymap of varying a single LIV parameter.}} In this plot, the only nonzero LIV parameter is the CPT-odd parameter $(c^{(4)}_\text{eff})^{\tau \tau}_{2 2}$, whose value we vary to illustrate the transition from dominant standard oscillations, at low values, to dominant LIV, at high values.  Our constraints on LIV come from the interplay between standard oscillations and LIV, which happens when they are comparable in size (\ie, intermediate values of $(c^{(4)}_\text{eff})^{\tau \tau}_{2 2}$) and which induces flavor anisotropy, and from the case when LIV is dominant (\ie, large $(c^{(4)}_\text{eff})^{\tau \tau}_{2 2}$), which yields flavor isotropy, but with an expectation that is different from standard oscillations.  The disfavored range of $(c^{(4)}_\text{eff})^{\tau \tau}_{2 2}$ in this plot is from our analysis (Table~\ref{tab:constraint_tables_d4}).   \textbf{\textit{Our constraints on LIV parameters come  flavor skymaps similar to the ones in this figure vs.~observed ones derived from the public 7.5-year IceCube HESE sample.}}  See Secs.~\ref{sec:liv_strategies-sensitivity_origin}, \ref{sec:liv_strategies-single_parameter}, and \ref{sec:stat_methods} for details.} 
 \label{fig:1param_pred}
\end{figure*}


\subsection{Constraining a single LIV parameter at a time}
\label{sec:liv_strategies-single_parameter}

This is our default LIV exploration strategy.  When placing constraints on a single LIV parameter (sp), we pick a single choice of operator dimension, $d$, of harmonic mode, $\ell$ and $m$, and of flavor indices, $\alpha$ and $\beta$, at a time, \ie, we place constraints on a single CPT-odd parameter $(a^{(d)}_\text{eff})^{\alpha \beta}_{\ell m}$ or CPT-even parameter $(c^{(d)}_\text{eff})^{\alpha \beta}_{\ell m}$ at a time.  By singling out one LIV parameter, the CPT-odd LIV Hamiltonian, \equ{hamiltonian_liv} with $\hat{c}_{\rm eff} = 0$, reduces to
\begin{equation}
 \label{equ:liv_hamiltonian_sp}
 H_{\text{LIV}, \text{sp}}
 =
 E^{d-2}
 Y_\ell^{m}
 \left[
 (a^{(d)}_\text{eff})^{\alpha \beta}_{\ell m}
 I^{\alpha \beta} +
 \text{h.c.}
 \right] \;,
\end{equation}
using the spherical harmonic properties and the Hermiticity conditions on the LIV operators to simplify the expressions in \equ{h_liv_matrix}.  In \equ{liv_hamiltonian_sp}, the LIV parameter is complex and $I^{\alpha\beta}$ is a $3\times 3$ single-entry matrix with the only one nonzero entry in row $\alpha$ and column $\beta$, where it takes the value $1$.  For the CPT-even case, the expression is similar to \equ{liv_hamiltonian_sp}, with $(a^{(d)}_\text{eff})^{\alpha \beta}_{\ell m} \to (c^{(d)}_\text{eff})^{\alpha \beta}_{\ell m}$.

Thus, the total Hamiltonian (Sec.~\ref{sec:nu_oscillations-prob}), including standard plus LIV effects, becomes $H_\text{tot,sp} = H_\text{vac} + H_\text{LIV,sp}$.  With it, we compute the energy-averaged neutrino flavor composition across the sky (Sec.~\ref{sec:astro_nu-flavor}), $f_{\alpha, \oplus}$ in \equ{flavor_ratio}, as a function of the single live LIV parameter, $(a^{(d)}_\text{eff})^{\alpha \beta}_{\ell m}$ or $(c^{(d)}_\text{eff})^{\alpha \beta}_{\ell m}$.  Later (Sec.~\ref{sec:stat_methods}), we constrain the value of this parameter by comparing the $f_{\alpha, \oplus}$ so predicted to pre-computed allowed regions of energy-averaged flavor composition across a tessellated sky~\cite{Telalovic:2023tcb}.

\smallskip

\begin{figure}[t!]
 \centering
 \includegraphics[trim={0.25cm 0cm 0cm 0cm}, clip, width=0.75\textwidth]{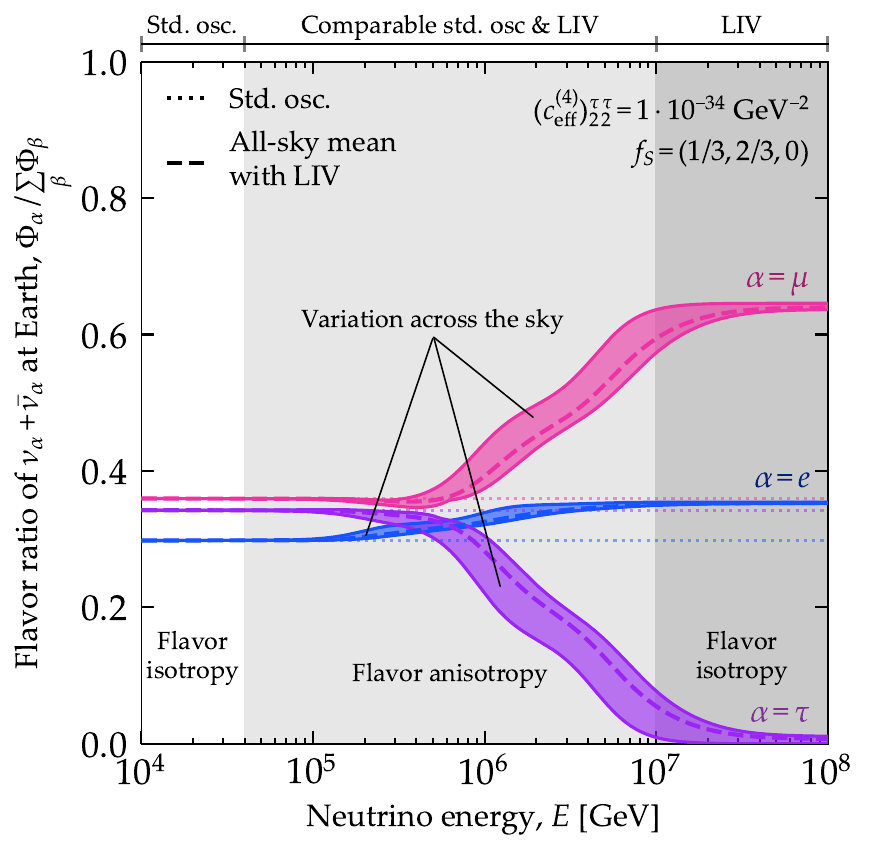}
 \caption{\textbf{\textit{Variation of the flavor composition with energy.}}  In this plot, the only nonzero LIV parameter is $(c_\textrm{eff}^{(4})_{22}^{\tau\tau}$, which is kept fixed at an illustrative value.  All other model parameters, LIV and otherwise, are fixed to the central values of their prior distributions in Table~\ref{tab:fit_params}, and the flavor composition at the sources, $f_{\rm S}$, is fixed to pion decay.  (When computing limits on LIV parameters, we do vary the values of other parameters.)  The evolution of the flavor ratios with energy reflects the interplay between standard-oscillation and LIV effects.  The transition energies between regimes of flavor isotropy and anisotropy depends on the choice of LIV parameter and its value.  The widths of the flavor-ratio curves reflect their variation across the sky; they are widest at intermediate energies, where these effects are comparable in size.  \textbf{\textit{Our constraints on LIV come the regimes where standard oscillations and LIV effects are comparable and where LIV effects are dominant.}}  See Sec.~\ref{sec:liv_strategies-single_parameter} for details.}
 \label{fig:energy_resonance}
\end{figure}

\textbf{\textit{Three regimes.---}}Figure \ref{fig:1param_pred} illustrates the effect of LIV on the sky distribution of $f_{\alpha, \oplus}$ by varying the value of the single  parameter $(c^{(4)}_\text{eff})^{\tau\tau}_{22}$.  Three different regimes of LIV dominance are apparent, depending on its value:
\begin{enumerate}[(a)]
 \item 
  \textbf{Dominant standard oscillations:} When the LIV parameter is small, standard oscillations drive flavor transitions, making the flavor-composition skymap isotropic or nearly so.  In this case, $H_\text{LIV,sp} \ll H_\text{vac}$, implying (Sec.~\ref{sec:nu_oscillations}) that $(a^{(d)}_\text{eff})^{\alpha \beta}_{\ell m} \ll \Delta m^2_{ij}E^{1-d}$.  (We show later how flavor anisotropy changes with energy, in \figu{energy_resonance}.) 
 \item
  \textbf{Comparable standard oscillations and LIV:} When $H_\text{LIV,sp} \sim H_\text{vac}$, implying that $(a^{(d)}_\text{eff})^{\alpha \beta}_{\ell m} \sim \Delta m^2_{ij}E^{1-d}$, the interplay between standard and LIV effects induces visible anisotropy in the flavor-composition sky.  Only LIV parameters with $\ell > 0$ induce anisotropy; those with $\ell = 0$ change the mean all-sky flavor composition, but keep it isotropic.  The shape of the anisotropy pattern is determined by the directions along which $Y_\ell^m$ and, therefore, $H_\text{LIV,sp}$, vanishes.  Broadly stated, LIV parameters with larger values of $\ell$ generate smaller angular structures on the $f_{\alpha, \oplus}$ sky.  Also, as illustrated in \figu{1param_pred}, larger $(a^{(d)}_\text{eff})^{\alpha \beta}_{\ell m}$ induce stronger anisotropies (as long as standard oscillations and LIV remain relatively comparable in size).  
 \item
  \textbf{Dominant LIV:} When the LIV parameter is large, LIV effects drive flavor transitions.  Because we consider a single nonzero LIV parameter at a time, when it is dominant, \ie, when $H_\text{LIV,sp} \gg H_\text{vac}$, it makes the total Hamiltonian a single-entry matrix.  This suppresses flavor transitions, making the flavor-composition sky isotropic or nearly so, but not necessarily with the same value as from standard oscillations, as illustrated in \figu{1param_pred}.
\end{enumerate}
Similar behavior as in \figu{1param_pred} occurs for other choices of $(a^{(d)}_\text{eff})^{\alpha \beta}_{\ell m}$ or $(c^{(d)}_\text{eff})^{\alpha \beta}_{\ell m}$, with the anisotropy patterns of $f_{\alpha, \oplus}$ depending on the choice of parameter. 

\smallskip

\textbf{\textit{Flavor anisotropy vs.~energy.---}}Figure~\ref{fig:energy_resonance} illustrates how the degree of flavor anisotropy induced by a single LIV parameter---the same one as in \figu{1param_pred}---changes with neutrino energy, for a fixed value of the parameter and for the nominal expectation of $f_{\rm S} = (1/3, 2/3, 0)$ for the flavor composition at the sources (Sec.~\ref{sec:astro_nu-flavor}).  Figure~\ref{fig:energy_resonance} underlies the results shown in \figu{1param_pred} for the energy-averaged flavor composition.

Figure~\ref{fig:energy_resonance} shows that, for a fixed value of the LIV parameter, the three regimes of standard-oscillation vs.~LIV dominance that we identified above in \figu{1param_pred} appear at different energies:
\begin{itemize}
 \item
  \textbf{Low energies:} Below about 100~TeV, standard oscillations dominate because its effects are $\propto 1/E$, and so the resulting flavor composition is largely isotropic.  In \figu{1param_pred}, this is evidenced by the narrow spread of $f_{\alpha, \oplus}$ along different directions compared to the all-sky mean.  In this case, the all-sky mean is at around $(0.30, 0.36, 0.34)_\oplus$, as predicted by standard oscillations.  This corresponds to the standard-oscillation skymaps in \figu{1param_pred}.
 \item
  \textbf{Intermediate energies:} Between about 100~TeV and 2~PeV, standard oscillations and LIV are comparable, and so the resulting flavor composition is anisotropic.  This is evidenced by the asymmetric spread of $f_{\alpha, \oplus}$ around the all-sky mean, which is close, but no longer equal to the value predicted by standard oscillations.  This corresponds to the anisotropic skymaps in \figu{1param_pred}.
 \item
  \textbf{High energies:} Above about 2~PeV, LIV is dominant because its effects are $\propto E^{d-3}$ (\ie, $\propto E^2$ in \figu{energy_resonance}).  This manifests in two ways: the all-sky mean is longer close to the standard-oscillation value and the flavor sky is again isotropic.  This corresponds to the LIV-dominant skymaps in \figu{1param_pred}.
\end{itemize}

The energies demarcating the above three regimes are fixed by our choice of value for the LIV parameter.  Because anisotropy occurs when standard oscillations and LIV effects are comparable, \ie, when $(a^{(d)}_\text{eff})^{\alpha \beta}_{\ell m} \sim \Delta m^2_{ij}E^{1-d}$ (see above), lower values of $(a^{(d)}_\text{eff})^{\alpha \beta}_{\ell m}$ shift the transitions between regimes to higher energies, and higher values of $(a^{(d)}_\text{eff})^{\alpha \beta}_{\ell m}$ shift them to lower energies.  The range of LIV parameters to which our analysis is sensitive is determined by the condition that the flavor-anisotropy regime falls within the IceCube HESE energy, \ie, between tens of~TeV and a few PeV.  

Further, higher values of the LIV operator dimension, $d$, narrow down the intermediate-energy interval where there is flavor anisotropy, since they make the LIV effects grow faster with energy relative to the standard-oscillation effects, \ie, $E^{d-3}$ vs.~$1/E$.  As a result, for high-dimension operators, the contribution of this energy interval to the energy-averaged flavor composition becomes comparatively less important.

When placing our constraints on the LIV parameters based on $f_{\alpha, \oplus}$, all of the above effects are taken into account tacitly as part of our statistical methods.


\subsection{Constraining multiple LIV parameters jointly}
\label{sec:liv_strategies-multiple_parameters}

\begin{figure*}[t!]
 \centering
 \includegraphics[trim={0.2cm 0cm 0.5cm 0cm}, clip, width=\textwidth]{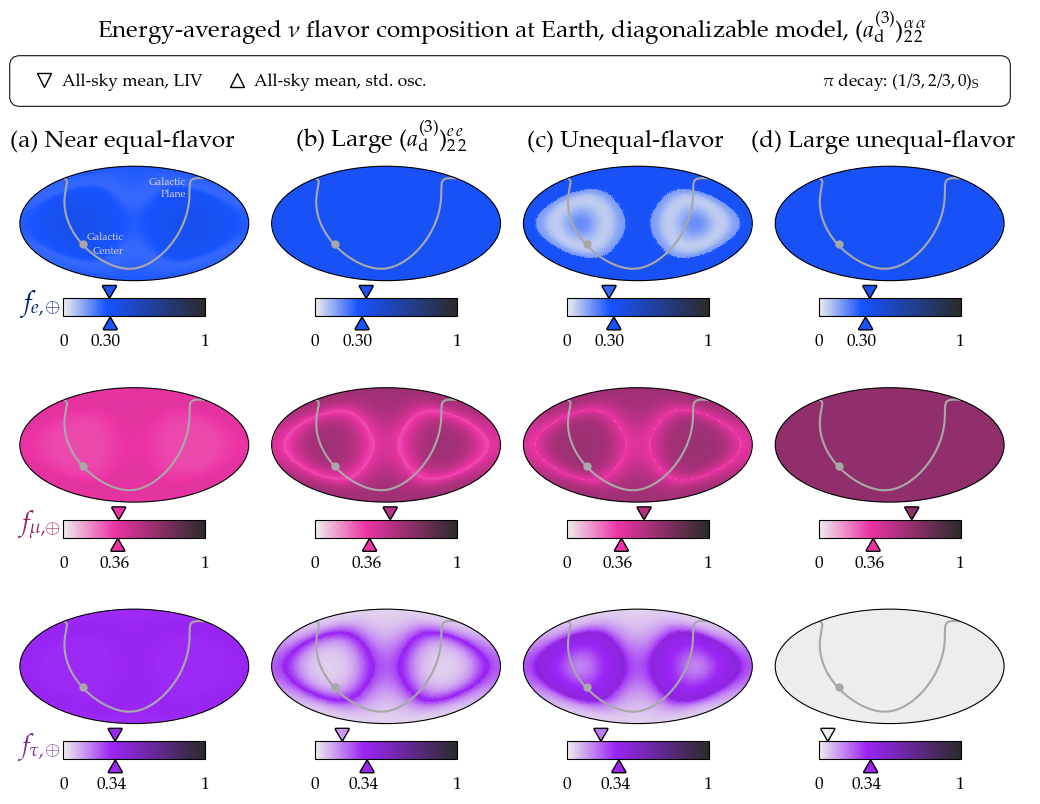}
 \caption{\textbf{\textit{Effect on the flavor-composition skymap of jointly varying multiple LIV parameters.}}  
 Similar to \figu{1param_pred}, but simultaneously varying the LIV parameters $(a_{\rm d}^{(3)})_{22}^{ee}$, $(a_{\rm d}^{(3)})_{22}^{\mu\mu}$, and $(a_{\rm d}^{(3)})_{22}^{\tau\tau}$ within a diagonalizable model [\equ{h_liv_d_diag}] containing only CPT-odd LIV operators. 
 The illustrative values of the LIV parameters used to create these maps are: (a) $(a_{\rm d}^{(3)})_{22}^{ee} = -2.99 \cdot 10^{-25}$~GeV, $(a_{\rm d}^{(3)})_{22}^{\mu\mu} = -3 \cdot 10^{-25}$~GeV, and $(a_{\rm d}^{(3)})_{22}^{\tau\tau} = -3.01 \cdot 10^{-25}$~GeV; (b) $(a_{\rm d}^{(3)})_{22}^{ee} = -3 \cdot 10^{-20}~\text{GeV} = 3 \cdot 10^5 \times (a_{\rm d}^{(3)})_{22}^{\mu\mu} = 1.5 \cdot 10^5 \times (a_{\rm d}^{(3)})_{22}^{\tau\tau}$; (c) $(a_{\rm d}^{(3)})_{22}^{ee} = -3 \cdot 10^{-25}~\text{GeV}= -3~(a_{\rm d}^{(3)})_{22}^{\mu\mu} = -1.5~(a_{\rm d}^{(3)})_{22}^{\tau\tau}$; and (d) $(a_{\rm d}^{(3)})_{22}^{ee} = -3 \cdot 10^{-20}~\text{GeV}= -3~(a_{\rm d}^{(3)})_{22}^{\mu\mu} = -1.5~(a_{\rm d}^{(3)})_{22}^{\tau\tau}$.
  \textbf{\textit{The interplay between multiple nonzero LIV parameters induces structures in the flavor-composition skymap beyond what single LIV parameters can induce.}}  See Sec.~\ref{sec:liv_strategies-multiple_parameters} for details.}
 \label{fig:diag_pred}
\end{figure*}

So far, we have followed our default LIV exploration strategy of considering a single nonzero LIV parameter at a time (Sec.~\ref{sec:liv_strategies-single_parameter}).  This strategy, however, has inherent limitations, as pointed out in Sec.~\ref{sec:liv_strategies-high_dim}.  Below, we illustrate these limitations by showing how the alternative strategy of allowing multiple nonzero LIV parameters simultaneously yields forms of flavor anisotropy different from those achievable by a single parameter.  

To do this, we consider a simple model with only CPT-odd LIV operators where, for each operator dimension $d$, all flavor-diagonal LIV parameters, $(a^{(d)}_{\rm d})^{\alpha \alpha}_{\ell m}$, are allowed to be nonzero, and all off-diagonal parameters are set to zero, \ie, $(a^{(d)}_{\rm d})^{\alpha \beta}_{\ell m} = 0$, for $\alpha \neq \beta$.  This choice reduces the number of free CPT-odd LIV parameters in $H_\textrm{LIV}^{(d)}$, \equ{h_liv_d}, to $3d^2$.
This model is called \textit{diagonalizable} (d) because the resulting LIV Hamiltonian is diagonal in the flavor basis, \ie, for dimension $d$,
\begin{equation}
 \label{equ:h_liv_d_diag}
 H_\textrm{LIV,d}^{(d)}
 =
 E^{d-3}
 \sum_{\ell=0}^{d-1}
 \sum_{m=-\ell}^{\ell} 
 Y_\ell^m(\hat{\boldsymbol{p}})
 \begin{pmatrix}
  (a_{\rm d}^{(d)})_{\ell m}^{ee}     & 0 & 0   \\
  0  & (a_{\rm d}^{(d)})_{\ell m}^{\mu\mu}  & 0  \\
  0 & 0 & (a_{\rm d}^{(d)})_{\ell m}^{\tau\tau}
 \end{pmatrix} \;,
\end{equation}
with odd-valued $d \geq 3$.  As a result, the LIV Hamiltonian can be trivially diagonalized jointly with the neutrino mass matrix, $M^2$, by the PMNS mixing matrix [see \equ{hamiltonian_total}].  
(It is straightforward to build a diagonalizable model for CPT-even LIV operators in a similar manner.  But, since it leads to broadly similar results as in the CPT-odd case, we do not explore it here.)

Figure~\ref{fig:diag_pred} shows the sky distribution of $f_{\alpha, \oplus}$ for an example diagonalizable model with only $d = 3$ LIV parameters and where the only nonzero parameters are the ones with $l = m = 2$, \ie, $(a^{(3)}_\text{d})^{ee}_{22}$, $(a^{(3)}_\text{d})^{\mu\mu}_{22}$, and $(a^{(3)}_\text{d})^{\tau\tau}_{22}$.  Similar observations apply to other LIV parameters.  We identify four regimes of LIV dominance, depending on the values of these parameters:
\begin{enumerate}[(a)]
 \item
   \textbf{Near flavor equality:} When $(a^{(3)}_\text{d})_{22}^{ee} \approx (a^{(3)}_\text{d})_{22}^{\mu\mu} \approx (a^{(3)}_\text{d})_{22}^{\tau\tau}$, $H_\textrm{LIV,d}$ is proportional to the identity matrix.  Therefore, it merely introduces an approximately global phase (dependent on the neutrino direction) into the evolution of the neutrino states that has little effect on flavor transitions.  As a result, standard oscillations dominate and the distributions of $f_{e,\oplus}$, $f_{\mu,\oplus}$, and $f_{\tau,\oplus}$ are nearly isotropic, as shown in \figu{diag_pred}.
 \item
  \textbf{One flavor dominant:} When the LIV parameter for one flavor, $(a^{(3)}_\text{d})_{22}^{ee}$ in \figu{diag_pred}, dominates over the other two, that flavor, $\nu_e$, effectively no longer mixes with the other two, $\nu_\mu$ and $\nu_\tau$ in \figu{diag_pred}.  As a result, $f_{e,\oplus}$ is isotropic, and only $f_{\mu, \oplus}$ and $f_{\tau, \oplus}$ exhibit anisotropy.
 \item
  \textbf{Comparable contribution per flavor:} Unequal LIV parameters that produce a diagonal Hamiltonian comparable in strength to the standard-oscillation Hamiltonian produce the most anisotropy.
 \item
  \textbf{Dominant LIV:} Similarly to what we observed earlier in \figu{1param_pred} (Sec.~\ref{sec:liv_strategies-single_parameter}), when all of the LIV parameters are large, LIV effects drive flavor transitions and the resulting skymaps of flavor composition are isotropic for all flavors.
\end{enumerate}

Thus, \figu{diag_pred} establishes that, as expected, \textbf{\textit{when there are multiple nonzero LIV parameters, the resulting flavor anisotropy can be different from the anisotropy that single nonzero LIV parameters are able to induce.}} Later (Sec.~\ref{sec:results}), we show that this yields appreciably different allowed regions of the LIV parameters.  
Nevertheless, due to the computational cost of varying many free parameters simultaneously (Sec.~\ref{sec:liv_strategies-single_parameter}), our main constraints are computed assuming a single nonzero parameter at a time.


\section{Statistical Methods}
\label{sec:stat_methods}

To constrain the values of the LIV parameters, we follow the procedure introduced in \Refe~\cite{Telalovic:2023tcb}, with a few improvements.  The constraints are based on the measurement of the directional flavor composition of the high-energy astrophysical neutrino flux using 7.5 years of public IceCube HESE events~\cite{IceCube:2020wum, IC75yrHESEPublicDataRelease}.  We summarize the procedure below and defer to \Refe~\cite{Telalovic:2023tcb} for details.  


\subsection{Inferring the directional flavor composition}
\label{sec:stat_methods-inferring_flavor_composition}

We start by inferring the flavor composition of the high-energy astrophysical neutrino flux along different directions in the sky.  Afterward, we use these results to constrain the LIV parameters.

First, we tessellate the sky into $N_{\rm pix} = 12$ pixels of equal area using \textsc{HEALpix}~\cite{Gorski:2004by, healpix_url} (\ie, we set the \textsc{HEALpix} parameter $n_{\rm side} = 1$).  Figures~\ref{fig:probability_vs_energy} and \ref{fig:all_triangles} show the resulting tessellated sky.  Into these pixels, we bin the events from the IceCube HESE sample according to their best-fit reconstructed direction~\cite{IC75yrHESEPublicDataRelease}.  The binning is coarse, with each pixel about $90^\circ$ wide, to reflect the large uncertainty in the reconstructed direction of HESE events, and to avoid having empty pixels in our analysis.

We model the all-flavor flux of high-energy neutrinos at Earth as an isotropic power law, $\Phi_\text{all} = [\Phi_0/(100~\text{TeV})] E^{-\gamma}$, where $\Phi_0$ and $\gamma$ are free parameters.  The flavor composition in each of the 12 pixels is also free: in the $i$-th pixel, it is described by the flavor fractions $f_{e,i}$ and $f_{\mu,i}$.  (The fraction of $\nu_\tau$ is not independent, since $f_{\tau,i} = 1-f_{e,i}-f_{\mu,i}$.)  We take the flavor composition in each pixel as independent from that of the other pixels.  Thus, while the all-flavor flux is isotropic in our treatment, the individual fluxes of $\nu_e$, $\nu_\mu$, and $\nu_\tau$ are allowed to be anisotropic, with a common shape for their energy spectrum.

Further, when extracting the flavor composition from IceCube data, we treat $\nu_\alpha$ and $\bar{\nu}_\alpha$ separately, assuming they have equal fluxes.  This assumption, often made in the literature, is motivated by the expectation that the diffuse neutrino flux is dominated by neutrino production in proton-proton interactions at the astrophysical sources, which produce negatively and positively charged pions in comparable numbers~\cite{Kelner:2006tc}.  This is true also in neutrino production by proton-photon interactions at high energies, due to the onset of multi-pion production~\cite{Kelner:2008ke, Hummer:2010vx}.  In addition, In IceCube, the detector responses for $\nu_\alpha$ and $\bar{\nu}_\alpha$ are nearly equal because their deep-inelastic-scattering cross sections on nucleons are nearly equal at these energies~\cite{CTEQ:1993hwr, Conrad:1997ne, Formaggio:2012cpf}, and their final states are indistinguishable.  The exception is the contribution from the Glashow resonance~\cite{Glashow:1960zz} of exclusively $\bar{\nu}_e$ scattering on electrons, which matters only at energies around 6.3~PeV~\cite{IceCube:2021rpz}.

To determine the allowed values of the free parameters, $\Phi_0$, $\gamma$, $\left\{ f_{e,i} \right\}_{i=1}^{12}$, and $\left\{ f_{\mu,i} \right\}_{i=1}^{12}$, we generate mock event distributions, each for different test values of these parameters, bin the mock events according to their direction in the same way as we bin the IceCube events above, and compare them.  To generate the mock event distributions, we reweigh the Monte Carlo event sample provided by the IceCube Collaboration together with their 7.5-year public HESE data release~\cite{IC75yrHESEPublicDataRelease}.  By using this Monte Carlo sample, our analysis tacitly incorporates a detailed detector response.  This is key to accurately describing the connection between the flavors of the interacting neutrinos and the signatures---cascade, track, double cascade---that they create, which varies with neutrino energy and direction.  

\begingroup
\begin{table*}[t!]
 \caption{\label{tab:fit_params}\textbf{\textit{Free model parameters and their priors.}}  The LIV parameters varied in our analysis are different depending on the choice of exploration strategy of the LIV parameter space.  When using our default strategy, we consider a single nonzero complex LIV parameter at a time and constrain its real and imaginary parts separately.  We explore LIV operators with dimension from $d = 2$ to 8, where $(a_{\rm eff}^{(d)})_{\ell m}^{\alpha \beta}$ are CPT-odd coefficients (for odd-valued $d$) and $(c_{\rm eff}^{(d)})_{\ell m}^{\alpha \beta}$ are CPT-even coefficients (for even-valued $d$).  When using instead our alternative strategy of varying multiple LIV parameters at a time, we simultaneously constrain all of the LIV parameters in an example diagonalizable model with only CPT-odd operators, with real-valued coefficients $(a_{\rm d}^{(3)})_{\ell m}^{\alpha\alpha}$. 
 See Sec.~\ref{sec:liv_strategies} for details on our LIV parameter-space exploration strategies and Sec.~\ref{sec:stat_methods} for details on the statistical procedure we use to constrain the LIV parameters.}
  \centering
  \renewcommand{\arraystretch}{1}
\begin{adjustbox}{max width=\textwidth}\normalsize
  \begin{tabular}{cccccc}
  \hline\\[-0.8em]
  \multicolumn{3}{c}{Parameter} &
  \multicolumn{2}{c}{Prior}     &
  \multirow{2}{*}{\makecell{\\[-0.5em]Constraints\\[0.5em](this work)}} 
  \\[0.5em] 
  \cline{1-3}
  \cline{4-5}
  \\[-0.8em]
  Name         &
  Units        &
  Description  &
  Distribution &
  Source       &  
  \\[0.5em]
  \hline
  \\[-0.8em]
  \multicolumn{6}{c}{LIV coefficients: single-parameter fits (default)} 
  \\[0.5em]
  \hline 
  \\[-0.8em]
  \multirow{2}{*}{Re$(a_{\text{eff}}^{(d)})^{\alpha\beta}_{\ell m}$ or Re$(c_{\text{eff}}^{(d)})^{\alpha\beta}_{\ell m}$} &
  \multirow{2}{*}{GeV$^{4-d}$} &
  \multirow{2}{*}{Real part} &
  \multirow{2}{*}{Uniform $\in[- 10^{-(4d+10)}, 10^{-(4d+10)}]$}  &
  \multirow{2}{*}{This work} &
  \multirow{2}{*}{\makecell{Figure~\ref{fig:limits_overview}\\[0.2em]
  Tables \ref{tab:constraint_tables_d2}--\ref{tab:constraint_tables_d8}}}
  \\[1.3em]
  \multirow{2}{*}{Im$(a_{\text{eff}}^{(d)})^{\alpha\beta}_{\ell m}$ or Im$(c_{\text{eff}}^{(d)})^{\alpha\beta}_{\ell m}$} &
  \multirow{2}{*}{GeV$^{4-d}$} &
  \multirow{2}{*}{Imaginary part} &
  \multirow{2}{*}{Uniform $\in[- 10^{-(4d+10)}, 10^{-(4d+10)}]$} &
  \multirow{2}{*}{This work} &
  \multirow{2}{*}{\makecell{Figure~\ref{fig:limits_overview}\\[0.2em]
  Tables \ref{tab:constraint_tables_d2}--\ref{tab:constraint_tables_d8}}}
  \\[1.5em]
  \hline
  \\[-0.8em]
  \multicolumn{6}{c}{LIV coefficients: joint multiple-parameter fit (example CPT-odd diagonalizable model for $d = 3$)} 
  \\[0.5em]
  \hline 
  \\[-0.8em]
  \multirow{2}{*}{Re$(a_{\text{d}}^{(3)})^{\alpha \alpha}_{\ell m}$} &
  \multirow{2}{*}{GeV$^{4-d}$} &
  \multirow{2}{*}{\makecell{Real component,\\$1\leq\ell\leq d-1$, $\alpha=e,\mu,\tau$}}  &
  \multirow{2}{*}{Uniform $\in[- 10^{-(4d+7)}, 10^{-(4d+7)}]$ } &
  \multirow{2}{*}{This work} &
  \multirow{2}{*}{Figure~\ref{fig:indep_vs_combined_cornerplot}}
  \\[1.6em]
  \hline
  \\[-0.8em]
  \multicolumn{6}{c}{Nuisance parameters (common to both fits)} 
  \\[0.5em]
  \hline
  \\[-0.8em]
  \multirow{2}{*}{$\Delta m^2_{21}$} &
  \multirow{2}{*}{eV$^2$} &
  \multirow{2}{*}{\makecell{Mass difference\\$m^2_2-m^2_1$}}
  &
  \multirow{2}{*}{\makecell{Normal,\\$\mu = 7.14\cdot10^{-5}$, $\sigma = 0.019\cdot 10^{-5}$}} &
  \multirow{2}{*}{NuFIT~5.2~\cite{Esteban:2020cvm}} &
  ---
  \\[1.6em] 
  \multirow{2}{*}{$\Delta m^2_{31}$} &
  \multirow{2}{*}{eV$^2$} &
  \multirow{2}{*}{\makecell{Mass difference\\$m^2_3-m^2_1$}s}
  &
  \multirow{2}{*}{\makecell{Normal,\\$\mu = 2.507\cdot 10^{-3}$, $\sigma = 0.026\cdot 10^{-3}$}} &
  \multirow{2}{*}{NuFIT~5.2~\cite{Esteban:2020cvm}} &
  ---
  \\[1.6em] 
  \multirow{2}{*}{$\sin \theta_{12}$} &
  \multirow{2}{*}{---} &
  \multirow{2}{*}{\makecell{Solar\\mixing angle}}
  &
  \multirow{2}{*}{\makecell{Normal,\\$\mu = 0.303$, $\sigma = 0.012$}} &
  \multirow{2}{*}{NuFIT~5.2~\cite{Esteban:2020cvm}} &
  ---
  \\[1.6em] 
  \multirow{2}{*}{$\sin \theta_{13}$} &
  \multirow{2}{*}{---} &
  \multirow{2}{*}{\makecell{Reactor\\mixing angle}}
  &
  \multirow{2}{*}{\makecell{Normal,\\$\mu = 0.02225$, $\sigma = 0.00056$}} &
  \multirow{2}{*}{NuFIT~5.2~\cite{Esteban:2020cvm}} &
  ---
  \\[1.6em] 
  \multirow{2}{*}{$\sin \theta_{23}$} &
  \multirow{2}{*}{---} &
  \multirow{2}{*}{\makecell{Atmospheric\\mixing angle}}
  &
  \multirow{2}{*}{\makecell{Joint 2D posterior\\with $\delta_{\rm CP}$}} &
  \multirow{2}{*}{NuFIT~5.2~\cite{Esteban:2020cvm}} &
  ---
  \\[1.6em] 
  \multirow{2}{*}{$\delta_\text{CP}$} &
  \multirow{2}{*}{---} &
  \multirow{2}{*}{\makecell{CP-violation\\phase}}
  &
  \multirow{2}{*}{\makecell{Joint 2D posterior\\with $\sin^2 \theta_{23}$}} &
  \multirow{2}{*}{NuFIT~5.2~\cite{Esteban:2020cvm}} &
  ---
  \\[1.6em] 
  \multirow{2}{*}{$\gamma$} &
  \multirow{2}{*}{---} &
  \multirow{2}{*}{\makecell{Astrophysical $\nu$ flux\\spectral index}} &
  \multirow{2}{*}{\makecell{Normal,\\$\mu = 2.89$, $\sigma = 0.23$}} &
  \multirow{2}{*}{\makecell{IceCube\\7.5-yr HESE~\cite{IceCube:2020wum}}}  &
  ---
  \\ [1.6em]
  $f_{e, {\rm S}}$ &
  --- &
  Fraction of flux produced as $\nu_e$ &
  Fixed $f_{e, {\rm S}} = 1/3$ &
  Pion decay &
  --- 
  \\[0.7em]
  &
  &
  &
  Fixed $f_{e, {\rm S}} = 0$ &
  Muon-damped &
  --- 
  \\[0.7em]
  &
  &
  &
  Fixed $f_{e, {\rm S}} = 1$ &
  Neutron decay &
  --- 
  \\[0.7em]
  &
  &
  &
  \multirow{1}{*}{\makecell{Uniform $\in[0, 1]$}} &
  Flavor agnostic &
  --- 
  \\[0.3em]
  \hline
  \end{tabular}
  \end{adjustbox}
\end{table*}

We compare mock and real event distributions by contrasting, via a Poisson likelihood function, the number of observed \textit{vs.}~expected cascades, tracks, and double cascades---each separately---in each of the pixels, and computing the product of the likelihoods in all of the pixels; see \Refe~\cite{Telalovic:2023tcb} for details.  We adopt a Bayesian approach and use \textsc{UltraNest}~\cite{Buchner:2021cql}, an efficient importance nested sampler~\cite{Buchner:2014, Buchner:2017}, to vary simultaneously all of the above free physical parameters, plus three nuisance parameters describing the background flux of atmospheric neutrinos and muons.  In total, there are 29 free model parameters that are floated simultaneously.  To avoid introducing bias, we sample their values from wide priors; see Table~2 in \Refe~\cite{Telalovic:2023tcb}.

The result of the optimization procedure is a joint posterior probability distribution of all 29 physical and nuisance parameters; see, \eg, Figs.~B1 and B2 in \Refe~\cite{Telalovic:2023tcb}.  We find the allowed flavor composition in the $i$-th pixel by marginalizing this full posterior over all free parameters other than $f_{e,i}$ and $f_{\mu,i}$.  This yields the two-parameter joint posterior $\mathcal{P}_i(f_{e,i}, f_{\mu,i})$ for each pixel, whose $1\sigma$ and $2\sigma$ contours are shown in \figu{all_triangles} (and also in Fig.~3 in \Refe~\cite{Telalovic:2023tcb}).  Below, we use these posteriors to constrain the values of the LIV parameters.


\subsection{Constraining the LIV parameters}
\label{sec:stat_methods-constraining_liv_parameters}

To constrain the values of the LIV parameters, we again adopt a Bayesian approach.  First, we reinterpret the posterior on the flavor composition in each pixel that we obtained above, $\mathcal{P}_i$, as a joint likelihood function on $f_{e,\oplus}$ and $f_{\mu,\oplus}$, \ie, $\mathcal{L}_i \equiv \mathcal{P}_i$.  Motivated by our benchmark scenarios of high-energy neutrino production (Sec.~\ref{sec:astro_nu-nuisance_fS}), we make the simplifying, but reasonable assumption that no $\nu_\tau$ are produced, \ie, that $f_{\tau, {\rm S}}$.  As a result, the flavor composition at the sources is completely determined by the single free parameter, which we choose to be $f_{e, {\rm S}}$, \ie, it is $\left( f_{e,{\rm S}}, 1-f_{e,{\rm S}}, 0 \right)$.  (We either fix the value of $f_{e, {\rm S}}$ or let it float between 0 and 1; see Table~\ref{tab:fit_params}.)

Given test values of our free model parameters---the standard mixing parameters, $\boldsymbol{\theta}_{\rm std}$; the astrophysical spectral index, $\gamma$; the flavor composition at the sources, $f_{e, {\rm S}}$; and the LIV parameters, $\boldsymbol{\Lambda}_{\rm LIV}$, we compute the predicted energy-averaged flavor composition at the Earth using \equ{flavor_ratio}, $f_{\alpha,\oplus}(\hat{\boldsymbol{p}}, \boldsymbol{\theta}_{\rm std}, \gamma, f_{e,{\rm S}}, \boldsymbol{\Lambda}_{\rm LIV})$, for $\alpha = e, \mu$, which depends on the neutrino direction, $\hat{\boldsymbol{p}}$.  

Next, we compute the direction-averaged value of the flavor composition in the $i$-th pixel, $f_{\alpha, \oplus}^{(i)}$, where $i = 1, \ldots, N_{\rm pix} = 12$.  To do this, we discretize the skymaps of flavor composition by using \textsc{HEALpix} to tessellate the sky finely into 1728 pixels ($n_\text{side} = 12$), each one about $4^\circ$ wide.  We evaluate the direction-dependent flavor composition, $f_{\alpha,\oplus}(\hat{\boldsymbol{p}}, \boldsymbol{\theta}_{\rm std}, \gamma, f_{e,{\rm S}}, \boldsymbol{\Lambda}_{\rm LIV})$, at values of $\hat{\boldsymbol{p}}$ corresponding to the center of each of these pixels.  Each of the original $N_{\rm pix} = 12$ large pixels of our analysis (Sec.~\ref{sec:stat_methods-inferring_flavor_composition}) is made up of 144 of the small ones.  We compute the direction-averaged flavor composition in the $i$-th pixel, $f_{\alpha, \oplus}^{(i)}(\boldsymbol{\theta}_{\rm std}, \gamma, f_{e,{\rm S}}, \boldsymbol{\Lambda}_{\rm LIV})$, as the average of the flavor composition across all its constituent 144 pixels.

(For all the values of LIV operator dimension $d \leq 8$ that we explore, the angular scale of the anisotropic features introduced by LIV on the flavor skymaps is larger than the size of each of our finer pixels, so that averaging the flavor composition over 144 of them accurately captures its directional dependence.)

Finally, we quantify the compatibility of the $f_{\alpha, \oplus}^{(i)}$ predicted above with the flavor composition inferred from the IceCube HESE data in Sec.~\ref{sec:stat_methods-inferring_flavor_composition} by evaluating the likelihood function at the predictions, \ie, $\mathcal{L}_i (f_{e, \oplus}^{(i)}, f_{\mu, \oplus}^{(i)})$ in the $i$-th pixel.  The total likelihood is the product of $\mathcal{L}_i$ in all of the  $N_{\rm pix}$ skymap pixels.  Using Bayes' theorem, the joint posterior of all the free model parameters is
\begin{eqnarray}
 \label{equ:likelihood}
 &&
 \mathcal{P}
 (
 \boldsymbol{\theta}_{\rm std},
 \gamma,
 f_{e, {\rm S}},
 \boldsymbol{\Lambda}_{\rm LIV}
 ) 
 =
 \pi(\boldsymbol{\theta}_{\rm std})~
 \pi(\gamma)~
 \pi(f_{e, {\rm S}})
 \nonumber \\
 && \qquad\qquad \times
 \prod_{i=1}^{N_{\rm pix} = 12} 
 \mathcal{L}_i
 \left( 
 f_{e,\oplus}^{(i)}, f_{\mu,\oplus}^{(i)}
 \right)
 \pi(\boldsymbol{\Lambda}_{\rm LIV}) \;,
\end{eqnarray}
where $f_{\alpha,\oplus}^{(i)} \equiv
f_{\alpha,\oplus}^{(i)}(\boldsymbol{\theta}_{\rm std}, \gamma, f_{e,{\rm S}}, \boldsymbol{\Lambda}_{\rm LIV})$, with $\alpha = e, \mu$, and the different $\pi$ functions denote the prior probability distributions of the different model parameters.  To constrain the value of $\boldsymbol{\Lambda}_{\rm LIV}$---the LIV parameters that we are interested in---we marginalize this joint posterior over all other free model parameters, which we treat as nuisance, to obtain a posterior only on $\boldsymbol{\Lambda}_{\rm LIV}$, \ie,
\begin{equation}
 \label{equ:likelihood_marg}
 \mathcal{P}(\boldsymbol{\Lambda}_{\rm LIV})
 =
 \int {\rm d}\boldsymbol{\theta}_{\rm std}~
 {\rm d}\gamma~
 {\rm d}f_{r,{\rm S}}~
 \mathcal{P}
 (
 \boldsymbol{\theta}_{\rm std},
 \gamma,
 f_{e, {\rm S}},
 \boldsymbol{\Lambda}_{\rm LIV}
 ) \;.
\end{equation}

Earlier, \Refe~\cite{Telalovic:2023tcb} presented a first attempt at constraining LIV parameters by looking for flavor anisotropies in the same IceCube data sample, using methods very similar to the ones above.  The key difference compared to the present analysis is that \Refe~\cite{Telalovic:2023tcb} used separate one-dimensional likelihood functions for $f_{e, \oplus}^{(i)}$ and $f_{\mu, \oplus}^{(i)}$, \ie, $\mathcal{L}(f_{e, \oplus}^{(i)}) \cdot \mathcal{L}(f_{\mu, \oplus}^{(i)})$ instead of $\mathcal{L}(f_{e, \oplus}^{(i)}, f_{\mu, \oplus}^{(i)})$ in \equ{likelihood}, which disregarded the correlations between the allowed values of $f_{e, \oplus}^{(i)}$ and $f_{\mu, \oplus}^{(i)}$ that are evident in \figu{all_triangles}.  Because our analysis does not suffer from that limitation, the constraints we set later on the LIV parameters are more accurate than the early estimates in \Refe~\cite{Telalovic:2023tcb}.


\subsection{Model parameter priors}
\label{sec:stat_methods-priors}

Table~\ref{tab:fit_params} collects all the free model parameters and our choice of their priors.  For each of the standard oscillation parameters in $\boldsymbol{\theta}_{\rm std}$ we adopt a normal prior with central value and width given by its one-dimensional allowed range from NuFIT~5.2~\cite{Esteban:2020cvm}, assuming normal neutrino mass ordering and including Super-Kamiokande data.  The present-day uncertainty on the standard neutrino mixing parameters $\theta_{12}$, $\theta_{23}$, $\theta_{13}$, and $\delta_{\rm CP}$, is still large enough to result in large uncertainties in the predicted flavor composition at Earth, even in the absence of LIV; see, \eg, \Refes~\cite{Bustamante:2015waa, Song:2020nfh}.  Further, the uncertainties on $\Delta m_{21}^2$ and $\Delta m_{31}^2$ blur the value of the neutrino energy above which LIV effects may become dominant over standard oscillations and, as a consequence, affect the constraints we place on the LIV parameters.

[Because the allowed regions of the standard oscillation parameters are similar in the normal and inverted ordering, assuming the latter in our analysis instead would yield similar results.  Also, later versions of  NuFIT~\cite{Esteban:2024eli} have somewhat different allowed parameter ranges, but are still closely compatible with NuFIT 5.2, so using them instead would also not affect our results appreciably.  Finally, we disregard correlations that exist in NuFIT~5.2 between the allowed regions of different parameters (which are expected to shrink in the coming years~\cite{Song:2020nfh}). The only exception is that we use the NuFIT two-dimensional joint allowed prior for $\theta_{23}$ and $\delta_{\rm CP}$, between which correlations are presently appreciable.]

For the astrophysical spectral index, $\gamma$, we adopt a normal prior with central value and width set to those reported by the IceCube Collaboration in their analysis of the 7.5-year HESE sample~\cite{IceCube:2020wum}.  (Other IceCube analyses based on different event data samples report values for the spectral index different than the one we have adopted, but largely compatible with it; see, \eg, \Refe~\cite{Naab:2023xcz}.)  The uncertainty in $\gamma$ introduces blurs the relative number of low-energy \textit{vs.}~high-energy neutrinos.

For the $\nu_e$ fraction emitted by the sources, $f_{e, {\rm S}}$, we explore two possibilities.   First, we conservatively fix its value to each one of our benchmark neutrino production scenarios (Sec.~\ref{sec:astro_nu-nuisance_fS}) in turn---pion decay, muon-damped, and beta decay---so that the prior on $f_{e, {\rm S}}$ is a delta function.  Then, we also explore our flavor-agnostic scenario by sampling the value of $f_{e, {\rm S}}$ between 0 and 1 using a uniform prior, treating all possible realizations of the flavor composition at the sources (without $\nu_\tau$ production) as equally likely.

Finally, for the LIV parameters---the physical parameters that we are interested in---we treat their real and imaginary parts as independent parameters.  [When an LIV parameter is real due to Hermiticity (Sec.~\ref{sec:nu_oscillations-liv}), we only vary its real part, keeping its imaginary part null.]  To avoid introducing bias in our analysis, we adopt for the real and imaginary parts uniform priors over vast intervals that span negative and positive values.  We adjust the size of the interval depending on the operator dimension, $d$, in order to capture the fact that the value at which the LIV parameters become comparable to the standard oscillation parameters (\ie, where $H_{\rm vac} \sim H_{\rm LIV}$) shifts appreciably with $d$. 

\begin{figure*}[t!]
 \centering
 \includegraphics[trim={0.2cm 0.1cm 0.2cm 0.1cm}, clip, width=\textwidth]{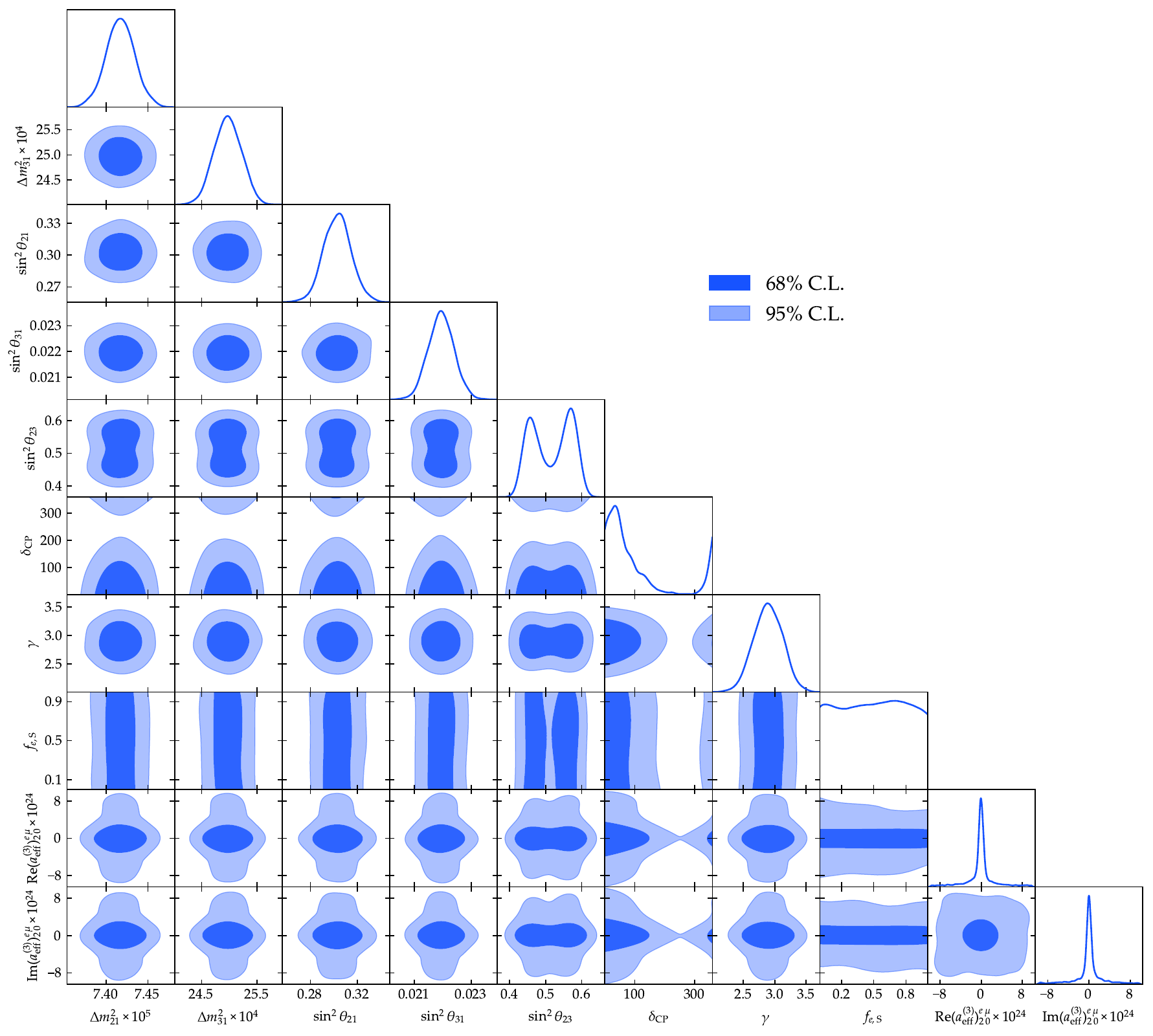}
 \caption{\textbf{\textit{Example pairwise joint posterior used to constrain an LIV parameter.}}  The LIV parameter in question is $(a^{(3)}_{\rm eff})^{e\mu}_{20}$, and we fit its real and imaginary parts separately.  The fit is produced assuming that this is the only nonzero LIV parameter---our default strategy---and under our flavor-agnostic prior on $f_{e, {\rm S}}$.  See Table~\ref{tab:fit_params} for the priors used and the units of the parameters.  We generate a similar posterior to constrain each of the LIV parameters in Tables \ref{tab:constraint_tables_d2}--\ref{tab:constraint_tables_d8}. }
 \label{fig:full_cornerplot}
\end{figure*}

\begin{figure*}[t!]
  \centering
  \includegraphics[width=\textwidth]{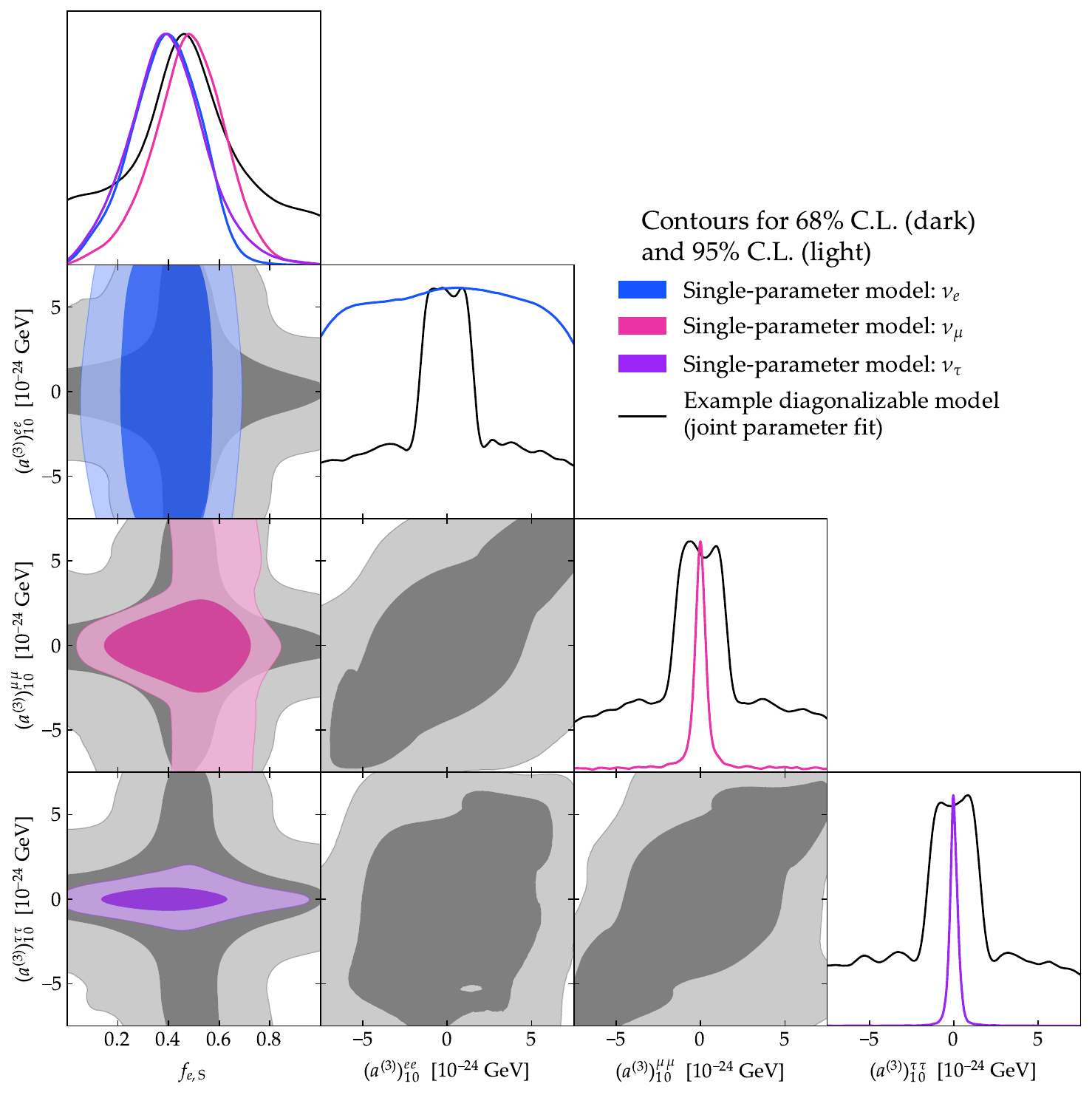}
  \caption{\textbf{\textit{Parameter posteriors computed under our two LIV exploration strategies: single-parameter and joint multiple-parameter.}}  For the latter, we use the example diagonalizable model introduced in Table~\ref{tab:fit_params}.  We show the posteriors only for a subset of four of the model parameters, all real-valued, produced assuming our flavor-agnostic prior on $f_{e, {\rm S}}$ (Table~\ref{tab:fit_params}).  In the single-parameter fits---our default strategy, which we use to produce the constraints in Table \ref{tab:constraint_tables_d2}--\ref{tab:constraint_tables_d8}---the LIV parameters are $(a_{\rm eff}^{(3)})_{10}^{ee}$, $(a_{\rm eff}^{(3)})_{10}^{\mu\mu}$, and $(a_{\rm eff}^{(3)})_{10}^{\tau\tau}$.  In the diagonalizable model, the LIV parameters are $(a_{\rm d}^{(3)})_{10}^{ee}$, $(a_{\rm d}^{(3)})_{10}^{\mu\mu}$, and $(a_{\rm d}^{(3)})_{10}^{\tau\tau}$, and they are fit jointly. \textbf{\textit{Jointly fitting multiple LIV parameters removes the constraints that arise when fitting them individually.}}  See Sec.~\ref{sec:results-posteriors} for details.
  }
  \label{fig:indep_vs_combined_cornerplot}
\end{figure*}


\subsection{Computing the posterior}
\label{sec:stat_methods-computing_posterior}

We use \textsc{UltraNest} to maximize the full joint posterior, \equ{likelihood}, compute the marginalized posterior in the LIV parameters, \equ{likelihood_marg}, and extract from them the best-fit and allowed ranges of the model parameters.

As stated earlier (Sec.~\ref{sec:liv_strategies-high_dim}), we pursue two strategies to explore the LIV parameter space.  Our default strategy, with which we produce or main results, is to consider a single nonzero LIV parameter at a time (Sec.~\ref{sec:liv_strategies-single_parameter}).   Under this strategy, we explore a 10-dimensional parameter space to maximize the full posterior (9-dimensional if the LIV parameter is real); see Table~\ref{tab:fit_params}.  

Separately, we adopt the alternative strategy where we jointly consider multiple nonzero LIV parameters at a time (Sec.~\ref{sec:liv_strategies-multiple_parameters}).  Since our goal with this is merely to illustrate the influence of correlations between LIV parameters, we explore the single case of the diagonalizable model introduced in Sec.~\ref{sec:liv_strategies-multiple_parameters}.  Under this strategy, we explore a 23-dimensional parameter space to maximize the full posterior; see Table~\ref{tab:fit_params}.


\section{Results}
\label{sec:results}

Reference~\cite{Telalovic:2023tcb} found no statistically significant evidence for the presence of flavor compass anisotropies in the 7.5-year public IceCube HESE sample.  In other words, using the HESE sample, the high-energy neutrino flavor composition inferred for each the 12 pixels into which we tessellate the sky is compatible with each other.  Therefore, we use the statistical procedure above (Sec.~\ref{sec:stat_methods-constraining_liv_parameters}) to set upper limits on the values of the LIV parameters.  (Reference~\cite{Telalovic:2023tcb} pointed out that, if flavor anisotropy does exist, it could be detectable in the next 10--20 years using the combined results of multiple neutrino telescopes.)  


\subsection{Joint posteriors}
\label{sec:results-posteriors}

Figure \ref{fig:full_cornerplot} shows the full posterior [\equ{likelihood}], arranged pairwise for all of the free parameters, and obtained by allowing a single nonzero example LIV parameter, under our flavor-agnostic prior on $f_{e, {\rm S}}$.  We produce one such posterior for each of the LIV parameters that we constrain; results for other LIV parameters are similar.  Predictably, \figu{full_cornerplot} shows that the posteriors of the nuisance parameters are dominated by their priors, which are already relative narrow (Table~\ref{tab:fit_params}).  (The double-peak structure in the posterior of $\sin^2 \theta_{23}$ comes from its prior and reflects the uncertainty in the quadrant of $\theta_{23}$~\cite{Esteban:2020cvm}.)   For the specific LIV parameter constrained in \figu{full_cornerplot}, the flavor composition at the source, $f_{e, {\rm S}}$, is unconstrained.  However, for other LIV parameters, $f_{e, {\rm S}}$ can be constrained, albeit weakly, as shown in \Refes~\cite{Bustamante:2019sdb, Song:2020nfh, Liu:2023flr}; see also the joint posterior on $f_{e, {\rm S}}$ and $(a_{\rm eff}^{(3)})_{10}^{ee}$ for the single-parameter fit in \figu{indep_vs_combined_cornerplot}.

Figure \ref{fig:indep_vs_combined_cornerplot} illustrates how our choice of exploration strategy of the LIV parameter space impacts the constraints that we can place on the LIV parameters.  Specifically, we contrast the constraints on $(a_{\rm eff}^{(3)})_{10}^{ee}$, $(a_{\rm eff}^{(3)})_{10}^{\mu\mu}$, and $(a_{\rm eff}^{(3)})_{10}^{\tau\tau}$ obtained in their single-parameter fits (Sec.~\ref{sec:liv_strategies-single_parameter}) against the constraints on the corresponding $(a_{\rm d}^{(3)})_{10}^{ee}$, $(a_{\rm d}^{(3)})_{10}^{\mu\mu}$, and $(a_{\rm d}^{(3)})_{10}^{\tau\tau}$ obtained by jointly fitting them in our example diagonalizable model (Sec.~\ref{sec:liv_strategies-multiple_parameters}).  [Figure \ref{fig:indep_vs_combined_cornerplot} displays only a subset of four model parameters, but all the model parameters of each exploration strategy were varied (Table~\ref{tab:fit_params}).] 

Figure \ref{fig:indep_vs_combined_cornerplot} makes explicitly the important caveat pointed out in Sec.~\ref{sec:liv_strategies-high_dim}:  \textbf{\textit{while the LIV parameters can be constrained in single-parameter fits, fitting several of them jointly may preclude constraining any one individually.}}  Moreover, the constraint on $f_{e, {\rm S}}$ that exists in the single-parameter fit of $(a_{\rm eff}^{(3)})_{10}^{ee}$ disappears in the fit to the diagonalizable model.  Other than in \figu{indep_vs_combined_cornerplot}, all the constraints on LIV parameters that we report, including in Tables \ref{tab:constraint_tables_d2}--\ref{tab:constraint_tables_d8}, are obtained using single-parameter fits, and we warn against interpreting them otherwise (\eg, building models with multiple nonzero LIV parameters where the value of each one is constrained by our results).

\begin{figure*}[t!]
 \centering
 \includegraphics[trim={0.25cm 0cm 0.2cm 0cm}, clip, width=\textwidth]{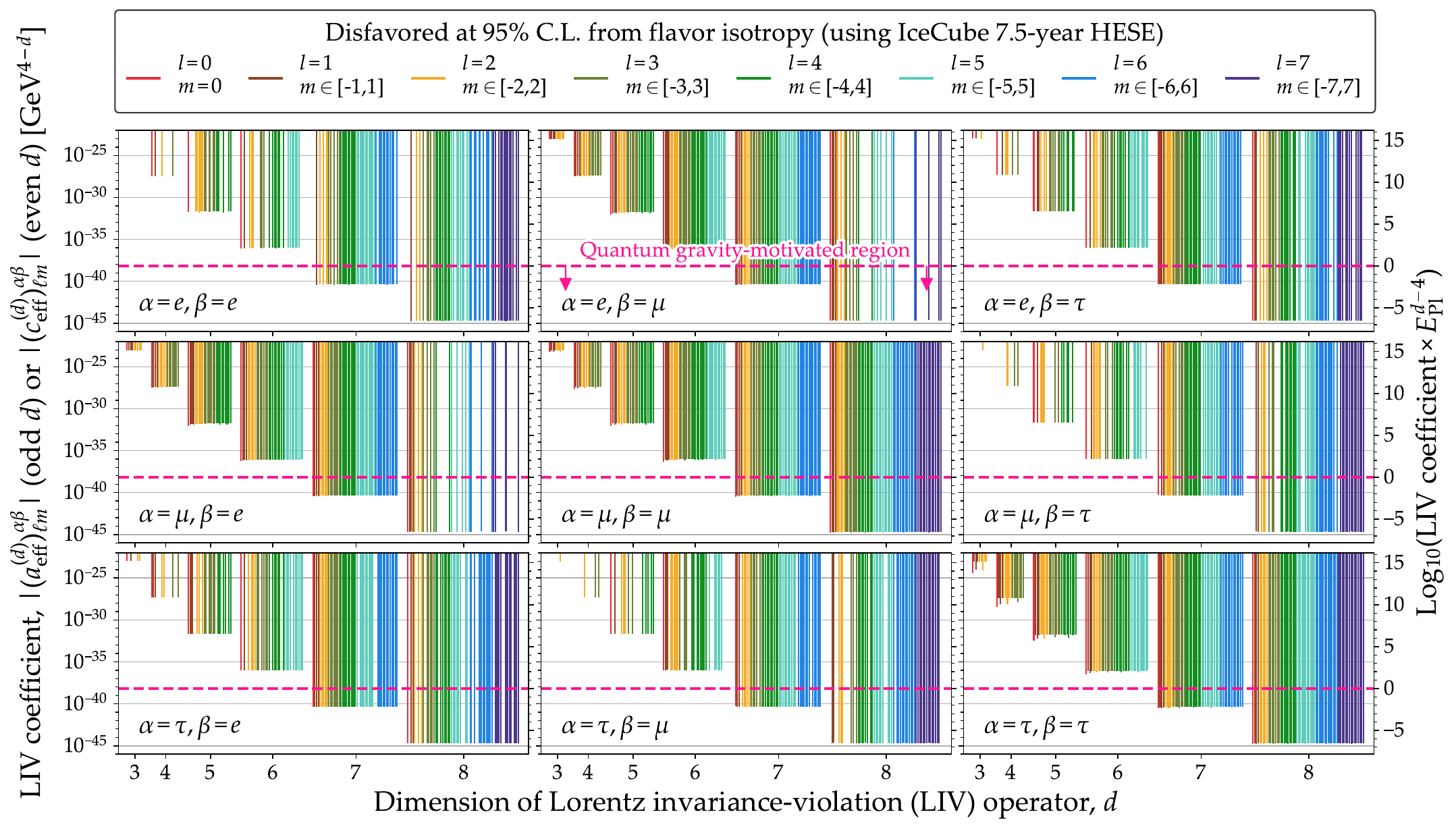}
 \caption{\textbf{\textit{Overview of our new upper limits on LIV parameters.}}  The LIV parameters appear in the spherical-harmonic expansion of CPT-odd and CPT-even LIV operators, of dimension $d$ and with harmonic modes $\ell$ and $m$, that affect transitions between $\nu_\alpha$ and $\nu_\beta$, and that are proposed by the Standard Model Extension~\cite{Kostelecky:2003fs, Kostelecky:2011gq}; see Eqs.~(\ref{equ:sph_exp_a}) and (\ref{equ:sph_exp_c}).
 The limits placed on them come from searching for LIV-induced anisotropies in the flavor-composition skymaps of high-energy astrophysical neutrinos, inferred from 7.5 years of public IceCube HESE data~\cite{IceCube:2020wum, IC75yrHESEPublicDataRelease}.   The limits shown here are for single-parameter fits for $d > 2$ and are grouped according to the value of $d$.  They were obtained by varying a single nonzero LIV parameter at a time and using a flat prior for $f_{e,{\rm S}}$ (flavor-agnostic prior in Table~\ref{tab:fit_params}).  LIV parameters unconstrained by our analysis are omitted; they appear as blank vertical lines.   The constraints improve for higher operator dimensions, due to their growing energy dependence; they enter the quantum-gravity motivated region for operator dimensions $d = 7, 8$.  \textbf{\textit{We place upper limits on 1071 LIV parameters; for 816 of them, we improve limits or place them for the first time.}}  See Tables \ref{tab:constraint_tables_d2}--\ref{tab:constraint_tables_d8} for the numerical values of the limits, Sec.~\ref{sec:stat_methods} for details on our limit-setting procedure, and Sec.~\ref{sec:results} for a discussion of our results. }
 \label{fig:limits_overview}
\end{figure*}

\begin{figure*}[t!]
 \centering
 \includegraphics[width=\textwidth]{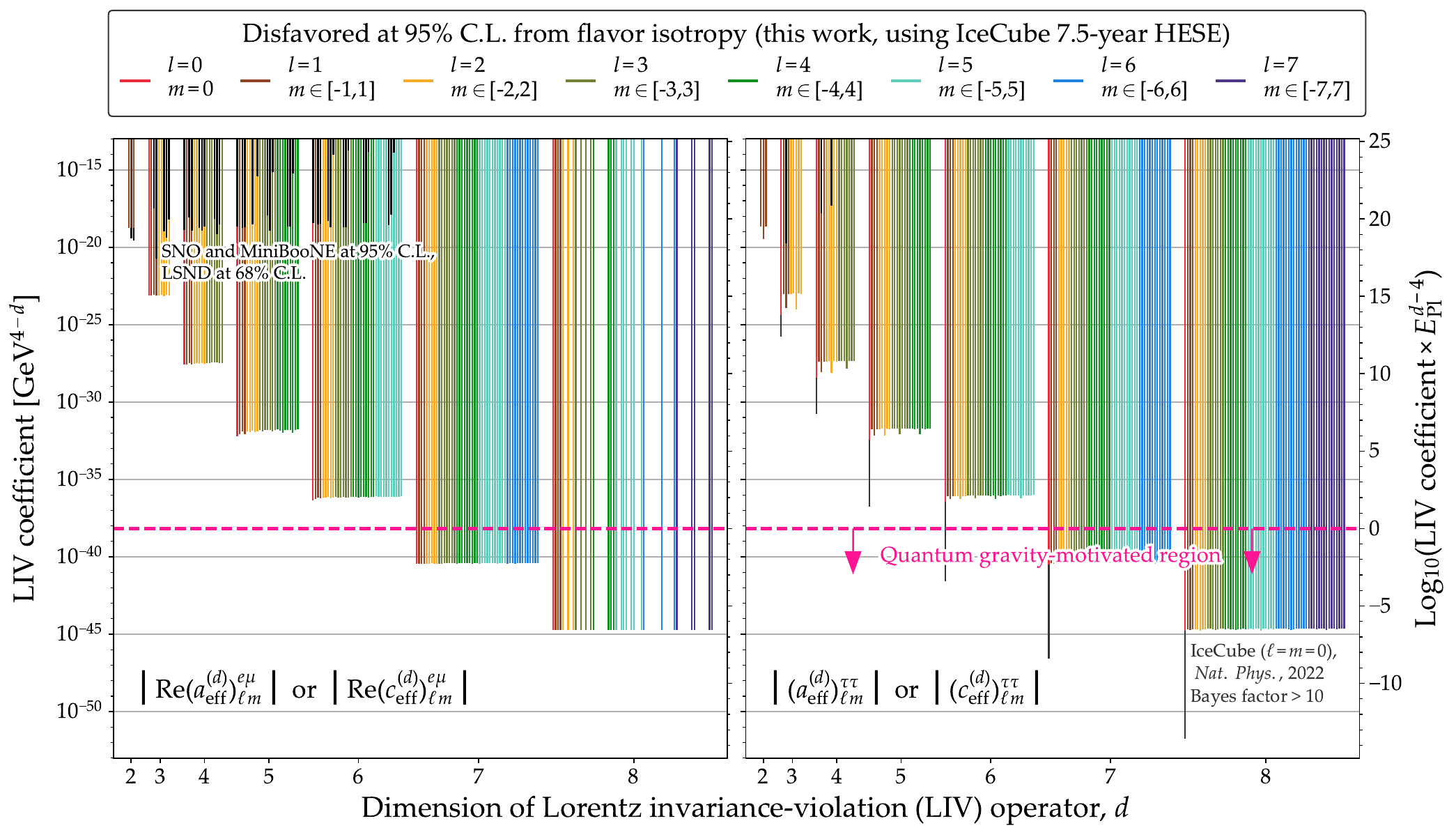}
 \caption{\textbf{\textit{Comparison between our new upper limits on LIV parameters and previous limits.}}  In this figure, we show only limits for the example parameters $(a_{\rm eff}^{(d)})_{\ell m}^{e \mu}$ and $(c_{\rm eff}^{(d)})_{\ell m}^{e \mu}$; \figu{limits_overview} shows limits for all parameters.  Limits from previous analyses are from solar neutrinos in SNO~\cite{SNO:2018mge}, accelerator neutrinos from LSND~\cite{LSND:2005oop} and MiniBooNE~\cite{MiniBooNE:2011pix}, and high-energy astrophysical neutrinos in IceCube~\cite{IceCube:2021tdn}.  \textit{Left:} Limits on the real parts of $(a^{(d)}_{\rm eff})^{e\mu}_{\ell m}$ and $(c^{(d)}_{\rm eff})^{e\mu}_{\ell m}$.  \textit{Right:} Limits on the norms of $(a^{(d)}_{\rm eff})^{\tau\tau}_{\ell m}$ and $(c^{(d)}_{\rm eff})^{\tau\tau}_{\ell m}$.  \textbf{\textit{Our limits improve on previous limits derived from lower-energy neutrinos in SNO, MiniBooNE, and LSND.}} Our limits on isotropic LIV parameters fall short of those placed by IceCube due to the difference in the minimum neutrino energy used in the analyses, which makes our limits conservative.  See Sec.~\ref{sec:results-comparison_previous_limits} for details.}
 \label{fig:limits_comparison}
\end{figure*}


\subsection{New upper limits on LIV parameters}
\label{sec:results-limits}

Tables \ref{tab:constraint_tables_d2}--\ref{tab:constraint_tables_d8} in Appendix~\ref{sec:constraint_tables} contain our main results: new upper limits on the LIV parameters from operators dimension $d = 2$ to 8.  These are one-dimensional marginalized upper limits on $(a_{\rm eff}^{(d)})_{\ell m}^{\alpha\beta}$ and $(c_{\rm eff}^{(d)})_{\ell m}^{\alpha\beta}$, obtained using the posterior in \equ{likelihood_marg} by adopting our default strategy of considering a single nonzero LIV parameter at a time.  We present results for our four choices of prior on $f_{e, {\rm S}}$ in Table~\ref{tab:fit_params}.  Results obtained assuming neutrino production via the full pion decay chain ($f_{e, {\rm S}} = 1/3$) are arguably the most physically motivated ones, while results obtained by marginalizing over $f_{e, {\rm S}}$ (flavor-agnostic) are the most conservative ones.  

We constrain a vastly larger number of LIV parameters in the neutrino sector than ever before.  In total, we set out to constrain 1818 separate parameters from $d = 2$ to 8, some real (\ie, those with $m=0$ and $\alpha = \beta$), some complex.  \textbf{\textit{We successfully constrain 1071 of them}}---either their real part, imaginary part, or their norm---under at least one of our benchmark choices of flavor composition at the sources, or under our flavor agnostic scheme (see Table~\ref{tab:fit_params}).  Out of them, \textbf{\textit{we either constrain for the first time or improve on the constraints of 816--905 LIV parameters}}, the exact number depending on whether we consider constraints on their real part, imaginary part, or their norm.  Of the 747 parameters that we are unable to constrain, only 15--18 of them had been constrained by previous analyses.  

Below, we summarize trends in our limits and comment on cases where are not able to place a limit.

Figure~\ref{fig:limits_overview} shows a summary of the limits contained in Tables \ref{tab:constraint_tables_d2}--\ref{tab:constraint_tables_d8}, obtained assuming our flavor-agnostic prior on $f_{e, {\rm S}}$.  Because the limits on flavor anisotropy come from the regime where the standard oscillations and LIV contributions are comparable in size, \ie, $H_{\rm vac} \sim H_{\rm LIV}$ in \equ{hamiltonian_total} (see Sec.~\ref{sec:liv_strategies-single_parameter}) and because the LIV Hamiltonian grows $\propto E^{d-3}$ [\equ{h_liv_d}], we set stronger limits on the LIV parameter values as $d$ grows, dipping below the quantum gravity-motivated Planck scale for $d \geq 7$.

For each value of $d$, most LIV parameters are constrained to within an order of magnitude of each other.  The constraints on parameters with different values of $\alpha$ and $\beta$ are similar because the size of the allowed region of flavor composition in each of the pixels is comparable along the $f_{e, \oplus}$, $f_{\mu, \oplus}$, and $f_{\tau, \oplus}$ axes (see \figu{all_triangles}).  

We see slightly better constraints in lower-$\ell$ parameters, since they generate large-scale angular structures that are more easily captured by our coarse tessellation of the sky.  Not all low-$\ell$ modes generate anisotropies that are visible in our coarse tessellation, but even in some the cases where they do not we can constrain the value of the LIV parameter via the modification of the all-sky flavor composition that they induce, as pointed out in Sec.~\ref{sec:liv_strategies-sensitivity_origin}.  This is because the region of allowed flavor composition expected at Earth---even under our flavor-agnostic prior---is significantly smaller under standard oscillations than under LIV; see, \eg, \Refes~\cite{Arguelles:2015dca, Bustamante:2015waa, Song:2020nfh, Liu:2023flr}.


\subsection{Comparison to previous limits}
\label{sec:results-comparison_previous_limits}

\textbf{\textit{Our new limits on anisotropic LIV parameters (i.e., those with $\mathbf{\ell \neq 0}$) are the strongest ones to date;}} \cf~the limits from MiniBooNE placed in \Refe~\cite{Kostelecky:2011gq}.  \textbf{\textit{For most of the parameters we explore, our limits are the first ever to be placed,}} as seen in a comparison to the limit tables in \Refe~\cite{Kostelecky:2008ts}.

Reference~\cite{Kostelecky:2008ts} (Tables D32--D39) maintains an updated list of limits on LIV parameters within the SME.  Below, we contrast our limits to previous ones from IceCube and lower-energy experiments.  Although limits exist for LIV parameters for operator dimensions up to $d = 10$~\cite{Kostelecky:2008ts}, the coverage of the different harmonic modes, $\ell$ and $m$, and flavors, $\alpha$ and $\beta$, is sparse.  This is in contrast to our results, where we have comprehensively explored all CPT-odd and CPT-even LIV parameters at every value from $d = 2$ to 8. 

Figure~\ref{fig:limits_comparison} shows a comparison between the constraints we place on a selection of LIV parameters in this work and constraints placed by previous analyses, taken from \Refe~\cite{Kostelecky:2008ts}.  We elaborate on their differences below. 

\smallskip

\textbf{\textit{Limits from lower energy.---}}In the neutrino sector, other than the limits on LIV parameters derived from IceCube high-energy astrophysical~\cite{IceCube:2021tdn} and atmospheric~\cite{IceCube:2017qyp} neutrinos (more on them below), existing limits come from experiments that detect lower-energy neutrinos, \eg, SNO~\cite{SNO:2018mge}, LSND~\cite{LSND:2005oop}, MiniBooNE~\cite{MiniBooNE:2011pix} as shown in \figu{limits_comparison}.

For the CPT-even LIV parameters with $d=2$, $(c_{\rm eff}^{(2)})_{\ell m}^{\alpha\beta}$, the operators are energy-independent, setting our analysis on equal ground with said neutrino experiments.  Our limits (Table~\ref{tab:constraint_tables_d2}) are in the range $10^{-19}$--$10^{-22}~{\rm GeV}^2$; previous upper limits are in Table~D32 of \Refe~\cite{Kostelecky:2008ts}.  Limits from LSND and MiniBooNE~\cite{Kostelecky:2011gq} are comparable to ours, at $10^{-19}$--$10^{-20}~{\rm GeV}^2$.  
Limits from tritium decay~\cite{KATRIN:2022qou} are weaker, at about $10^{-17}~{\rm GeV}^2$ (when placed on the absolute value of the parameter).

For larger values of $d$, the LIV operators are energy-dependent.  Previous limits on the LIV parameters are in Tables D32--D38 in \Refe~\cite{Kostelecky:2008ts}.  They predominantly come from LSND and MiniBooNE, as derived in Ref.~\cite{Kostelecky:2011gq}, from SNO~\cite{SNO:2018mge}, and from IceCube (on isotropic LIV parameters only) using atmospheric~\cite{IceCube:2017qyp} and astrophysical neutrinos~\cite{IceCube:2021tdn}.  The IceCube limits, similar to the ones presented here, are stronger than the others due to the higher neutrino energies [\equ{liv_coeff_sensitivity}].

Figure~\ref{fig:limits_comparison} shows that, for some of the LIV parameters previously constrained, our limits do not show a stark improvement.  For instance, SNO placed an upper limit on $(a_{\rm eff}^{(3)})_{11}^{e\mu}$ of $10^{-21}$~GeV, whereas we place it at $10^{-23}$~GeV (Table~\ref{tab:constraint_tables_d3}).  This is weaker than our expected sensitivity estimated in \equ{liv_coeff_sensitivity}.  This is because, in this and similar cases in our analysis, the parameter in question either shows a strong degeneracy with the flavor composition at the sources, $f_{e,{\rm S}}$, or does not generate a significant departure in the flavor composition at the Earth relative to the standard-oscillation expectation.  This is particularly common in isotropic ($\ell = 0$) or higher-$\ell$ parameters, where there are either no anisotropies in the flavor-composition skymap or they are predominantly at smaller angular scales, and so are washed out by our coarse sky tessellation (Sec.~\ref{sec:astro_nu-flavor}).

\smallskip

\textbf{\textit{Comparison to previous IceCube limits.---}}In \Refe~\cite{IceCube:2021tdn}, the IceCube Collaboration used the measurement of the all-sky flavor composition in the 7.5-year HESE sample to constrain exclusively isotropic LIV parameters (\ie, those with $\ell = m = 0$), for operator dimensions $d = 3$ to 8.   Figure~\ref{fig:limits_comparison} illustrates that our constraints on the same isotropic LIV parameters constrained by IceCube are weaker than theirs.  For the lower dimensions, $d = 3, 4$, our constraints are up to two orders of magnitude weaker; for the highest dimension constrained, $d = 8$, our constraints are 5--10 orders of magnitude weaker. 

The primary cause of this difference is that the IceCube analysis from \Refe~\cite{IceCube:2021tdn} and ours use different values of the minimum astrophysical neutrino energy.  This matters because, as pointed out in Sec.~\ref{sec:nu_oscillations-liv_sensitivity}, because the astrophysical neutrino spectrum is a steeply falling function of energy, the constraints on the LIV parameters come predominantly from the lowest-energy neutrinos, which are vastly more abundant. 

Via \equ{liv_coeff_sensitivity}, both the analysis by the IceCube Collaboration and ours are sensitive to values of the LIV parameter $(a_{\rm eff}^{(d)})_{\ell m}^{\alpha\beta} \gtrsim 10^{-23}~(E_{\rm min}/{\rm GeV})^{2-d}~{\rm GeV}^{4-d}$, where $E_{\rm min}$ is the minimum astrophysical neutrino energy of the analysis.  In the IceCube analysis from \Refe~\cite{IceCube:2021tdn}, this is close to $E_{\rm min} = 60$~TeV, yielding sensitivity to $(a_{\rm eff}^{(d)})_{\ell m}^{\alpha\beta} \gtrsim 10^{-14-4d-0.4d^{1.3}}~{\rm GeV}^{4-d}$. (In full rigor, 60~TeV is the energy of the lowest-energy detected event in the HESE sample used in \Refe~\cite{IceCube:2021tdn}, not the minimum neutrino energy, but they are similar.)  In contrast, in our analysis, the minimum energy is lower, $E_{\rm min} = 10$~TeV, yielding sensitivity to larger values of $(a_{\rm eff}^{(d)})_{\ell m}^{\alpha\beta} \gtrsim 10^{-15-4d}~{\rm GeV}^{4-d}$.  The difference between the sensitivity of \Refe~\cite{IceCube:2021tdn} and ours matches the differences in the upper limits on isotropic LIV parameters in \Refe~\cite{IceCube:2021tdn} and Tables~\ref{tab:constraint_tables_d2}--\ref{tab:constraint_tables_d8}.  The difference grows with $d$ because higher-dimension LIV operators have a stronger energy dependence [\equ{h_liv_matrix}].

The IceCube analysis in \Refe~\cite{IceCube:2021tdn} and ours use the same 7.5-year sample of HESE events.  The choice of $E_{\rm min} = 60$~TeV in \Refe~\cite{IceCube:2021tdn} was made to mitigate the impact of the atmospheric neutrino and muon background.  Our choice of $E_{\rm min} = 10$~TeV follows from the choice made in \Refe~\cite{Telalovic:2023tcb} when extracting the directional flavor composition that we use as the basis of our analysis (Sec.~\ref{sec:stat_methods-inferring_flavor_composition}).  The decision to use a lower value of 10~TeV was made to increase the available number of events from which to extract the directional flavor composition.  In doing so, \Refe~\cite{Telalovic:2023tcb} also mitigated the contamination of the atmospheric background, using similar, though less comprehensive background modeling as \Refe~\cite{IceCube:2021tdn}.  

\textbf{\textit{Ultimately, our decision to use a lower value of $\mathbf{E_{\rm min}}$ makes our upper limits on the LIV parameters conservative.}}  For the isotropic LIV parameters, our limits are unsurprisingly outclassed by the limits placed by the IceCube Collaboration, for the reason above.  In contrast, for the anisotropic LIV parameters, our limits, though conservative, outclass existing ones coming from lower-energy neutrinos and, for many parameters, are the first ones to be placed.  Future searches for directional flavor anisotropy, performed on larger data sets, could consider using a higher value of $E_{\rm min}$, possibly leading to more stringent limits.

There are further differences between the IceCube analysis in \Refe~\cite{IceCube:2021tdn} and ours.  First, the IceCube analysis constrains the LIV parameters by comparing mock HESE event samples that include the effects of LIV {\it vs.}~the observed 7.5-year HESE sample.  In contrast, our analysis does so by comparing mock distributions of the flavor composition in the sky that include the effects of LIV {\it vs.} the directional flavor composition extracted in \Refe~\cite{Telalovic:2023tcb}.  The latter, however, {\it was} derived from an analysis of the 7.5-year HESE events, as explained in Sec.~\ref{sec:stat_methods-inferring_flavor_composition}, using a detailed detector response similar to the one used in \Refe~\cite{IceCube:2021tdn}, which grants credibility to our results.  We perform our analysis at the flavor-composition level rather than at the event level because the latter would be too computationally demanding, given the large number of LIV parameters that we study.

Second, when computing the diffuse flux of high-energy neutrinos at Earth, we account for the evolution of the number density of sources with redshift, unlike \Refe~\cite{IceCube:2021tdn}.  (For this, we adopted the star-formation rate, but in Sec.~\ref{sec:astro_nu-nuisance_redshift}, we showed that deviations from it do not impact the expected flavor composition at Earth significantly.) As a result, in our analysis, the LIV-induced features in the flavor composition are smeared in energy due to the redshifting of neutrino energy from adiabatic cosmological expansion, making them harder to meaningfully affect the energy-averaged flavor composition, \equ{flavor_ratio}, that we use in our analysis.  Neglecting this smearing would make the LIV-induced features in the flavor composition sharper than they ought to be.

Third, and finally, the IceCube analysis in \Refe~\cite{IceCube:2021tdn} constrains only the real part of the LIV parameters, even if in reality they are complex numbers, arguing that \Refe~\cite{IceCube:2017qyp} found there is little sensitivity to the phase of the parameters.  In contrast, for complex-valued LIV parameters, we constrain their real and imaginary parts separately, and, from them, also their norm.  We derive the posterior of the norm from the posteriors of the real and imaginary parts, and extract the constraint from it.  Because of this, for some of the LIV parameters we are able to constrain either their real or imaginary parts, but not their norm; see, \eg, parameter $(c_{\rm eff}^{(2)})_{10}^{e\tau}$ under our flavor-agnostic prior in Table~\ref{tab:constraint_tables_d2}.


\section{Summary and outlook}
\label{sec:summary}

High-energy cosmic neutrinos, with energies between tens of TeV and a few PeV, provide us with novel opportunities to discover new neutrino physics. Because of their high energies, they may manifest new-physics effects that are suppressed at lower neutrino energies and that would otherwise go unnoticed.  We have used high-energy cosmic neutrinos to look for signs of the violation of Lorentz invariance, one of the pillars of the Standard Model.  Lorentz invariance may manifest in multiple forms, including, as we have explored here, as there being preferred directions for the propagation of neutrinos.

We have taken this one step further and looked for the possibility that different neutrino flavors have different preferred directions, opening up a vast sector of anisotropic Lorentz-invariance violation (LIV) that hitherto was poorly studied.  This form of LIV would manifest as anisotropies in the distributions of the arrival directions of high-energy cosmic neutrinos of different flavors to Earth.  We have devised sensitivity to these anisotropies by combining existing detection capabilities of the IceCube neutrino telescope: the distribution of neutrino energies, their flavor composition---\ie, the proportion of $\nu_e$, $\nu_\mu$, and $\nu_\tau$ that reaches Earth---and the distribution of their arrival directions.  

Reference~\cite{Telalovic:2023tcb} introduced methods to combine these capabilities in order to infer the high-energy neutrino flavor composition along different directions in the sky, relying on detailed modeling of the IceCube neutrino telescope.  Reference~\cite{Telalovic:2023tcb} found no evidence for a directional variation in the flavor composition in the 7.5-year public IceCube sample of High-Energy Starting Events (HESE)~\cite{IceCube:2020wum, IC75yrHESEPublicDataRelease}.  Building on that, we have looked for evidence of the specific patterns of flavor anisotropy in the sky predicted by LIV.  As expected, we found none and, as a result, we have placed limits on the parameters that control the shape and size of LIV.  

In computing our new limits on LIV parameters, we have accounted for the most relevant uncertainties from particle physics---the values of the standard neutrino oscillation parameters---and from astrophysics---the shape of the neutrino energy spectrum and the flavor composition with which neutrinos are produced at their astrophysical sources.  In our statistical analysis, we have used informed priors on all of these parameter---based on a global fit to oscillation data and on IceCube results.  The exception is the flavor composition at the sources, about which, by default, we have assumed no knowledge other than $\nu_\tau$ not being produced at the sources.

Tables \ref{tab:constraint_tables_d2}--\ref{tab:constraint_tables_d8} show our main results: upper limits on 1071 complex-valued parameters that regulate LIV, both CPT-preserving and CPT-violating, associated to LIV operators with dimensions from 2 to 8 posited by the Standard Model Extension~\cite{Kostelecky:2003fs, Kostelecky:2011gq}.  Different operators induce different patterns of flavor anisotropy in the high-energy neutrino sky and have different energy dependence.  For 816 of these parameters, we tighten existing constraints by orders of magnitude or constrain them for the first time.

The main obstacle to looking for flavor anisotropies is the larger error in the reconstructed direction of many HESE events, especially in shower events, which can be as large as tens of degrees.  This restricts our analysis to searching for anisotropies on large angular scales and requires careful statistical treatment to avoid overstating the flavor isotropy of the sky, which remains untested at small angular scales.  We expect that this will be overcome in the near future by the operation of new neutrino telescopes KM3NeT~\cite{KM3Net:2016zxf} and Baikal-GVD~\cite{Baikal-GVD:2019kwy}, expected to have better angular resolution for shower events. 

In the next 10--20 years, an analysis of data collected by a global network of upcoming neutrino TeV--PeV telescopes could improve the limits on anisotropic LIV parameters by about one order of magnitude~\cite{Telalovic:2023tcb} (see also \Refe~\cite{Schumacher:2025qca} for a broader perspective).  Progress may be swifter thanks to the current fast-paced advances in improving energy and angular reconstruction, aided by machine learning~\cite{IceCube:2023ame, Abbasi:2022ypr, Sogaard:2022qgg, Bukhari:2023ezc}, and in new techniques for flavor identification, like the use of dedicated templates~\cite{IceCube:2020fpi, IceCube:2023fgt} and muon and neutron echoes~\cite{Li:2016kra, Steuer:2017tca, Farrag:2023jut}.  In addition, the observation of large samples of ultra-high-energy neutrinos~\cite{Valera:2022wmu, MammenAbraham:2022xoc, Ackermann:2022rqc}, with energies in the EeV range---as hinted by the recent observation by KM3NeT~\cite{KM3NeT:2025npi}---coupled with newly proposed techniques to infer their flavor composition~\cite{Testagrossa:2023ukh, Coleman:2024scd},  would considerably tighten the limits on the LIV parameters

\textbf{\textit{The present-day lack of evidence for neutrinos of different flavors having preferred directions of propagation has allowed us to strongly constrain the values of hundreds of LIV parameters.}} As the field of high-energy cosmic neutrinos matures, so will the tests of new physics that we are able to perform.

\smallskip

\textbf{\textit{Availability of digitized limit tables.---}} Appendix~\ref{sec:constraint_tables} contains tables of limits at the 95\%~confidence level (C.L.).  They are available digitally on GitHub~\href{https://github.com/BernieTelalovic/LIV_constraints_from_HESE_flavour_isotropy}{\faGithubSquare}, together with limits at the 68\% and 99\% C.L.  

\vspace*{-0.5cm}


\section*{Acknowledgments}

MB and BT are supported by {\sc Villum Fonden} under project no.~29388.  This work used resources provided by the High-Performance Computing Center at the University of Copenhagen. 

\appendix

\section{Tables of constraints on Lorentz-invariance-violating parameters}
\label{sec:constraint_tables}

\renewcommand{\theequation}{A\arabic{equation}}
\renewcommand{\thefigure}{A\arabic{figure}}
\renewcommand{\thetable}{A\arabic{table}}
\setcounter{figure}{0} 
\setcounter{table}{0} 

Tables \ref{tab:constraint_tables_d2}--\ref{tab:constraint_tables_d8} contain our new one-dimensional marginalized upper limits on the Standard Model Extension~\cite{Kostelecky:2003fs, Kostelecky:2011gq} parameters that describe CPT-odd and CPT-even Lorentz-invariance violation (LIV) in neutrinos.  The limits come from searching for LIV-induced ``flavor compass anisotropies'' in high-energy (TeV--PeV) astrophysical neutrinos.  These are anisotropies in the flavor-composition skymaps of $\nu_e$, $\nu_\mu$, and $\nu_\tau$, \ie, variations in the proportion of each neutrino flavor detected across the sky.  We search for them in data collected by the IceCube neutrino telescope; specifically, in the public 7.5-year sample of High-Energy Starting Events (HESE)~\cite{IceCube:2020wum, IC75yrHESEPublicDataRelease}.  See the main text (also \Refe~\cite{Telalovic:2023tcb}) for details on the data and our procedure to compute limits.

Each limit in Tables \ref{tab:constraint_tables_d2}--\ref{tab:constraint_tables_d8} was obtained assuming that the only nonzero LIV parameter is the one being constrained.  The constraints are shown at the 95\%~C.L.  Blank entries, marked with ``---'', denote parameters that cannot be constrained with this statistical significance.  See below for limits at a different significance.  (Sections~\ref{sec:liv_strategies-multiple_parameters} and \ref{sec:results-posteriors} in the main text contain constraints obtained by varying multiple LIV parameters jointly.)

The Hermiticity of the LIV operators (Sec.~\ref{sec:nu_oscillations-liv}) enforces that $(a_{\rm eff}^{(d)})^{\alpha\beta}_{\ell m} = (-1)^{m} (a_{\rm eff}^{(d)})^{\beta\alpha}_{\ell (-m)}$, and $(c_{\rm eff}^{(d)})^{\alpha\beta}_{\ell m} = (-1)^{m} (c_{\rm eff}^{(d)})^{\beta\alpha}_{\ell (-m)}$.  Thus, we show limits only for the LIV parameters with $m \geq 0$.  Because of the Hermiticity condition, $(a^{(d)}_{\rm eff})^{\alpha\alpha}_{\ell 0}$ and $(c^{(d)}_{\rm eff})^{\alpha\alpha}_{\ell 0}$ are real-valued. In these cases, the constraint on the real part and on the norm are the same.  The imaginary parts of real-valued parameters are null---and, therefore, not free parameters in our analysis---and are denoted by ``N/A'' in the tables.

If an LIV parameter has both a real and an imaginary part, and either one cannot be constrained individually, then the norm cannot be constrained either; see, \eg, parameter $(a_{\rm eff}^{(3)})_{00}^{\tau\mu}$ under $(1/3, 2/3, 0)_{\rm S}$ in Table~\ref{tab:constraint_tables_d3}. Section~\ref{sec:stat_methods} in the main text contains details about the procedure we use to compute limits. 

As shown in Sec.~\ref{sec:astro_nu-nuisance_fS} in the main text, the flavor composition of the flux of high-energy neutrinos that reaches Earth depends on the flavor composition with which the neutrinos were created in their astrophysical sources, which is unknown.  Thus, we have produced limits under different well-motivated possibilities of the flavor composition at the sources, expressed as $\left( f_{e, {\rm S}}, f_{\mu, {\rm S}}, f_{\tau, {\rm S}} \right)$, where $f_{\alpha, {\rm S}}$ is the fraction of $\nu_\alpha + \bar{\nu}_\alpha$ in the total neutrino flux.  Motivated by theory expectations (Sec.~\ref{sec:astro_nu-flavor}), we assume no $\nu_\tau$ are produced in high-energy astrophysical neutrino sources, \ie, $f_{\tau, {\rm S}} = 0$, so that the flavor composition is specified solely by $f_{e, {\rm S}}$, with $f_{\mu, {\rm S}} = 1-f_{e, {\rm S}}$.

The tables below contain limits obtained using four choices of prior on $f_{e, {\rm S}}$ (Table~\ref{tab:fit_params}): for each of our three benchmark scenarios of neutrino production (Sec.~\ref{sec:astro_nu-flavor})---where $f_{e, {\rm S}}$ is fixed to $1/3$ (pion decay), 0 (muon-damped), or 1 (beta decay)---and for a ``flavor-agnostic'' scenario where we fully account for our ignorance on $f_{e, {\rm S}}$ by varying it uniformly between 0 and 1.

The limits that are astrophysically most likely are those generated assuming production by pion decay, $\left( \frac{1}{3}, \frac{2}{3}, 0 \right)_{\rm S}$, while the limits that are most conservative are those generated under the flavor-agnostic scenario.

The tables group parameters by the dimension, $d$, of the LIV operators (Sec.~\ref{sec:nu_oscillations-liv}):
\begin{description}
 \item[Table~\ref{tab:constraint_tables_d2}]
  CPT-even operators with $d = 2$, $(c_\textrm{eff}^{(2)})_{\ell m}^{\alpha \beta}$
 \item[Table~\ref{tab:constraint_tables_d3}]
  CPT-odd operators with $d = 3$, $(a_\textrm{eff}^{(3)})_{\ell m}^{\alpha \beta}$
 \item[Table~\ref{tab:constraint_tables_d4}]
  CPT-even operators with $d = 4$, $(c_\textrm{eff}^{(4)})_{\ell m}^{\alpha \beta}$
 \item[Table~\ref{tab:constraint_tables_d5}]
  CPT-odd operators with $d = 5$, $(a_\textrm{eff}^{(5)})_{\ell m}^{\alpha \beta}$
 \item[Table~\ref{tab:constraint_tables_d6}]
  CPT-even operators with $d = 6$, $(c_\textrm{eff}^{(6)})_{\ell m}^{\alpha \beta}$
 \item[Table~\ref{tab:constraint_tables_d7}]
  CPT-odd operators with $d = 7$, $(a_\textrm{eff}^{(7)})_{\ell m}^{\alpha \beta}$
 \item[Table~\ref{tab:constraint_tables_d8}]
  CPT-even operators with $d = 8$, $(c_\textrm{eff}^{(8)})_{\ell m}^{\alpha \beta}$
\end{description}
These tables are available in digital form for download on GitHub \href{https://github.com/BernieTelalovic/LIV_constraints_from_HESE_flavour_isotropy}{\faGithubSquare}, including constraints also at 68\% and 99\%~C.L.

\begin{table}[t!]
    \caption{Upper limits ($95\%$ C.L.) on CPT-even LIV coefficients for $d=2$, $(c^{(2)}_{\rm eff})^{\alpha\beta}_{\ell m}$. The $\ell=0$ mode is not present in the SME for $d=2$, since its parameters are isotropic and share the same dimension as the standard oscillation Hamiltonian.\\ \label{tab:constraint_tables_d2}}
    \renewcommand{\arraystretch}{1}
    \begin{adjustbox}{max width=\textwidth}\normalsize
\begin{tabular}{c c | c c c | c c c | c c c | c c c}
\\ & LIV & \multicolumn{12}{c}{Neutrino flavor composition at the astrophysical sources, $(f_e,f_\mu,f_\tau)_{\rm S}$} \\
 $\ell$ & coefficient & \multicolumn{3}{c}{Any $(f_{e, {\rm S}}, 1-f_{e, {\rm S}}, 0)$} & \multicolumn{3}{c}{$\left(1/3,1/3,0\right)_{\rm S}$} & \multicolumn{3}{c}{$\left(1,0,0\right)_{\rm S}$}&\multicolumn{3}{c}{$\left(0,1,0\right)_{\rm S}$}\\
  & [GeV$^2$] & Norm & Re & Im & Norm & Re & Im & Norm & Re & Im & Norm & Re & Im
 \\ \hline
\\[-1.25em]
1 & $(c^{(2)}_{\rm eff})^{e e}_{10}$ & --- & --- & N/A & --- & --- & N/A & $1 \cdot 10^{-22}$ & $1 \cdot 10^{-22}$ & N/A & $6 \cdot 10^{-21}$ & $6 \cdot 10^{-21}$ & N/A\\[0.3em]
 & $(c^{(2)}_{\rm eff})^{e \mu}_{10}$ & $2 \cdot 10^{-19}$ & $2 \cdot 10^{-19}$ & $2 \cdot 10^{-19}$ & $3 \cdot 10^{-19}$ & $2 \cdot 10^{-19}$ & $2 \cdot 10^{-19}$ & $2 \cdot 10^{-19}$ & $2 \cdot 10^{-19}$ & $2 \cdot 10^{-19}$ & $3 \cdot 10^{-19}$ & $2 \cdot 10^{-19}$ & $2 \cdot 10^{-19}$\\[0.3em]
 & $(c^{(2)}_{\rm eff})^{e \tau}_{10}$ & --- & --- & $2 \cdot 10^{-19}$ & $3 \cdot 10^{-19}$ & $2 \cdot 10^{-19}$ & $2 \cdot 10^{-19}$ & $5 \cdot 10^{-20}$ & $4 \cdot 10^{-20}$ & $4 \cdot 10^{-20}$ & $5 \cdot 10^{-21}$ & $4 \cdot 10^{-21}$ & $4 \cdot 10^{-21}$\\
\end{tabular}
 \end{adjustbox}
\end{table}
\newpage
\begin{table}[t!]
    \caption{Upper limits ($95\%$ C.L.) on CPT-odd LIV coefficients for $d=3$, $(a^{(3)}_{\rm eff})^{\alpha\beta}_{\ell m}$. \label{tab:constraint_tables_d3}}
    \renewcommand{\arraystretch}{1}
    \begin{adjustbox}{max width=\textwidth}\normalsize
    \begin{tabular}{c c | c c c | c c c | c c c | c c c}
\\[-1.25em] & LIV & \multicolumn{12}{c}{Neutrino flavor composition at the astrophysical sources, $(f_e,f_\mu,f_\tau)_{\rm S}$} \\
 $\ell$ & coefficient & \multicolumn{3}{c}{Any $(f_{e, {\rm S}}, 1-f_{e, {\rm S}}, 0)$} & \multicolumn{3}{c}{$\left(1/3,2/3,0\right)_{\rm S}$} & \multicolumn{3}{c}{$\left(1,0,0\right)_{\rm S}$}&\multicolumn{3}{c}{$\left(0,1,0\right)_{\rm S}$}\\
  & [GeV] & Norm & Re & Im & Norm & Re & Im & Norm & Re & Im & Norm & Re & Im
 \\[0.25em] \hline
\\[-1.25em]
0 & $(a^{(3)}_{\rm eff})^{e e}_{00}$ & --- & --- & N/A & --- & --- & N/A & $4 \cdot 10^{-25}$ & $4 \cdot 10^{-25}$ & N/A & $5 \cdot 10^{-25}$ & $5 \cdot 10^{-25}$ & N/A\\[0.3em]
 & $(a^{(3)}_{\rm eff})^{e \mu}_{00}$ & $1 \cdot 10^{-23}$ & $9 \cdot 10^{-24}$ & $9 \cdot 10^{-24}$ & $1 \cdot 10^{-23}$ & $9 \cdot 10^{-24}$ & $9 \cdot 10^{-24}$ & $1 \cdot 10^{-23}$ & $9 \cdot 10^{-24}$ & $9 \cdot 10^{-24}$ & $1 \cdot 10^{-23}$ & $9 \cdot 10^{-24}$ & $9 \cdot 10^{-24}$\\[0.3em]
 & $(a^{(3)}_{\rm eff})^{e \tau}_{00}$ & $1 \cdot 10^{-23}$ & $1 \cdot 10^{-23}$ & $1 \cdot 10^{-23}$ & $1 \cdot 10^{-23}$ & $9 \cdot 10^{-24}$ & $1 \cdot 10^{-23}$ & $2 \cdot 10^{-24}$ & $2 \cdot 10^{-24}$ & $2 \cdot 10^{-24}$ & $3 \cdot 10^{-25}$ & $2 \cdot 10^{-25}$ & $2 \cdot 10^{-25}$\\
\end{tabular}
\end{adjustbox}
\end{table}
\addtocounter{table}{-1}
\begin{table}[t!]
    \caption{Upper limits ($95\%$ C.L.) on CPT-odd LIV coefficients for $d=3$, $(a^{(3)}_{\rm eff})^{\alpha\beta}_{\ell m}$ (continued). 
    }
    \renewcommand{\arraystretch}{1}
    \begin{adjustbox}{max width=\textwidth}\normalsize
    \begin{tabular}{c c | c c c | c c c | c c c | c c c}
\\[-1.25em] & LIV & \multicolumn{12}{c}{Neutrino flavor composition at the astrophysical sources, $(f_e,f_\mu,f_\tau)_{\rm S}$} \\
 $\ell$ & coefficient & \multicolumn{3}{c}{Any $(f_{e, {\rm S}}, 1-f_{e, {\rm S}}, 0)$} & \multicolumn{3}{c}{$\left(1/3,2/3,0\right)_{\rm S}$} & \multicolumn{3}{c}{$\left(1,0,0\right)_{\rm S}$}&\multicolumn{3}{c}{$\left(0,1,0\right)_{\rm S}$}\\
  & [GeV] & Norm & Re & Im & Norm & Re & Im & Norm & Re & Im & Norm & Re & Im
 \\[0.25em] \hline
\\[-1.25em]
2 & $(a^{(3)}_{\rm eff})^{\mu e}_{22}$ & $1 \cdot 10^{-23}$ & $9 \cdot 10^{-24}$ & $9 \cdot 10^{-24}$ & $1 \cdot 10^{-23}$ & $8 \cdot 10^{-24}$ & $8 \cdot 10^{-24}$ & $1 \cdot 10^{-23}$ & $8 \cdot 10^{-24}$ & $8 \cdot 10^{-24}$ & $1 \cdot 10^{-23}$ & $9 \cdot 10^{-24}$ & $9 \cdot 10^{-24}$\\[0.3em]
 & $(a^{(3)}_{\rm eff})^{\mu \mu}_{22}$ & $1 \cdot 10^{-23}$ & $9 \cdot 10^{-24}$ & $9 \cdot 10^{-24}$ & $1 \cdot 10^{-23}$ & $9 \cdot 10^{-24}$ & $9 \cdot 10^{-24}$ & $2 \cdot 10^{-24}$ & $1 \cdot 10^{-24}$ & $2 \cdot 10^{-24}$ & $4 \cdot 10^{-25}$ & $2 \cdot 10^{-25}$ & $3 \cdot 10^{-25}$\\[0.3em]
 & $(a^{(3)}_{\rm eff})^{\mu \tau}_{22}$ & --- & --- & --- & --- & --- & $1 \cdot 10^{-23}$ & $2 \cdot 10^{-25}$ & $2 \cdot 10^{-25}$ & $2 \cdot 10^{-25}$ & $5 \cdot 10^{-25}$ & $3 \cdot 10^{-25}$ & $4 \cdot 10^{-25}$\\
\end{tabular}
\end{adjustbox}
\end{table}
\newpage
\begin{table}[t!]
    \caption{Upper limits ($95\%$ C.L.) on CPT-even LIV coefficients for $d=4$, $(c^{(4)}_{\rm eff})^{\alpha\beta}_{\ell m}$. \label{tab:constraint_tables_d4}}
    \renewcommand{\arraystretch}{1}
    \begin{adjustbox}{max width=\textwidth}\normalsize
    \begin{tabular}{c c | c c c | c c c | c c c | c c c}
\\[-1.25em] & LIV & \multicolumn{12}{c}{Neutrino flavor composition at the astrophysical sources, $(f_e,f_\mu,f_\tau)_{\rm S}$} \\
$\ell$ & coefficient & \multicolumn{3}{c}{Any $(f_{e, {\rm S}}, 1-f_{e, {\rm S}}, 0)$} & \multicolumn{3}{c}{$\left(1/3,2/3,0\right)_{\rm S}$} & \multicolumn{3}{c}{$\left(1,0,0\right)_{\rm S}$}&\multicolumn{3}{c}{$\left(0,1,0\right)_{\rm S}$}\\
  & [unitless] & Norm & Re & Im & Norm & Re & Im & Norm & Re & Im & Norm & Re & Im
 \\[0.25em] \hline
\\[-1.25em]
0 & $(c^{(4)}_{\rm eff})^{e e}_{00}$ & $4 \cdot 10^{-28}$ & $4 \cdot 10^{-28}$ & N/A & --- & --- & N/A & $4 \cdot 10^{-29}$ & $4 \cdot 10^{-29}$ & N/A & $5 \cdot 10^{-29}$ & $5 \cdot 10^{-29}$ & N/A\\[0.3em]
 & $(c^{(4)}_{\rm eff})^{e \mu}_{00}$ & $4 \cdot 10^{-28}$ & $3 \cdot 10^{-28}$ & $3 \cdot 10^{-28}$ & $4 \cdot 10^{-28}$ & $3 \cdot 10^{-28}$ & $3 \cdot 10^{-28}$ & $4 \cdot 10^{-28}$ & $3 \cdot 10^{-28}$ & $3 \cdot 10^{-28}$ & $4 \cdot 10^{-28}$ & $3 \cdot 10^{-28}$ & $3 \cdot 10^{-28}$\\[0.3em]
 & $(c^{(4)}_{\rm eff})^{e \tau}_{00}$ & $6 \cdot 10^{-28}$ & $4 \cdot 10^{-28}$ & $4 \cdot 10^{-28}$ & $5 \cdot 10^{-28}$ & $4 \cdot 10^{-28}$ & $4 \cdot 10^{-28}$ & $2 \cdot 10^{-28}$ & $2 \cdot 10^{-28}$ & $2 \cdot 10^{-28}$ & $3 \cdot 10^{-29}$ & $2 \cdot 10^{-29}$ & $2 \cdot 10^{-29}$\\
\end{tabular}
\end{adjustbox}
\end{table}
\addtocounter{table}{-1}
 \begin{table}[t!]
    \caption{Upper limits ($95\%$ C.L.) on CPT-even LIV coefficients for $d=4$, $(c^{(4)}_{\rm eff})^{\alpha\beta}_{\ell m}$ (continued). 
    }
    \renewcommand{\arraystretch}{1}
    \begin{adjustbox}{max width=\textwidth}\normalsize
    \begin{tabular}{c c | c c c | c c c | c c c | c c c}
\\[-1.25em] & LIV & \multicolumn{12}{c}{Neutrino flavor composition at the astrophysical sources, $(f_e,f_\mu,f_\tau)_{\rm S}$} \\
$\ell$ & coefficient & \multicolumn{3}{c}{Any $(f_{e, {\rm S}}, 1-f_{e, {\rm S}}, 0)$} & \multicolumn{3}{c}{$\left(1/3,2/3,0\right)_{\rm S}$} & \multicolumn{3}{c}{$\left(1,0,0\right)_{\rm S}$}&\multicolumn{3}{c}{$\left(0,1,0\right)_{\rm S}$}\\
  & [unitless] & Norm & Re & Im & Norm & Re & Im & Norm & Re & Im & Norm & Re & Im
 \\[0.25em] \hline
\\[-1.25em]
2 & $(c^{(4)}_{\rm eff})^{\mu e}_{22}$ & $4 \cdot 10^{-28}$ & $3 \cdot 10^{-28}$ & $3 \cdot 10^{-28}$ & $4 \cdot 10^{-28}$ & $3 \cdot 10^{-28}$ & $3 \cdot 10^{-28}$ & $4 \cdot 10^{-28}$ & $3 \cdot 10^{-28}$ & $3 \cdot 10^{-28}$ & $5 \cdot 10^{-28}$ & $3 \cdot 10^{-28}$ & $4 \cdot 10^{-28}$\\[0.3em]
 & $(c^{(4)}_{\rm eff})^{\mu \mu}_{22}$ & $5 \cdot 10^{-28}$ & $4 \cdot 10^{-28}$ & $4 \cdot 10^{-28}$ & $5 \cdot 10^{-28}$ & $4 \cdot 10^{-28}$ & $4 \cdot 10^{-28}$ & $2 \cdot 10^{-29}$ & $1 \cdot 10^{-29}$ & $1 \cdot 10^{-29}$ & $3 \cdot 10^{-29}$ & $2 \cdot 10^{-29}$ & $3 \cdot 10^{-29}$\\[0.3em]
 & $(c^{(4)}_{\rm eff})^{\mu \tau}_{22}$ & --- & $4 \cdot 10^{-28}$ & --- & --- & --- & --- & $2 \cdot 10^{-29}$ & $2 \cdot 10^{-29}$ & $1 \cdot 10^{-29}$ & $4 \cdot 10^{-29}$ & $3 \cdot 10^{-29}$ & $4 \cdot 10^{-29}$\\
\end{tabular}
\end{adjustbox}
\end{table}
\newpage
\begin{table}[t!]
    \caption{Upper limits ($95\%$ C.L.) on CPT-odd LIV coefficients for $d=5$, $(a^{(5)}_{\rm eff})^{\alpha\beta}_{\ell m}$. \label{tab:constraint_tables_d5}}
    \renewcommand{\arraystretch}{1}     \begin{adjustbox}{max width=\textwidth}\normalsize
    \begin{tabular}{c c | c c c | c c c | c c c | c c c}
\\[-1.25em] & LIV & \multicolumn{12}{c}{Neutrino flavor composition at the astrophysical sources, $(f_e,f_\mu,f_\tau)_{\rm S}$)} \\
$\ell$ & coefficient & \multicolumn{3}{c}{Any $(f_{e, {\rm S}}, 1-f_{e, {\rm S}}, 0)$} & \multicolumn{3}{c}{$\left(1/3,2/3,0\right)_{\rm S}$} & \multicolumn{3}{c}{$\left(1,0,0\right)_{\rm S}$}&\multicolumn{3}{c}{$\left(0,1,0\right)_{\rm S}$}\\
  & [GeV$^{-1}$] & Norm & Re & Im & Norm & Re & Im & Norm & Re & Im & Norm & Re & Im
 \\[0.25em] \hline
\\[-1.25em]
0 & $(a^{(5)}_{\rm eff})^{e e}_{00}$ & $2 \cdot 10^{-32}$ & $2 \cdot 10^{-32}$ & N/A & $2 \cdot 10^{-32}$ & $2 \cdot 10^{-32}$ & N/A & $4 \cdot 10^{-33}$ & $4 \cdot 10^{-33}$ & N/A & $5 \cdot 10^{-33}$ & $5 \cdot 10^{-33}$ & N/A\\[0.3em]
 & $(a^{(5)}_{\rm eff})^{e \mu}_{00}$ & $1 \cdot 10^{-32}$ & $7 \cdot 10^{-33}$ & $7 \cdot 10^{-33}$ & $1 \cdot 10^{-32}$ & $7 \cdot 10^{-33}$ & $7 \cdot 10^{-33}$ & $1 \cdot 10^{-32}$ & $7 \cdot 10^{-33}$ & $7 \cdot 10^{-33}$ & $1 \cdot 10^{-32}$ & $8 \cdot 10^{-33}$ & $8 \cdot 10^{-33}$\\[0.3em]
 & $(a^{(5)}_{\rm eff})^{e \tau}_{00}$ & $3 \cdot 10^{-32}$ & $2 \cdot 10^{-32}$ & $2 \cdot 10^{-32}$ & $2 \cdot 10^{-32}$ & $2 \cdot 10^{-32}$ & $2 \cdot 10^{-32}$ & $2 \cdot 10^{-32}$ & $2 \cdot 10^{-32}$ & $2 \cdot 10^{-32}$ & $3 \cdot 10^{-33}$ & $2 \cdot 10^{-33}$ & $2 \cdot 10^{-33}$\\
\end{tabular}
\end{adjustbox}
\end{table}
\addtocounter{table}{-1}
\begin{table}[t!]
    \caption{Upper limits ($95\%$ C.L.) on CPT-odd LIV coefficients for $d=5$, $(a^{(5)}_{\rm eff})^{\alpha\beta}_{\ell m}$ (continued). 
    }
    \renewcommand{\arraystretch}{1}     \begin{adjustbox}{max width=\textwidth}\normalsize
    \begin{tabular}{c c | c c c | c c c | c c c | c c c}
\\[-1.25em] & LIV & \multicolumn{12}{c}{Neutrino flavor composition at the astrophysical sources, $(f_e,f_\mu,f_\tau)_{\rm S}$)} \\
$\ell$ & coefficient & \multicolumn{3}{c}{Any $(f_{e, {\rm S}}, 1-f_{e, {\rm S}}, 0)$} & \multicolumn{3}{c}{$\left(1/3,2/3,0\right)_{\rm S}$} & \multicolumn{3}{c}{$\left(1,0,0\right)_{\rm S}$}&\multicolumn{3}{c}{$\left(0,1,0\right)_{\rm S}$}\\
  & [GeV$^{-1}$] & Norm & Re & Im & Norm & Re & Im & Norm & Re & Im & Norm & Re & Im
 \\[0.25em] \hline
\\[-1.25em]
2 & $(a^{(5)}_{\rm eff})^{\mu e}_{22}$ & $2 \cdot 10^{-32}$ & $1 \cdot 10^{-32}$ & $1 \cdot 10^{-32}$ & $2 \cdot 10^{-32}$ & $1 \cdot 10^{-32}$ & $1 \cdot 10^{-32}$ & $2 \cdot 10^{-32}$ & $1 \cdot 10^{-32}$ & $1 \cdot 10^{-32}$ & $2 \cdot 10^{-32}$ & $1 \cdot 10^{-32}$ & $1 \cdot 10^{-32}$\\[0.3em]
 & $(a^{(5)}_{\rm eff})^{\mu \mu}_{22}$ & $2 \cdot 10^{-32}$ & $2 \cdot 10^{-32}$ & $2 \cdot 10^{-32}$ & $2 \cdot 10^{-32}$ & $2 \cdot 10^{-32}$ & $2 \cdot 10^{-32}$ & $7 \cdot 10^{-34}$ & $5 \cdot 10^{-34}$ & $4 \cdot 10^{-34}$ & $3 \cdot 10^{-33}$ & $2 \cdot 10^{-33}$ & $3 \cdot 10^{-33}$\\[0.3em]
 & $(a^{(5)}_{\rm eff})^{\mu \tau}_{22}$ & --- & --- & $2 \cdot 10^{-32}$ & $3 \cdot 10^{-32}$ & $2 \cdot 10^{-32}$ & $2 \cdot 10^{-32}$ & $2 \cdot 10^{-33}$ & $2 \cdot 10^{-33}$ & $1 \cdot 10^{-33}$ & $4 \cdot 10^{-33}$ & $2 \cdot 10^{-33}$ & $4 \cdot 10^{-33}$\\
\end{tabular}
\end{adjustbox}
\end{table}
\addtocounter{table}{-1}
\begin{table}[t!]
    \caption{Upper limits ($95\%$ C.L.) on CPT-odd LIV coefficients for $d=5$, $(a^{(5)}_{\rm eff})^{\alpha\beta}_{\ell m}$ (continued). 
    }
    \renewcommand{\arraystretch}{1}     \begin{adjustbox}{max width=\textwidth}\normalsize
    \begin{tabular}{c c | c c c | c c c | c c c | c c c}
\\[-1.25em] & LIV & \multicolumn{12}{c}{Neutrino flavor composition at the astrophysical sources, $(f_e,f_\mu,f_\tau)_{\rm S}$)} \\
$\ell$ & coefficient & \multicolumn{3}{c}{Any $(f_{e, {\rm S}}, 1-f_{e, {\rm S}}, 0)$} & \multicolumn{3}{c}{$\left(1/3,2/3,0\right)_{\rm S}$} & \multicolumn{3}{c}{$\left(1,0,0\right)_{\rm S}$}&\multicolumn{3}{c}{$\left(0,1,0\right)_{\rm S}$}\\
  & [GeV$^{-1}$] & Norm & Re & Im & Norm & Re & Im & Norm & Re & Im & Norm & Re & Im
 \\[0.25em] \hline
\\[-1.25em]
4 & $(a^{(5)}_{\rm eff})^{e e}_{41}$ & --- & $2 \cdot 10^{-32}$ & --- & --- & --- & --- & $3 \cdot 10^{-33}$ & $2 \cdot 10^{-33}$ & $2 \cdot 10^{-33}$ & $1 \cdot 10^{-32}$ & $9 \cdot 10^{-33}$ & $1 \cdot 10^{-32}$\\[0.3em]
 & $(a^{(5)}_{\rm eff})^{e \mu}_{41}$ & $2 \cdot 10^{-32}$ & $2 \cdot 10^{-32}$ & $2 \cdot 10^{-32}$ & $2 \cdot 10^{-32}$ & $2 \cdot 10^{-32}$ & $2 \cdot 10^{-32}$ & $2 \cdot 10^{-32}$ & $2 \cdot 10^{-32}$ & $2 \cdot 10^{-32}$ & $2 \cdot 10^{-32}$ & $2 \cdot 10^{-32}$ & $2 \cdot 10^{-32}$\\[0.3em]
 & $(a^{(5)}_{\rm eff})^{e \tau}_{41}$ & --- & --- & --- & $2 \cdot 10^{-32}$ & $2 \cdot 10^{-32}$ & $2 \cdot 10^{-32}$ & $3 \cdot 10^{-32}$ & $2 \cdot 10^{-32}$ & $2 \cdot 10^{-32}$ & $3 \cdot 10^{-33}$ & $2 \cdot 10^{-33}$ & $3 \cdot 10^{-33}$\\
\end{tabular}
\end{adjustbox}
\end{table}
\newpage
\begin{table}[t!]
    \caption{Upper limits ($95\%$ C.L.) on CPT-even LIV coefficients for $d=6$, $(c^{(6)}_{\rm eff})^{\alpha\beta}_{\ell m}$. \label{tab:constraint_tables_d6}}
    \renewcommand{\arraystretch}{1}     \begin{adjustbox}{max width=\textwidth}\normalsize
    \begin{tabular}{c c | c c c | c c c | c c c | c c c}
\\[-1.25em] & LIV & \multicolumn{12}{c}{Neutrino flavor composition at the astrophysical sources, $(f_e,f_\mu,f_\tau)_{\rm S}$)} \\
$\ell$ & coefficient & \multicolumn{3}{c}{Any $(f_{e, {\rm S}}, 1-f_{e, {\rm S}}, 0)$} & \multicolumn{3}{c}{$\left(1/3,2/3,0\right)_{\rm S}$} & \multicolumn{3}{c}{$\left(1,0,0\right)_{\rm S}$}&\multicolumn{3}{c}{$\left(0,1,0\right)_{\rm S}$}\\
  & [GeV$^{-2}$] & Norm & Re & Im & Norm & Re & Im & Norm & Re & Im & Norm & Re & Im
 \\[0.25em] \hline
\\[-1.25em]
0 & $(c^{(6)}_{\rm eff})^{e e}_{00}$ & $9 \cdot 10^{-37}$ & $9 \cdot 10^{-37}$ & N/A & --- & --- & N/A & $4 \cdot 10^{-37}$ & $4 \cdot 10^{-37}$ & N/A & $5 \cdot 10^{-37}$ & $5 \cdot 10^{-37}$ & N/A\\[0.3em]
 & $(c^{(6)}_{\rm eff})^{e \mu}_{00}$ & $7 \cdot 10^{-37}$ & $5 \cdot 10^{-37}$ & $5 \cdot 10^{-37}$ & $7 \cdot 10^{-37}$ & $5 \cdot 10^{-37}$ & $5 \cdot 10^{-37}$ & $7 \cdot 10^{-37}$ & $5 \cdot 10^{-37}$ & $5 \cdot 10^{-37}$ & $7 \cdot 10^{-37}$ & $5 \cdot 10^{-37}$ & $5 \cdot 10^{-37}$\\[0.3em]
 & $(c^{(6)}_{\rm eff})^{e \tau}_{00}$ & $1 \cdot 10^{-36}$ & $9 \cdot 10^{-37}$ & $9 \cdot 10^{-37}$ & $1 \cdot 10^{-36}$ & $8 \cdot 10^{-37}$ & $8 \cdot 10^{-37}$ & $1 \cdot 10^{-36}$ & $9 \cdot 10^{-37}$ & $9 \cdot 10^{-37}$ & $3 \cdot 10^{-37}$ & $2 \cdot 10^{-37}$ & $2 \cdot 10^{-37}$\\
\end{tabular}
\end{adjustbox}
\end{table}
\addtocounter{table}{-1}
\begin{table}[t!]
    \caption{Upper limits ($95\%$ C.L.) on CPT-even LIV coefficients for $d=6$, $(c^{(6)}_{\rm eff})^{\alpha\beta}_{\ell m}$ (continued). 
    }
    \renewcommand{\arraystretch}{1}     \begin{adjustbox}{max width=\textwidth}\normalsize
    \begin{tabular}{c c | c c c | c c c | c c c | c c c}
\\[-1.25em] & LIV & \multicolumn{12}{c}{Neutrino flavor composition at the astrophysical sources, $(f_e,f_\mu,f_\tau)_{\rm S}$)} \\
$\ell$ & coefficient & \multicolumn{3}{c}{Any $(f_{e, {\rm S}}, 1-f_{e, {\rm S}}, 0)$} & \multicolumn{3}{c}{$\left(1/3,2/3,0\right)_{\rm S}$} & \multicolumn{3}{c}{$\left(1,0,0\right)_{\rm S}$}&\multicolumn{3}{c}{$\left(0,1,0\right)_{\rm S}$}\\
  & [GeV$^{-2}$] & Norm & Re & Im & Norm & Re & Im & Norm & Re & Im & Norm & Re & Im
 \\[0.25em] \hline
\\[-1.25em]
2 & $(c^{(6)}_{\rm eff})^{\mu e}_{22}$ & $1 \cdot 10^{-36}$ & $7 \cdot 10^{-37}$ & $8 \cdot 10^{-37}$ & $1 \cdot 10^{-36}$ & $8 \cdot 10^{-37}$ & $8 \cdot 10^{-37}$ & $9 \cdot 10^{-37}$ & $7 \cdot 10^{-37}$ & $7 \cdot 10^{-37}$ & $9 \cdot 10^{-37}$ & $8 \cdot 10^{-37}$ & $7 \cdot 10^{-37}$\\[0.3em]
 & $(c^{(6)}_{\rm eff})^{\mu \mu}_{22}$ & $1 \cdot 10^{-36}$ & $9 \cdot 10^{-37}$ & $9 \cdot 10^{-37}$ & $1 \cdot 10^{-36}$ & $8 \cdot 10^{-37}$ & $9 \cdot 10^{-37}$ & $5 \cdot 10^{-38}$ & $4 \cdot 10^{-38}$ & $3 \cdot 10^{-38}$ & $3 \cdot 10^{-37}$ & $2 \cdot 10^{-37}$ & $3 \cdot 10^{-37}$\\[0.3em]
 & $(c^{(6)}_{\rm eff})^{\mu \tau}_{22}$ & $1 \cdot 10^{-36}$ & $9 \cdot 10^{-37}$ & $1 \cdot 10^{-36}$ & --- & --- & $9 \cdot 10^{-37}$ & $2 \cdot 10^{-37}$ & $2 \cdot 10^{-37}$ & $1 \cdot 10^{-37}$ & $4 \cdot 10^{-37}$ & $2 \cdot 10^{-37}$ & $4 \cdot 10^{-37}$\\
\end{tabular}
\end{adjustbox}
\end{table}
\addtocounter{table}{-1}
\begin{table}[t!]
    \caption{Upper limits ($95\%$ C.L.) on CPT-even LIV coefficients for $d=6$, $(c^{(6)}_{\rm eff})^{\alpha\beta}_{\ell m}$ (continued). 
    }
    \renewcommand{\arraystretch}{1}     \begin{adjustbox}{max width=\textwidth}\normalsize
    \begin{tabular}{c c | c c c | c c c | c c c | c c c}
\\[-1.25em] & LIV & \multicolumn{12}{c}{Neutrino flavor composition at the astrophysical sources, $(f_e,f_\mu,f_\tau)_{\rm S}$)} \\
$\ell$ & coefficient & \multicolumn{3}{c}{Any $(f_{e, {\rm S}}, 1-f_{e, {\rm S}}, 0)$} & \multicolumn{3}{c}{$\left(1/3,2/3,0\right)_{\rm S}$} & \multicolumn{3}{c}{$\left(1,0,0\right)_{\rm S}$}&\multicolumn{3}{c}{$\left(0,1,0\right)_{\rm S}$}\\
  & [GeV$^{-2}$] & Norm & Re & Im & Norm & Re & Im & Norm & Re & Im & Norm & Re & Im
 \\[0.25em] \hline
\\[-1.25em]
4 & $(c^{(6)}_{\rm eff})^{e e}_{41}$ & $1 \cdot 10^{-36}$ & $9 \cdot 10^{-37}$ & $1 \cdot 10^{-36}$ & --- & --- & --- & $3 \cdot 10^{-37}$ & $2 \cdot 10^{-37}$ & $2 \cdot 10^{-37}$ & $9 \cdot 10^{-37}$ & $7 \cdot 10^{-37}$ & $8 \cdot 10^{-37}$\\[0.3em]
 & $(c^{(6)}_{\rm eff})^{e \mu}_{41}$ & $1 \cdot 10^{-36}$ & $8 \cdot 10^{-37}$ & $8 \cdot 10^{-37}$ & $1 \cdot 10^{-36}$ & $8 \cdot 10^{-37}$ & $8 \cdot 10^{-37}$ & $1 \cdot 10^{-36}$ & $8 \cdot 10^{-37}$ & $8 \cdot 10^{-37}$ & $1 \cdot 10^{-36}$ & $8 \cdot 10^{-37}$ & $9 \cdot 10^{-37}$\\[0.3em]
 & $(c^{(6)}_{\rm eff})^{e \tau}_{41}$ & $1 \cdot 10^{-36}$ & $9 \cdot 10^{-37}$ & $9 \cdot 10^{-37}$ & $1 \cdot 10^{-36}$ & $9 \cdot 10^{-37}$ & $9 \cdot 10^{-37}$ & $1 \cdot 10^{-36}$ & $9 \cdot 10^{-37}$ & $9 \cdot 10^{-37}$ & $3 \cdot 10^{-37}$ & $2 \cdot 10^{-37}$ & $2 \cdot 10^{-37}$\\
\end{tabular}
\end{adjustbox}
\end{table}
\addtocounter{table}{-1}
\begin{table}[t!]
    \caption{Upper limits ($95\%$ C.L.) on CPT-even LIV coefficients for $d=6$, $(c^{(6)}_{\rm eff})^{\alpha\beta}_{\ell m}$ (continued). 
    }
    \renewcommand{\arraystretch}{1}     \begin{adjustbox}{max width=\textwidth}\normalsize
    \begin{tabular}{c c | c c c | c c c | c c c | c c c}
\\[-1.25em] & LIV & \multicolumn{12}{c}{Neutrino flavor composition at the astrophysical sources, $(f_e,f_\mu,f_\tau)_{\rm S}$)} \\
$\ell$ & coefficient & \multicolumn{3}{c}{Any $(f_{e, {\rm S}}, 1-f_{e, {\rm S}}, 0)$} & \multicolumn{3}{c}{$\left(1/3,2/3,0\right)_{\rm S}$} & \multicolumn{3}{c}{$\left(1,0,0\right)_{\rm S}$}&\multicolumn{3}{c}{$\left(0,1,0\right)_{\rm S}$}\\
  & [GeV$^{-2}$] & Norm & Re & Im & Norm & Re & Im & Norm & Re & Im & Norm & Re & Im
 \\[0.25em] \hline
\\[-1.25em]
5 & $(c^{(6)}_{\rm eff})^{\tau e}_{51}$ & $1 \cdot 10^{-36}$ & $1 \cdot 10^{-36}$ & $9 \cdot 10^{-37}$ & $1 \cdot 10^{-36}$ & $9 \cdot 10^{-37}$ & $9 \cdot 10^{-37}$ & $1 \cdot 10^{-36}$ & $9 \cdot 10^{-37}$ & $9 \cdot 10^{-37}$ & $3 \cdot 10^{-37}$ & $3 \cdot 10^{-37}$ & $3 \cdot 10^{-37}$\\[0.3em]
 & $(c^{(6)}_{\rm eff})^{\tau \mu}_{51}$ & $1 \cdot 10^{-36}$ & $1 \cdot 10^{-36}$ & $9 \cdot 10^{-37}$ & $1 \cdot 10^{-36}$ & $1 \cdot 10^{-36}$ & $9 \cdot 10^{-37}$ & $3 \cdot 10^{-37}$ & $2 \cdot 10^{-37}$ & $2 \cdot 10^{-37}$ & $1 \cdot 10^{-36}$ & $9 \cdot 10^{-37}$ & $9 \cdot 10^{-37}$\\[0.3em]
 & $(c^{(6)}_{\rm eff})^{\tau \tau}_{51}$ & $1 \cdot 10^{-36}$ & $8 \cdot 10^{-37}$ & $8 \cdot 10^{-37}$ & $1 \cdot 10^{-36}$ & $8 \cdot 10^{-37}$ & $8 \cdot 10^{-37}$ & $3 \cdot 10^{-38}$ & $3 \cdot 10^{-38}$ & $2 \cdot 10^{-38}$ & $8 \cdot 10^{-37}$ & $6 \cdot 10^{-37}$ & $6 \cdot 10^{-37}$\\
\end{tabular}
\end{adjustbox}
\end{table}
\newpage
\begin{table}[t!]
    \caption{Upper limits ($95\%$ C.L.) on CPT-odd LIV coefficients for $d=7$, $(a^{(7)}_{\rm eff})^{\alpha\beta}_{\ell m}$. \label{tab:constraint_tables_d7}}
    \renewcommand{\arraystretch}{1}     \begin{adjustbox}{max width=\textwidth}\normalsize
    \begin{tabular}{c c | c c c | c c c | c c c | c c c}
\\[-1.25em] & LIV & \multicolumn{12}{c}{Neutrino flavor composition at the astrophysical sources, $(f_e,f_\mu,f_\tau)_{\rm S}$)} \\
$\ell$ & coefficient & \multicolumn{3}{c}{Any $(f_{e, {\rm S}}, 1-f_{e, {\rm S}}, 0)$} & \multicolumn{3}{c}{$\left(1/3,2/3,0\right)_{\rm S}$} & \multicolumn{3}{c}{$\left(1,0,0\right)_{\rm S}$}&\multicolumn{3}{c}{$\left(0,1,0\right)_{\rm S}$}\\
  & [GeV$^{-3}$] & Norm & Re & Im & Norm & Re & Im & Norm & Re & Im & Norm & Re & Im
 \\[0.25em] \hline
\\[-1.25em]
0 & $(a^{(7)}_{\rm eff})^{e e}_{00}$ & --- & --- & N/A & --- & --- & N/A & --- & --- & N/A & $4 \cdot 10^{-41}$ & $4 \cdot 10^{-41}$ & N/A\\[0.3em]
 & $(a^{(7)}_{\rm eff})^{e \mu}_{00}$ & $5 \cdot 10^{-41}$ & $4 \cdot 10^{-41}$ & $4 \cdot 10^{-41}$ & $5 \cdot 10^{-41}$ & $4 \cdot 10^{-41}$ & $4 \cdot 10^{-41}$ & $5 \cdot 10^{-41}$ & $4 \cdot 10^{-41}$ & $4 \cdot 10^{-41}$ & $5 \cdot 10^{-41}$ & $4 \cdot 10^{-41}$ & $4 \cdot 10^{-41}$\\[0.3em]
 & $(a^{(7)}_{\rm eff})^{e \tau}_{00}$ & $5 \cdot 10^{-41}$ & $4 \cdot 10^{-41}$ & $4 \cdot 10^{-41}$ & $5 \cdot 10^{-41}$ & $4 \cdot 10^{-41}$ & $4 \cdot 10^{-41}$ & $5 \cdot 10^{-41}$ & $4 \cdot 10^{-41}$ & $4 \cdot 10^{-41}$ & $3 \cdot 10^{-41}$ & $2 \cdot 10^{-41}$ & $2 \cdot 10^{-41}$\\
\end{tabular}
\end{adjustbox}
\end{table}
\addtocounter{table}{-1}
 \begin{table}[t!]
    \caption{Upper limits ($95\%$ C.L.) on CPT-odd LIV coefficients for $d=7$, $(a^{(7)}_{\rm eff})^{\alpha\beta}_{\ell m}$ (continued). 
    }
    \renewcommand{\arraystretch}{1}     \begin{adjustbox}{max width=\textwidth}\normalsize
    \begin{tabular}{c c | c c c | c c c | c c c | c c c}
\\[-1.25em] & LIV & \multicolumn{12}{c}{Neutrino flavor composition at the astrophysical sources, $(f_e,f_\mu,f_\tau)_{\rm S}$)} \\
$\ell$ & coefficient & \multicolumn{3}{c}{Any $(f_{e, {\rm S}}, 1-f_{e, {\rm S}}, 0)$} & \multicolumn{3}{c}{$\left(1/3,2/3,0\right)_{\rm S}$} & \multicolumn{3}{c}{$\left(1,0,0\right)_{\rm S}$}&\multicolumn{3}{c}{$\left(0,1,0\right)_{\rm S}$}\\
  & [GeV$^{-3}$] & Norm & Re & Im & Norm & Re & Im & Norm & Re & Im & Norm & Re & Im
 \\[0.25em] \hline
\\[-1.25em]
2 & $(a^{(7)}_{\rm eff})^{\mu e}_{22}$ & $5 \cdot 10^{-41}$ & $4 \cdot 10^{-41}$ & $4 \cdot 10^{-41}$ & $5 \cdot 10^{-41}$ & $4 \cdot 10^{-41}$ & $4 \cdot 10^{-41}$ & $5 \cdot 10^{-41}$ & $4 \cdot 10^{-41}$ & $4 \cdot 10^{-41}$ & $5 \cdot 10^{-41}$ & $4 \cdot 10^{-41}$ & $4 \cdot 10^{-41}$\\[0.3em]
 & $(a^{(7)}_{\rm eff})^{\mu \mu}_{22}$ & $5 \cdot 10^{-41}$ & $4 \cdot 10^{-41}$ & $4 \cdot 10^{-41}$ & $5 \cdot 10^{-41}$ & $4 \cdot 10^{-41}$ & $4 \cdot 10^{-41}$ & $5 \cdot 10^{-41}$ & $4 \cdot 10^{-41}$ & $4 \cdot 10^{-41}$ & $3 \cdot 10^{-41}$ & $2 \cdot 10^{-41}$ & $2 \cdot 10^{-41}$\\[0.3em]
 & $(a^{(7)}_{\rm eff})^{\mu \tau}_{22}$ & $5 \cdot 10^{-41}$ & $4 \cdot 10^{-41}$ & $4 \cdot 10^{-41}$ & $5 \cdot 10^{-41}$ & $4 \cdot 10^{-41}$ & $4 \cdot 10^{-41}$ & $2 \cdot 10^{-41}$ & $2 \cdot 10^{-41}$ & $1 \cdot 10^{-41}$ & $4 \cdot 10^{-41}$ & $2 \cdot 10^{-41}$ & $4 \cdot 10^{-41}$\\
\end{tabular}
\end{adjustbox}
\end{table}
\addtocounter{table}{-1}
 \begin{table}[t!]
    \caption{Upper limits ($95\%$ C.L.) on CPT-odd LIV coefficients for $d=7$, $(a^{(7)}_{\rm eff})^{\alpha\beta}_{\ell m}$ (continued). 
    }
    \renewcommand{\arraystretch}{1}     \begin{adjustbox}{max width=\textwidth}\normalsize
    \begin{tabular}{c c | c c c | c c c | c c c | c c c}
\\[-1.25em] & LIV & \multicolumn{12}{c}{Neutrino flavor composition at the astrophysical sources, $(f_e,f_\mu,f_\tau)_{\rm S}$)} \\
$\ell$ & coefficient & \multicolumn{3}{c}{Any $(f_{e, {\rm S}}, 1-f_{e, {\rm S}}, 0)$} & \multicolumn{3}{c}{$\left(1/3,2/3,0\right)_{\rm S}$} & \multicolumn{3}{c}{$\left(1,0,0\right)_{\rm S}$}&\multicolumn{3}{c}{$\left(0,1,0\right)_{\rm S}$}\\
  & [GeV$^{-3}$] & Norm & Re & Im & Norm & Re & Im & Norm & Re & Im & Norm & Re & Im
 \\[0.25em] \hline
\\[-1.25em]
4 & $(a^{(7)}_{\rm eff})^{e e}_{41}$ & $6 \cdot 10^{-41}$ & $4 \cdot 10^{-41}$ & $4 \cdot 10^{-41}$ & --- & --- & --- & $3 \cdot 10^{-41}$ & $2 \cdot 10^{-41}$ & $2 \cdot 10^{-41}$ & $5 \cdot 10^{-41}$ & $4 \cdot 10^{-41}$ & $4 \cdot 10^{-41}$\\[0.3em]
 & $(a^{(7)}_{\rm eff})^{e \mu}_{41}$ & $5 \cdot 10^{-41}$ & $4 \cdot 10^{-41}$ & $4 \cdot 10^{-41}$ & $5 \cdot 10^{-41}$ & $4 \cdot 10^{-41}$ & $4 \cdot 10^{-41}$ & $5 \cdot 10^{-41}$ & $4 \cdot 10^{-41}$ & $4 \cdot 10^{-41}$ & $5 \cdot 10^{-41}$ & $4 \cdot 10^{-41}$ & $4 \cdot 10^{-41}$\\[0.3em]
 & $(a^{(7)}_{\rm eff})^{e \tau}_{41}$ & $5 \cdot 10^{-41}$ & $4 \cdot 10^{-41}$ & $4 \cdot 10^{-41}$ & $5 \cdot 10^{-41}$ & $4 \cdot 10^{-41}$ & $4 \cdot 10^{-41}$ & $5 \cdot 10^{-41}$ & $4 \cdot 10^{-41}$ & $4 \cdot 10^{-41}$ & $3 \cdot 10^{-41}$ & $2 \cdot 10^{-41}$ & $2 \cdot 10^{-41}$\\
\end{tabular}
\end{adjustbox}
\end{table}
\addtocounter{table}{-1}
 \begin{table}[t!]
    \caption{Upper limits ($95\%$ C.L.) on CPT-odd LIV coefficients for $d=7$, $(a^{(7)}_{\rm eff})^{\alpha\beta}_{\ell m}$ (continued). 
    }
    \renewcommand{\arraystretch}{1}     \begin{adjustbox}{max width=\textwidth}\normalsize
    \begin{tabular}{c c | c c c | c c c | c c c | c c c}
\\[-1.25em] & LIV & \multicolumn{12}{c}{Neutrino flavor composition at the astrophysical sources, $(f_e,f_\mu,f_\tau)_{\rm S}$)} \\
$\ell$ & coefficient & \multicolumn{3}{c}{Any $(f_{e, {\rm S}}, 1-f_{e, {\rm S}}, 0)$} & \multicolumn{3}{c}{$\left(1/3,2/3,0\right)_{\rm S}$} & \multicolumn{3}{c}{$\left(1,0,0\right)_{\rm S}$}&\multicolumn{3}{c}{$\left(0,1,0\right)_{\rm S}$}\\
  & [GeV$^{-3}$] & Norm & Re & Im & Norm & Re & Im & Norm & Re & Im & Norm & Re & Im
 \\[0.25em] \hline
\\[-1.25em]
5 & $(a^{(7)}_{\rm eff})^{\tau e}_{51}$ & $5 \cdot 10^{-41}$ & $4 \cdot 10^{-41}$ & $4 \cdot 10^{-41}$ & $5 \cdot 10^{-41}$ & $4 \cdot 10^{-41}$ & $4 \cdot 10^{-41}$ & --- & --- & $4 \cdot 10^{-41}$ & $3 \cdot 10^{-41}$ & $3 \cdot 10^{-41}$ & $3 \cdot 10^{-41}$\\[0.3em]
 & $(a^{(7)}_{\rm eff})^{\tau \mu}_{51}$ & $5 \cdot 10^{-41}$ & $4 \cdot 10^{-41}$ & $4 \cdot 10^{-41}$ & --- & $4 \cdot 10^{-41}$ & --- & $3 \cdot 10^{-41}$ & $2 \cdot 10^{-41}$ & $2 \cdot 10^{-41}$ & $5 \cdot 10^{-41}$ & $4 \cdot 10^{-41}$ & $4 \cdot 10^{-41}$\\[0.3em]
 & $(a^{(7)}_{\rm eff})^{\tau \tau}_{51}$ & $5 \cdot 10^{-41}$ & $4 \cdot 10^{-41}$ & $4 \cdot 10^{-41}$ & $5 \cdot 10^{-41}$ & $4 \cdot 10^{-41}$ & $4 \cdot 10^{-41}$ & $4 \cdot 10^{-41}$ & $3 \cdot 10^{-41}$ & $3 \cdot 10^{-41}$ & $5 \cdot 10^{-41}$ & $4 \cdot 10^{-41}$ & $4 \cdot 10^{-41}$\\
\end{tabular}
\end{adjustbox}
\end{table}
\addtocounter{table}{-1}
 \begin{table}[t!]
    \caption{Upper limits ($95\%$ C.L.) on CPT-odd LIV coefficients for $d=7$, $(a^{(7)}_{\rm eff})^{\alpha\beta}_{\ell m}$ (continued). 
    }
    \renewcommand{\arraystretch}{1}     \begin{adjustbox}{max width=\textwidth}\normalsize
    \begin{tabular}{c c | c c c | c c c | c c c | c c c}
\\[-1.25em] & LIV & \multicolumn{12}{c}{Neutrino flavor composition at the astrophysical sources, $(f_e,f_\mu,f_\tau)_{\rm S}$)} \\
$\ell$ & coefficient & \multicolumn{3}{c}{Any $(f_{e, {\rm S}}, 1-f_{e, {\rm S}}, 0)$} & \multicolumn{3}{c}{$\left(1/3,2/3,0\right)_{\rm S}$} & \multicolumn{3}{c}{$\left(1,0,0\right)_{\rm S}$}&\multicolumn{3}{c}{$\left(0,1,0\right)_{\rm S}$}\\
  & [GeV$^{-3}$] & Norm & Re & Im & Norm & Re & Im & Norm & Re & Im & Norm & Re & Im
 \\[0.25em] \hline
\\[-1.25em]
6 & $(a^{(7)}_{\rm eff})^{\mu e}_{61}$ & $5 \cdot 10^{-41}$ & $4 \cdot 10^{-41}$ & $4 \cdot 10^{-41}$ & $5 \cdot 10^{-41}$ & $4 \cdot 10^{-41}$ & $4 \cdot 10^{-41}$ & $5 \cdot 10^{-41}$ & $4 \cdot 10^{-41}$ & $4 \cdot 10^{-41}$ & $5 \cdot 10^{-41}$ & $4 \cdot 10^{-41}$ & $4 \cdot 10^{-41}$\\[0.3em]
 & $(a^{(7)}_{\rm eff})^{\mu \mu}_{61}$ & $5 \cdot 10^{-41}$ & $4 \cdot 10^{-41}$ & $4 \cdot 10^{-41}$ & $5 \cdot 10^{-41}$ & $4 \cdot 10^{-41}$ & $4 \cdot 10^{-41}$ & $5 \cdot 10^{-41}$ & $4 \cdot 10^{-41}$ & $4 \cdot 10^{-41}$ & $3 \cdot 10^{-41}$ & $2 \cdot 10^{-41}$ & $2 \cdot 10^{-41}$\\[0.3em]
 & $(a^{(7)}_{\rm eff})^{\mu \tau}_{61}$ & $5 \cdot 10^{-41}$ & $4 \cdot 10^{-41}$ & $4 \cdot 10^{-41}$ & --- & --- & $4 \cdot 10^{-41}$ & $2 \cdot 10^{-41}$ & $2 \cdot 10^{-41}$ & $2 \cdot 10^{-41}$ & $5 \cdot 10^{-41}$ & $4 \cdot 10^{-41}$ & $4 \cdot 10^{-41}$\\
\end{tabular}
\end{adjustbox}
\end{table}
\newpage
\begin{table}[t!]
       
    \caption{Upper limits ($95\%$ C.L.) on CPT-even LIV coefficients for $d=8$, $(c^{(8)}_{\rm eff})^{\alpha\beta}_{\ell m}$. \label{tab:constraint_tables_d8}}
    \renewcommand{\arraystretch}{1}     \begin{adjustbox}{max width=\textwidth}\normalsize
    \begin{tabular}{c c | c c c | c c c | c c c | c c c}
\\[-1.25em] & LIV & \multicolumn{12}{c}{Neutrino flavor composition at the astrophysical sources, $(f_e,f_\mu,f_\tau)_{\rm S}$)} \\
 $\ell$ & coefficient & \multicolumn{3}{c}{Any $(f_{e, {\rm S}}, 1-f_{e, {\rm S}}, 0)$} & \multicolumn{3}{c}{$\left(1/3,2/3,0\right)_{\rm S}$} & \multicolumn{3}{c}{$\left(1,0,0\right)_{\rm S}$}&\multicolumn{3}{c}{$\left(0,1,0\right)_{\rm S}$}\\
  & [GeV$^{-4}$] & Norm & Re & Im & Norm & Re & Im & Norm & Re & Im & Norm & Re & Im
 \\[0.25em] \hline
\\[-1.25em]
 0 & $(c^{(8)}_{\rm eff})^{e \mu}_{00}$ & $3 \cdot 10^{-45}$ & $2 \cdot 10^{-45}$ & $2 \cdot 10^{-45}$ & $3 \cdot 10^{-45}$ & $2 \cdot 10^{-45}$ & $2 \cdot 10^{-45}$ & $3 \cdot 10^{-45}$ & $2 \cdot 10^{-45}$ & $2 \cdot 10^{-45}$ & $3 \cdot 10^{-45}$ & $2 \cdot 10^{-45}$ & $2 \cdot 10^{-45}$\\[0.3em]
 & $(c^{(8)}_{\rm eff})^{e \tau}_{00}$ & $3 \cdot 10^{-45}$ & $2 \cdot 10^{-45}$ & $2 \cdot 10^{-45}$ & $3 \cdot 10^{-45}$ & $2 \cdot 10^{-45}$ & $2 \cdot 10^{-45}$ & $3 \cdot 10^{-45}$ & $2 \cdot 10^{-45}$ & $2 \cdot 10^{-45}$ & $2 \cdot 10^{-45}$ & $2 \cdot 10^{-45}$ & $2 \cdot 10^{-45}$\\[0.3em]
 & $(c^{(8)}_{\rm eff})^{\mu e}_{00}$ & $3 \cdot 10^{-45}$ & $2 \cdot 10^{-45}$ & $2 \cdot 10^{-45}$ & $3 \cdot 10^{-45}$ & $2 \cdot 10^{-45}$ & $2 \cdot 10^{-45}$ & $3 \cdot 10^{-45}$ & $2 \cdot 10^{-45}$ & $2 \cdot 10^{-45}$ & $3 \cdot 10^{-45}$ & $2 \cdot 10^{-45}$ & $2 \cdot 10^{-45}$\\
\end{tabular}
\end{adjustbox}
\end{table}
\addtocounter{table}{-1}
 \begin{table}[t!]
       
    \caption{Upper limits ($95\%$ C.L.) on CPT-even LIV coefficients for $d=8$, $(c^{(8)}_{\rm eff})^{\alpha\beta}_{\ell m}$ (continued). 
    }
    \renewcommand{\arraystretch}{1}     \begin{adjustbox}{max width=\textwidth}\normalsize
    \begin{tabular}{c c | c c c | c c c | c c c | c c c}
\\[-1.25em] & LIV & \multicolumn{12}{c}{Neutrino flavor composition at the astrophysical sources, $(f_e,f_\mu,f_\tau)_{\rm S}$)} \\
 $\ell$ & coefficient & \multicolumn{3}{c}{Any $(f_{e, {\rm S}}, 1-f_{e, {\rm S}}, 0)$} & \multicolumn{3}{c}{$\left(1/3,2/3,0\right)_{\rm S}$} & \multicolumn{3}{c}{$\left(1,0,0\right)_{\rm S}$}&\multicolumn{3}{c}{$\left(0,1,0\right)_{\rm S}$}\\
  & [GeV$^{-4}$] & Norm & Re & Im & Norm & Re & Im & Norm & Re & Im & Norm & Re & Im
 \\[0.25em] \hline
\\[-1.25em]
2 & $(c^{(8)}_{\rm eff})^{\tau e}_{22}$ & $3 \cdot 10^{-45}$ & $2 \cdot 10^{-45}$ & $2 \cdot 10^{-45}$ & $2 \cdot 10^{-45}$ & $2 \cdot 10^{-45}$ & $2 \cdot 10^{-45}$ & --- & --- & --- & $2 \cdot 10^{-45}$ & $2 \cdot 10^{-45}$ & $2 \cdot 10^{-45}$\\[0.3em]
 & $(c^{(8)}_{\rm eff})^{\tau \mu}_{22}$ & --- & --- & $2 \cdot 10^{-45}$ & --- & --- & --- & $2 \cdot 10^{-45}$ & $2 \cdot 10^{-45}$ & $1 \cdot 10^{-45}$ & $2 \cdot 10^{-45}$ & $2 \cdot 10^{-45}$ & $2 \cdot 10^{-45}$\\[0.3em]
 & $(c^{(8)}_{\rm eff})^{\tau \tau}_{22}$ & $2 \cdot 10^{-45}$ & $2 \cdot 10^{-45}$ & $2 \cdot 10^{-45}$ & $2 \cdot 10^{-45}$ & $2 \cdot 10^{-45}$ & $2 \cdot 10^{-45}$ & $2 \cdot 10^{-45}$ & $2 \cdot 10^{-45}$ & $2 \cdot 10^{-45}$ & $2 \cdot 10^{-45}$ & $2 \cdot 10^{-45}$ & $2 \cdot 10^{-45}$\\
\end{tabular}
\end{adjustbox}
\end{table}
\addtocounter{table}{-1}
 \begin{table}[t!]
       
    \caption{Upper limits ($95\%$ C.L.) on CPT-even LIV coefficients for $d=8$, $(c^{(8)}_{\rm eff})^{\alpha\beta}_{\ell m}$ (continued). 
    }
    \renewcommand{\arraystretch}{1}     \begin{adjustbox}{max width=\textwidth}\normalsize
    \begin{tabular}{c c | c c c | c c c | c c c | c c c}
\\[-1.25em] & LIV & \multicolumn{12}{c}{Neutrino flavor composition at the astrophysical sources, $(f_e,f_\mu,f_\tau)_{\rm S}$)} \\
 $\ell$ & coefficient & \multicolumn{3}{c}{Any $(f_{e, {\rm S}}, 1-f_{e, {\rm S}}, 0)$} & \multicolumn{3}{c}{$\left(1/3,2/3,0\right)_{\rm S}$} & \multicolumn{3}{c}{$\left(1,0,0\right)_{\rm S}$}&\multicolumn{3}{c}{$\left(0,1,0\right)_{\rm S}$}\\
  & [GeV$^{-4}$] & Norm & Re & Im & Norm & Re & Im & Norm & Re & Im & Norm & Re & Im
 \\[0.25em] \hline
\\[-1.25em]
4 & $(c^{(8)}_{\rm eff})^{\mu e}_{41}$ & --- & --- & --- & --- & --- & --- & --- & --- & --- & --- & $2 \cdot 10^{-45}$ & ---\\[0.3em]
 & $(c^{(8)}_{\rm eff})^{\mu \mu}_{41}$ & $3 \cdot 10^{-45}$ & $2 \cdot 10^{-45}$ & $2 \cdot 10^{-45}$ & $2 \cdot 10^{-45}$ & $2 \cdot 10^{-45}$ & $2 \cdot 10^{-45}$ & $2 \cdot 10^{-45}$ & $2 \cdot 10^{-45}$ & $2 \cdot 10^{-45}$ & $2 \cdot 10^{-45}$ & $2 \cdot 10^{-45}$ & $2 \cdot 10^{-45}$\\[0.3em]
 & $(c^{(8)}_{\rm eff})^{\mu \tau}_{41}$ & $3 \cdot 10^{-45}$ & $2 \cdot 10^{-45}$ & $2 \cdot 10^{-45}$ & --- & --- & --- & $2 \cdot 10^{-45}$ & $2 \cdot 10^{-45}$ & $2 \cdot 10^{-45}$ & $2 \cdot 10^{-45}$ & $2 \cdot 10^{-45}$ & $2 \cdot 10^{-45}$\\
\end{tabular}
\end{adjustbox}
\end{table}
\addtocounter{table}{-1}
 \begin{table}[t!]
       
    \caption{Upper limits ($95\%$ C.L.) on CPT-even LIV coefficients for $d=8$, $(c^{(8)}_{\rm eff})^{\alpha\beta}_{\ell m}$ (continued). 
    }
    \renewcommand{\arraystretch}{1}     \begin{adjustbox}{max width=\textwidth}\normalsize
    \begin{tabular}{c c | c c c | c c c | c c c | c c c}
\\[-1.25em] & LIV & \multicolumn{12}{c}{Neutrino flavor composition at the astrophysical sources, $(f_e,f_\mu,f_\tau)_{\rm S}$)} \\
 $\ell$ & coefficient & \multicolumn{3}{c}{Any $(f_{e, {\rm S}}, 1-f_{e, {\rm S}}, 0)$} & \multicolumn{3}{c}{$\left(1/3,2/3,0\right)_{\rm S}$} & \multicolumn{3}{c}{$\left(1,0,0\right)_{\rm S}$}&\multicolumn{3}{c}{$\left(0,1,0\right)_{\rm S}$}\\
  & [GeV$^{-4}$] & Norm & Re & Im & Norm & Re & Im & Norm & Re & Im & Norm & Re & Im
 \\[0.25em] \hline
\\[-1.25em]
5 & $(c^{(8)}_{\rm eff})^{e \tau}_{52}$ & $3 \cdot 10^{-45}$ & $2 \cdot 10^{-45}$ & $2 \cdot 10^{-45}$ & --- & --- & $2 \cdot 10^{-45}$ & --- & --- & $2 \cdot 10^{-45}$ & $2 \cdot 10^{-45}$ & $2 \cdot 10^{-45}$ & $2 \cdot 10^{-45}$\\[0.3em]
 & $(c^{(8)}_{\rm eff})^{\mu e}_{52}$ & --- & $2 \cdot 10^{-45}$ & --- & --- & --- & $2 \cdot 10^{-45}$ & $3 \cdot 10^{-45}$ & $2 \cdot 10^{-45}$ & $2 \cdot 10^{-45}$ & --- & --- & ---\\[0.3em]
 & $(c^{(8)}_{\rm eff})^{\mu \mu}_{52}$ & $3 \cdot 10^{-45}$ & $2 \cdot 10^{-45}$ & $2 \cdot 10^{-45}$ & $3 \cdot 10^{-45}$ & $2 \cdot 10^{-45}$ & $2 \cdot 10^{-45}$ & $2 \cdot 10^{-45}$ & $2 \cdot 10^{-45}$ & $2 \cdot 10^{-45}$ & $2 \cdot 10^{-45}$ & $2 \cdot 10^{-45}$ & $2 \cdot 10^{-45}$\\
\end{tabular}
\end{adjustbox}
\end{table}
\addtocounter{table}{-1}
 \begin{table}[t!]
       
    \caption{Upper limits ($95\%$ C.L.) on CPT-even LIV coefficients for $d=8$, $(c^{(8)}_{\rm eff})^{\alpha\beta}_{\ell m}$ (continued). 
    }
    \renewcommand{\arraystretch}{1}     \begin{adjustbox}{max width=\textwidth}\normalsize
    \begin{tabular}{c c | c c c | c c c | c c c | c c c}
\\[-1.25em] & LIV & \multicolumn{12}{c}{Neutrino flavor composition at the astrophysical sources, $(f_e,f_\mu,f_\tau)_{\rm S}$)} \\
 $\ell$ & coefficient & \multicolumn{3}{c}{Any $(f_{e, {\rm S}}, 1-f_{e, {\rm S}}, 0)$} & \multicolumn{3}{c}{$\left(1/3,2/3,0\right)_{\rm S}$} & \multicolumn{3}{c}{$\left(1,0,0\right)_{\rm S}$}&\multicolumn{3}{c}{$\left(0,1,0\right)_{\rm S}$}\\
  & [GeV$^{-4}$] & Norm & Re & Im & Norm & Re & Im & Norm & Re & Im & Norm & Re & Im
 \\[0.25em] \hline
\\[-1.25em]
6 & $(c^{(8)}_{\rm eff})^{e e}_{62}$ & --- & --- & --- & --- & --- & $2 \cdot 10^{-45}$ & $2 \cdot 10^{-45}$ & $2 \cdot 10^{-45}$ & $2 \cdot 10^{-45}$ & $2 \cdot 10^{-45}$ & $2 \cdot 10^{-45}$ & $2 \cdot 10^{-45}$\\[0.3em]
 & $(c^{(8)}_{\rm eff})^{e \mu}_{62}$ & --- & --- & $2 \cdot 10^{-45}$ & --- & --- & --- & --- & --- & --- & --- & --- & ---\\[0.3em]
 & $(c^{(8)}_{\rm eff})^{e \tau}_{62}$ & --- & --- & $2 \cdot 10^{-45}$ & $3 \cdot 10^{-45}$ & $2 \cdot 10^{-45}$ & $2 \cdot 10^{-45}$ & --- & --- & $2 \cdot 10^{-45}$ & $2 \cdot 10^{-45}$ & $2 \cdot 10^{-45}$ & $2 \cdot 10^{-45}$\\
\end{tabular}
\end{adjustbox}
\end{table}
\addtocounter{table}{-1}
 \begin{table}[t!]
       
    \caption{Upper limits ($95\%$ C.L.) on CPT-even LIV coefficients for $d=8$, $(c^{(8)}_{\rm eff})^{\alpha\beta}_{\ell m}$ (continued). 
    }
    \renewcommand{\arraystretch}{1}     \begin{adjustbox}{max width=\textwidth}\normalsize
    \begin{tabular}{c c | c c c | c c c | c c c | c c c}
\\[-1.25em] & LIV & \multicolumn{12}{c}{Neutrino flavor composition at the astrophysical sources, $(f_e,f_\mu,f_\tau)_{\rm S}$)} \\
 $\ell$ & coefficient & \multicolumn{3}{c}{Any $(f_{e, {\rm S}}, 1-f_{e, {\rm S}}, 0)$} & \multicolumn{3}{c}{$\left(1/3,2/3,0\right)_{\rm S}$} & \multicolumn{3}{c}{$\left(1,0,0\right)_{\rm S}$}&\multicolumn{3}{c}{$\left(0,1,0\right)_{\rm S}$}\\
  & [GeV$^{-4}$] & Norm & Re & Im & Norm & Re & Im & Norm & Re & Im & Norm & Re & Im
 \\[0.25em] \hline
\\[-1.25em]
7 & $(c^{(8)}_{\rm eff})^{e e}_{71}$ & $3 \cdot 10^{-45}$ & $2 \cdot 10^{-45}$ & $2 \cdot 10^{-45}$ & --- & --- & $2 \cdot 10^{-45}$ & $2 \cdot 10^{-45}$ & $2 \cdot 10^{-45}$ & $2 \cdot 10^{-45}$ & $2 \cdot 10^{-45}$ & $2 \cdot 10^{-45}$ & $2 \cdot 10^{-45}$\\[0.3em]
 & $(c^{(8)}_{\rm eff})^{e \tau}_{71}$ & --- & $2 \cdot 10^{-45}$ & --- & $3 \cdot 10^{-45}$ & $2 \cdot 10^{-45}$ & $2 \cdot 10^{-45}$ & --- & --- & --- & $2 \cdot 10^{-45}$ & $2 \cdot 10^{-45}$ & $2 \cdot 10^{-45}$\\[0.3em]
 & $(c^{(8)}_{\rm eff})^{\mu e}_{71}$ & --- & $2 \cdot 10^{-45}$ & --- & --- & --- & --- & --- & --- & $2 \cdot 10^{-45}$ & $3 \cdot 10^{-45}$ & $2 \cdot 10^{-45}$ & $2 \cdot 10^{-45}$\\
\end{tabular}
\end{adjustbox}
\end{table}
\addtocounter{table}{-1}
 \begin{table}[t!]
       
    \caption{Upper limits ($95\%$ C.L.) on CPT-even LIV coefficients for $d=8$, $(c^{(8)}_{\rm eff})^{\alpha\beta}_{\ell m}$ (continued). 
    }
    \renewcommand{\arraystretch}{1}     \begin{adjustbox}{max width=\textwidth}\normalsize
    \begin{tabular}{c c | c c c | c c c | c c c | c c c}
\\[-1.25em] & LIV & \multicolumn{12}{c}{Neutrino flavor composition at the astrophysical sources, $(f_e,f_\mu,f_\tau)_{\rm S}$)} \\
 $\ell$ & coefficient & \multicolumn{3}{c}{Any $(f_{e, {\rm S}}, 1-f_{e, {\rm S}}, 0)$} & \multicolumn{3}{c}{$\left(1/3,2/3,0\right)_{\rm S}$} & \multicolumn{3}{c}{$\left(1,0,0\right)_{\rm S}$}&\multicolumn{3}{c}{$\left(0,1,0\right)_{\rm S}$}\\
  & [GeV$^{-4}$] & Norm & Re & Im & Norm & Re & Im & Norm & Re & Im & Norm & Re & Im
 \\[0.25em] \hline
\\[-1.25em]
7 & $(c^{(8)}_{\rm eff})^{e e}_{77}$ & $3 \cdot 10^{-45}$ & $2 \cdot 10^{-45}$ & $2 \cdot 10^{-45}$ & --- & --- & --- & $2 \cdot 10^{-45}$ & $2 \cdot 10^{-45}$ & $2 \cdot 10^{-45}$ & $2 \cdot 10^{-45}$ & $2 \cdot 10^{-45}$ & $2 \cdot 10^{-45}$\\[0.3em]
 & $(c^{(8)}_{\rm eff})^{e \mu}_{77}$ & $3 \cdot 10^{-45}$ & $2 \cdot 10^{-45}$ & $2 \cdot 10^{-45}$ & $3 \cdot 10^{-45}$ & $2 \cdot 10^{-45}$ & $2 \cdot 10^{-45}$ & $2 \cdot 10^{-45}$ & $2 \cdot 10^{-45}$ & $2 \cdot 10^{-45}$ & $3 \cdot 10^{-45}$ & $2 \cdot 10^{-45}$ & $2 \cdot 10^{-45}$\\[0.3em]
 & $(c^{(8)}_{\rm eff})^{e \tau}_{77}$ & --- & --- & $2 \cdot 10^{-45}$ & $3 \cdot 10^{-45}$ & $2 \cdot 10^{-45}$ & $2 \cdot 10^{-45}$ & --- & --- & --- & $2 \cdot 10^{-45}$ & $2 \cdot 10^{-45}$ & $2 \cdot 10^{-45}$\\
\end{tabular}
\end{adjustbox}
\end{table}



\begin{thebibliography}{100}

\bibitem{Ellis:2011ek}
J.~Ellis and N.E.~Mavromatos, \emph{{Probes of Lorentz Violation}},
  \href{https://doi.org/10.1016/j.astropartphys.2012.05.004}{\emph{Astropart.
  Phys.} {\bfseries 43} (2013) 50}
  [\href{https://arxiv.org/abs/1111.1178}{{\ttfamily 1111.1178}}].

\bibitem{Tasson:2014dfa}
J.D.~Tasson, \emph{{What Do We Know About Lorentz Invariance?}},
  \href{https://doi.org/10.1088/0034-4885/77/6/062901}{\emph{Rept. Prog. Phys.}
  {\bfseries 77} (2014) 062901}
  [\href{https://arxiv.org/abs/1403.7785}{{\ttfamily 1403.7785}}].

\bibitem{Carlip:2022pyh}
S.~Carlip, \emph{{Spacetime foam: a review}},
  \href{https://doi.org/10.1088/1361-6633/acceb4}{\emph{Rept. Prog. Phys.}
  {\bfseries 86} (2023) 066001}
  [\href{https://arxiv.org/abs/2209.14282}{{\ttfamily 2209.14282}}].

\bibitem{Basile:2024oms}
I.~Basile, L.~Buoninfante, F.~Di~Filippo, B.~Knorr, A.~Platania and
  A.~Tokareva, \emph{{Lectures in Quantum Gravity}} (12, 2024),
  [\href{https://arxiv.org/abs/2412.08690}{{\ttfamily 2412.08690}}].

\bibitem{Addazi:2021xuf}
A.~Addazi et~al., \emph{{Quantum gravity phenomenology at the dawn of the
  multi-messenger era\textemdash{}A review}},
  \href{https://doi.org/10.1016/j.ppnp.2022.103948}{\emph{Prog. Part. Nucl.
  Phys.} {\bfseries 125} (2022) 103948}
  [\href{https://arxiv.org/abs/2111.05659}{{\ttfamily 2111.05659}}].

\bibitem{AlvesBatista:2023wqm}
R.~Alves~Batista et~al., \emph{{White paper and roadmap for quantum gravity
  phenomenology in the multi-messenger era}},
  \href{https://doi.org/10.1088/1361-6382/ad605a}{\emph{Class. Quant. Grav.}
  {\bfseries 42} (2025) 032001}
  [\href{https://arxiv.org/abs/2312.00409}{{\ttfamily 2312.00409}}].

\bibitem{Anchordoqui:2013dnh}
L.A.~Anchordoqui et~al., \emph{{Cosmic Neutrino Pevatrons: A Brand New Pathway
  to Astronomy, Astrophysics, and Particle Physics}},
  \href{https://doi.org/10.1016/j.jheap.2014.01.001}{\emph{JHEAp} {\bfseries
  1-2} (2014) 1} [\href{https://arxiv.org/abs/1312.6587}{{\ttfamily
  1312.6587}}].

\bibitem{Ahlers:2018mkf}
M.~Ahlers, K.~Helbing and C.~P\'erez de~los Heros, \emph{{Probing Particle
  Physics with IceCube}},
  \href{https://doi.org/10.1140/epjc/s10052-018-6369-9}{\emph{Eur. Phys. J. C}
  {\bfseries 78} (2018) 924}
  [\href{https://arxiv.org/abs/1806.05696}{{\ttfamily 1806.05696}}].

\bibitem{Ackermann:2019cxh}
M.~Ackermann et~al., \emph{{Fundamental Physics with High-Energy Cosmic
  Neutrinos}}, {\emph{Bull. Am. Astron. Soc.} {\bfseries 51} (2019) 215}
  [\href{https://arxiv.org/abs/1903.04333}{{\ttfamily 1903.04333}}].

\bibitem{Arguelles:2019rbn}
C.A.~Arg\"uelles, M.~Bustamante, A.~Kheirandish, S.~Palomares-Ruiz,
  J.~Salvad\'o and A.C.~Vincent, \emph{{Fundamental physics with high-energy
  cosmic neutrinos today and in the future}},
  \href{https://doi.org/10.22323/1.358.0849}{\emph{PoS} {\bfseries ICRC2019}
  (2020) 849} [\href{https://arxiv.org/abs/1907.08690}{{\ttfamily
  1907.08690}}].

\bibitem{Ackermann:2022rqc}
M.~Ackermann et~al., \emph{{High-energy and ultra-high-energy neutrinos: A
  Snowmass white paper}},
  \href{https://doi.org/10.1016/j.jheap.2022.08.001}{\emph{JHEAp} {\bfseries
  36} (2022) 55} [\href{https://arxiv.org/abs/2203.08096}{{\ttfamily
  2203.08096}}].

\bibitem{MammenAbraham:2022xoc}
R.~Mammen~Abraham et~al., \emph{{Tau neutrinos in the next decade: from GeV to
  EeV}}, \href{https://doi.org/10.1088/1361-6471/ac89d2}{\emph{J. Phys. G}
  {\bfseries 49} (2022) 110501}
  [\href{https://arxiv.org/abs/2203.05591}{{\ttfamily 2203.05591}}].

\bibitem{Arguelles:2022tki}
C.A.~Arg\"uelles et~al., \emph{{Snowmass white paper: beyond the standard model
  effects on neutrino flavor: Submitted to the proceedings of the US community
  study on the future of particle physics (Snowmass 2021)}},
  \href{https://doi.org/10.1140/epjc/s10052-022-11049-7}{\emph{Eur. Phys. J. C}
  {\bfseries 83} (2023) 15} [\href{https://arxiv.org/abs/2203.10811}{{\ttfamily
  2203.10811}}].

\bibitem{Ackermann:2019ows}
M.~Ackermann et~al., \emph{{Astrophysics Uniquely Enabled by Observations of
  High-Energy Cosmic Neutrinos}}, {\emph{Bull. Am. Astron. Soc.} {\bfseries 51}
  (2019) 185} [\href{https://arxiv.org/abs/1903.04334}{{\ttfamily
  1903.04334}}].

\bibitem{AlvesBatista:2019tlv}
R.~Alves~Batista et~al., \emph{{Open Questions in Cosmic-Ray Research at
  Ultrahigh Energies}},
  \href{https://doi.org/10.3389/fspas.2019.00023}{\emph{Front. Astron. Space
  Sci.} {\bfseries 6} (2019) 23}
  [\href{https://arxiv.org/abs/1903.06714}{{\ttfamily 1903.06714}}].

\bibitem{AlvesBatista:2021eeu}
R.~Alves~Batista et~al., \emph{{EuCAPT White Paper: Opportunities and
  Challenges for Theoretical Astroparticle Physics in the Next Decade}},
  \href{https://arxiv.org/abs/2110.10074}{{\ttfamily 2110.10074}}.

\bibitem{Barenboim:2003jm}
G.~Barenboim and C.~Quigg, \emph{{Neutrino Observatories can Characterize
  Cosmic Sources and Neutrino Properties}},
  \href{https://doi.org/10.1103/PhysRevD.67.073024}{\emph{Phys. Rev. D}
  {\bfseries 67} (2003) 073024}
  [\href{https://arxiv.org/abs/hep-ph/0301220}{{\ttfamily hep-ph/0301220}}].

\bibitem{Christian:2004xb}
J.~Christian, \emph{{Testing quantum gravity via cosmogenic neutrino
  oscillations}}, \href{https://doi.org/10.1103/PhysRevD.71.024012}{\emph{Phys.
  Rev. D} {\bfseries 71} (2005) 024012}
  [\href{https://arxiv.org/abs/gr-qc/0409077}{{\ttfamily gr-qc/0409077}}].

\bibitem{Hooper:2005jp}
D.~Hooper, D.~Morgan and E.~Winstanley, \emph{{Lorentz and CPT invariance
  violation in high-energy neutrinos}},
  \href{https://doi.org/10.1103/PhysRevD.72.065009}{\emph{Phys. Rev. D}
  {\bfseries 72} (2005) 065009}
  [\href{https://arxiv.org/abs/hep-ph/0506091}{{\ttfamily hep-ph/0506091}}].

\bibitem{Bazo:2009en}
J.L.~Bazo, M.~Bustamante, A.M.~Gago and O.G.~Miranda, \emph{{High energy
  astrophysical neutrino flux and modified dispersion relations}},
  \href{https://doi.org/10.1142/S0217751X09047429}{\emph{Int. J. Mod. Phys. A}
  {\bfseries 24} (2009) 5819}
  [\href{https://arxiv.org/abs/0907.1979}{{\ttfamily 0907.1979}}].

\bibitem{Ando:2009ts}
S.~Ando, M.~Kamionkowski and I.~Mocioiu, \emph{{Neutrino Oscillations,
  Lorentz/CPT Violation, and Dark Energy}},
  \href{https://doi.org/10.1103/PhysRevD.80.123522}{\emph{Phys. Rev. D}
  {\bfseries 80} (2009) 123522}
  [\href{https://arxiv.org/abs/0910.4391}{{\ttfamily 0910.4391}}].

\bibitem{Bhattacharya:2009tx}
A.~Bhattacharya, S.~Choubey, R.~Gandhi and A.~Watanabe, \emph{{Diffuse
  Ultra-High Energy Neutrino Fluxes and Physics Beyond the Standard Model}},
  \href{https://doi.org/10.1016/j.physletb.2010.04.078}{\emph{Phys. Lett. B}
  {\bfseries 690} (2010) 42} [\href{https://arxiv.org/abs/0910.4396}{{\ttfamily
  0910.4396}}].

\bibitem{Bustamante:2010nq}
M.~Bustamante, A.M.~Gago and C.~Pe{\~n}a-Garay, \emph{{Energy-Independent New
  Physics in the Flavour Ratios of High-Energy Astrophysical Neutrinos}},
  \href{https://doi.org/10.1007/JHEP04(2010)066}{\emph{JHEP} {\bfseries 04}
  (2010) 066} [\href{https://arxiv.org/abs/1001.4878}{{\ttfamily 1001.4878}}].

\bibitem{Mehta:2011qb}
P.~Mehta and W.~Winter, \emph{{Interplay of energy dependent astrophysical
  neutrino flavor ratios and new physics effects}},
  \href{https://doi.org/10.1088/1475-7516/2011/03/041}{\emph{JCAP} {\bfseries
  03} (2011) 041} [\href{https://arxiv.org/abs/1101.2673}{{\ttfamily
  1101.2673}}].

\bibitem{Arguelles:2015dca}
C.A.~Arg\"uelles, T.~Katori and J.~Salvad\'o, \emph{{New Physics in
  Astrophysical Neutrino Flavor}},
  \href{https://doi.org/10.1103/PhysRevLett.115.161303}{\emph{Phys. Rev. Lett.}
  {\bfseries 115} (2015) 161303}
  [\href{https://arxiv.org/abs/1506.02043}{{\ttfamily 1506.02043}}].

\bibitem{Bustamante:2015waa}
M.~Bustamante, J.F.~Beacom and W.~Winter, \emph{{Theoretically palatable flavor
  combinations of astrophysical neutrinos}},
  \href{https://doi.org/10.1103/PhysRevLett.115.161302}{\emph{Phys. Rev. Lett.}
  {\bfseries 115} (2015) 161302}
  [\href{https://arxiv.org/abs/1506.02645}{{\ttfamily 1506.02645}}].

\bibitem{Rasmussen:2017ert}
R.W.~Rasmussen, L.~Lechner, M.~Ackermann, M.~Kowalski and W.~Winter,
  \emph{{Astrophysical neutrinos flavored with Beyond the Standard Model
  physics}}, \href{https://doi.org/10.1103/PhysRevD.96.083018}{\emph{Phys. Rev.
  D} {\bfseries 96} (2017) 083018}
  [\href{https://arxiv.org/abs/1707.07684}{{\ttfamily 1707.07684}}].

\bibitem{Moura:2022dev}
C.A.~Moura and F.~Rossi-Torres, \emph{{Searches for Violation of CPT Symmetry
  and Lorentz Invariance with Astrophysical Neutrinos}},
  \href{https://doi.org/10.3390/universe8010042}{\emph{Universe} {\bfseries 8}
  (2022) 42}.

\bibitem{PerezdelosHeros:2022izj}
C.~P\'erez de~los Heros and T.~Terzi\'c, \emph{{Cosmic Searches for Lorentz
  Invariance Violation}},
  \href{https://doi.org/10.1007/978-3-031-31520-6_6}{\emph{Lect. Notes Phys.}
  {\bfseries 1017} (2023) 241}
  [\href{https://arxiv.org/abs/2209.06531}{{\ttfamily 2209.06531}}].

\bibitem{Carmona:2023mzs}
J.M.~Carmona, J.L.~Cort\'es and M.A.~Reyes, \emph{{Consistency of
  Lorentz-invariance violation neutrino scenarios in time delay analyses}},
  \href{https://doi.org/10.1088/1361-6382/ad2f13}{\emph{Class. Quant. Grav.}
  {\bfseries 41} (2024) 075012}
  [\href{https://arxiv.org/abs/2310.12661}{{\ttfamily 2310.12661}}].

\bibitem{Diaz:2013wia}
J.S.~D{\'i}az, A.~Kosteleck{\'y} and M.~Mewes, \emph{{Testing Relativity with
  High-Energy Astrophysical Neutrinos}},
  \href{https://doi.org/10.1103/PhysRevD.89.043005}{\emph{Phys. Rev. D}
  {\bfseries 89} (2014) 043005}
  [\href{https://arxiv.org/abs/1308.6344}{{\ttfamily 1308.6344}}].

\bibitem{Wang:2016lne}
Z.-Y.~Wang, R.-Y.~Liu and X.-Y.~Wang, \emph{{Testing the equivalence principle
  and Lorentz invariance with PeV neutrinos from blazar flares}},
  \href{https://doi.org/10.1103/PhysRevLett.116.151101}{\emph{Phys. Rev. Lett.}
  {\bfseries 116} (2016) 151101}
  [\href{https://arxiv.org/abs/1602.06805}{{\ttfamily 1602.06805}}].

\bibitem{Wei:2016ygk}
J.-J.~Wei, X.-F.~Wu, H.~Gao and P.~M\'esz\'aros, \emph{{Limits on the Neutrino
  Velocity, Lorentz Invariance, and the Weak Equivalence Principle with TeV
  Neutrinos from Gamma-Ray Bursts}},
  \href{https://doi.org/10.1088/1475-7516/2016/08/031}{\emph{JCAP} {\bfseries
  08} (2016) 031} [\href{https://arxiv.org/abs/1603.07568}{{\ttfamily
  1603.07568}}].

\bibitem{IceCube:2017qyp}
{\scshape IceCube} collaboration, \emph{{Neutrino Interferometry for
  High-Precision Tests of Lorentz Symmetry with IceCube}},
  \href{https://doi.org/10.1038/s41567-018-0172-2}{\emph{Nature Phys.}
  {\bfseries 14} (2018) 961}
  [\href{https://arxiv.org/abs/1709.03434}{{\ttfamily 1709.03434}}].

\bibitem{Lai:2017bbl}
K.-C.~Lai, W.-H.~Lai and G.-L.~Lin, \emph{{Constraining the mass scale of a
  Lorentz-violating Hamiltonian with the measurement of astrophysical
  neutrino-flavor composition}},
  \href{https://doi.org/10.1103/PhysRevD.96.115026}{\emph{Phys. Rev. D}
  {\bfseries 96} (2017) 115026}
  [\href{https://arxiv.org/abs/1704.04027}{{\ttfamily 1704.04027}}].

\bibitem{Wang:2020tej}
K.~Wang, S.-Q.~Xi, L.~Shao, R.-Y.~Liu, Z.~Li and Z.-K.~Zhang, \emph{{Limiting
  Superluminal Neutrino Velocity and Lorentz Invariance Violation by Neutrino
  Emission from the Blazar TXS 0506+056}},
  \href{https://doi.org/10.1103/PhysRevD.102.063027}{\emph{Phys. Rev. D}
  {\bfseries 102} (2020) 063027}
  [\href{https://arxiv.org/abs/2009.05201}{{\ttfamily 2009.05201}}].

\bibitem{IceCube:2021tdn}
{\scshape IceCube} collaboration, \emph{{Search for quantum gravity using
  astrophysical neutrino flavour with IceCube}},
  \href{https://doi.org/10.1038/s41567-022-01762-1}{\emph{Nature Phys.}
  {\bfseries 18} (2022) 1287}
  [\href{https://arxiv.org/abs/2111.04654}{{\ttfamily 2111.04654}}].

\bibitem{Bustamante:2024fbj}
M.~Bustamante, J.~Ellis, R.~Konoplich and A.S.~Sakharov, \emph{{Probing Lorentz
  invariance with a high-energy neutrino flare}},
  \href{https://arxiv.org/abs/2408.15949}{{\ttfamily 2408.15949}}.

\bibitem{KM3NeT:2025mfl}
{\scshape KM3NeT} collaboration, \emph{{KM3NeT Constraint on Lorentz-Violating
  Superluminal Neutrino Velocity}},
  \href{https://arxiv.org/abs/2502.12070}{{\ttfamily 2502.12070}}.

\bibitem{Satunin:2025uui}
P.~Satunin, \emph{{Ultra-high-energy event KM3-230213A constraints on Lorentz
  Invariance Violation in neutrino sector}},
  \href{https://arxiv.org/abs/2502.09548}{{\ttfamily 2502.09548}}.

\bibitem{Amelino-Camelia:2025lqn}
G.~Amelino-Camelia, G.~D'Amico, G.~Fabiano, D.~Frattulillo, G.~Gubitosi,
  A.~Moia et~al., \emph{{On testing in-vacuo dispersion with the most energetic
  neutrinos: KM3-230213A case study}},
  \href{https://arxiv.org/abs/2502.13093}{{\ttfamily 2502.13093}}.

\bibitem{Yang:2025kfr}
Y.-M.~Yang, X.-J.~Lv, X.-J.~Bi and P.-F.~Yin, \emph{{Constraints on Lorentz
  invariance violation in neutrino sector from the ultra-high-energy event
  KM3-230213A}},  \href{https://arxiv.org/abs/2502.18256}{{\ttfamily
  2502.18256}}.

\bibitem{Kostelecky:2004hg}
V.A.~Kosteleck{\'y} and M.~Mewes, \emph{{Lorentz violation and short-baseline
  neutrino experiments}},
  \href{https://doi.org/10.1103/PhysRevD.70.076002}{\emph{Phys. Rev. D}
  {\bfseries 70} (2004) 076002}
  [\href{https://arxiv.org/abs/hep-ph/0406255}{{\ttfamily hep-ph/0406255}}].

\bibitem{Katori:2006mz}
T.~Katori, V.A.~Kosteleck{\'y} and R.~Tayloe, \emph{{Global three-parameter
  model for neutrino oscillations using Lorentz violation}},
  \href{https://doi.org/10.1103/PhysRevD.74.105009}{\emph{Phys. Rev. D}
  {\bfseries 74} (2006) 105009}
  [\href{https://arxiv.org/abs/hep-ph/0606154}{{\ttfamily hep-ph/0606154}}].

\bibitem{IceCube:2010fyu}
{\scshape IceCube} collaboration, \emph{{Search for a Lorentz-violating
  sidereal signal with atmospheric neutrinos in IceCube}},
  \href{https://doi.org/10.1103/PhysRevD.82.112003}{\emph{Phys. Rev. D}
  {\bfseries 82} (2010) 112003}
  [\href{https://arxiv.org/abs/1010.4096}{{\ttfamily 1010.4096}}].

\bibitem{Kostelecky:2011gq}
A.~Kosteleck{\'y} and M.~Mewes, \emph{{Neutrinos with Lorentz-violating
  operators of arbitrary dimension}},
  \href{https://doi.org/10.1103/PhysRevD.85.096005}{\emph{Phys. Rev. D}
  {\bfseries 85} (2012) 096005}
  [\href{https://arxiv.org/abs/1112.6395}{{\ttfamily 1112.6395}}].

\bibitem{Guo:2012mv}
Z.-K.~Guo, Q.-G.~Huang, R.-G.~Cai and Y.-Z.~Zhang, \emph{{Cosmological
  constraints on Lorentz invariance violation in the neutrino sector}},
  \href{https://doi.org/10.1103/PhysRevD.86.065004}{\emph{Phys. Rev. D}
  {\bfseries 86} (2012) 065004}
  [\href{https://arxiv.org/abs/1206.5588}{{\ttfamily 1206.5588}}].

\bibitem{DoubleChooz:2012eiq}
{\scshape Double Chooz} collaboration, \emph{{First Test of Lorentz Violation
  with a Reactor-based Antineutrino Experiment}},
  \href{https://doi.org/10.1103/PhysRevD.86.112009}{\emph{Phys. Rev. D}
  {\bfseries 86} (2012) 112009}
  [\href{https://arxiv.org/abs/1209.5810}{{\ttfamily 1209.5810}}].

\bibitem{Super-Kamiokande:2014exs}
{\scshape Super-Kamiokande} collaboration, \emph{{Test of Lorentz invariance
  with atmospheric neutrinos}},
  \href{https://doi.org/10.1103/PhysRevD.91.052003}{\emph{Phys. Rev. D}
  {\bfseries 91} (2015) 052003}
  [\href{https://arxiv.org/abs/1410.4267}{{\ttfamily 1410.4267}}].

\bibitem{T2K:2017ega}
{\scshape T2K} collaboration, \emph{{Search for Lorentz and CPT violation using
  sidereal time dependence of neutrino flavor transitions over a short
  baseline}}, \href{https://doi.org/10.1103/PhysRevD.95.111101}{\emph{Phys.
  Rev. D} {\bfseries 95} (2017) 111101}
  [\href{https://arxiv.org/abs/1703.01361}{{\ttfamily 1703.01361}}].

\bibitem{SNO:2018mge}
{\scshape SNO} collaboration, \emph{{Tests of Lorentz invariance at the Sudbury
  Neutrino Observatory}},
  \href{https://doi.org/10.1103/PhysRevD.98.112013}{\emph{Phys. Rev. D}
  {\bfseries 98} (2018) 112013}
  [\href{https://arxiv.org/abs/1811.00166}{{\ttfamily 1811.00166}}].

\bibitem{Mishra:2023tdj}
S.~Mishra, S.~Shukla, L.~Singh and V.~Singh, \emph{{Search for Lorentz
  violations through the sidereal effect at the NO\ensuremath{\nu}A
  experiment}}, \href{https://doi.org/10.1103/PhysRevD.109.075042}{\emph{Phys.
  Rev. D} {\bfseries 109} (2024) 075042}
  [\href{https://arxiv.org/abs/2309.01756}{{\ttfamily 2309.01756}}].

\bibitem{Torri:2024jwc}
M.D.C.~Torri and L.~Miramonti, \emph{{Neutrinos as possible probes for quantum
  gravity}}, \href{https://doi.org/10.1088/1361-6382/ad5825}{\emph{Class.
  Quant. Grav.} {\bfseries 41} (2024) 153001}
  [\href{https://arxiv.org/abs/2404.04076}{{\ttfamily 2404.04076}}].

\bibitem{Kostelecky:2008ts}
V.A.~Kosteleck{\'y} and N.~Russell, \emph{{Data Tables for Lorentz and CPT
  Violation}}, \href{https://doi.org/10.1103/RevModPhys.83.11}{\emph{Rev. Mod.
  Phys.} {\bfseries 83} (2011) 11}
  [\href{https://arxiv.org/abs/0801.0287}{{\ttfamily 0801.0287}}].

\bibitem{Diaz:2009qk}
J.S.~D{\'i}az, V.A.~Kosteleck{\'y} and M.~Mewes, \emph{{Perturbative Lorentz
  and CPT violation for neutrino and antineutrino oscillations}},
  \href{https://doi.org/10.1103/PhysRevD.80.076007}{\emph{Phys. Rev. D}
  {\bfseries 80} (2009) 076007}
  [\href{https://arxiv.org/abs/0908.1401}{{\ttfamily 0908.1401}}].

\bibitem{Diaz:2011ia}
J.S.~D{\'i}az and A.~Kosteleck{\'y}, \emph{{Lorentz- and CPT-violating models
  for neutrino oscillations}},
  \href{https://doi.org/10.1103/PhysRevD.85.016013}{\emph{Phys. Rev. D}
  {\bfseries 85} (2012) 016013}
  [\href{https://arxiv.org/abs/1108.1799}{{\ttfamily 1108.1799}}].

\bibitem{Klop:2017dim}
N.~Klop and S.~Ando, \emph{{Effects of a neutrino-dark energy coupling on
  oscillations of high-energy neutrinos}},
  \href{https://doi.org/10.1103/PhysRevD.97.063006}{\emph{Phys. Rev. D}
  {\bfseries 97} (2018) 063006}
  [\href{https://arxiv.org/abs/1712.05413}{{\ttfamily 1712.05413}}].

\bibitem{Lin:2025aym}
H.-X.~Lin, J.~Ren and J.~Tang, \emph{{Anisotropic Lorentz invariance violation
  in reactor neutrino experiments}},
  \href{https://arxiv.org/abs/2503.02305}{{\ttfamily 2503.02305}}.

\bibitem{Kostelecky:2003xn}
V.A.~Kosteleck{\'y} and M.~Mewes, \emph{{Lorentz and CPT violation in the
  neutrino sector}},
  \href{https://doi.org/10.1103/PhysRevD.70.031902}{\emph{Phys. Rev. D}
  {\bfseries 70} (2004) 031902}
  [\href{https://arxiv.org/abs/hep-ph/0308300}{{\ttfamily hep-ph/0308300}}].

\bibitem{Kostelecky:2003cr}
V.A.~Kosteleck{\'y} and M.~Mewes, \emph{{Lorentz and CPT violation in
  neutrinos}}, \href{https://doi.org/10.1103/PhysRevD.69.016005}{\emph{Phys.
  Rev. D} {\bfseries 69} (2004) 016005}
  [\href{https://arxiv.org/abs/hep-ph/0309025}{{\ttfamily hep-ph/0309025}}].

\bibitem{Kostelecky:2003fs}
V.A.~Kosteleck\'y, \emph{{Gravity, Lorentz violation, and the Standard Model}},
  \href{https://doi.org/10.1103/PhysRevD.69.105009}{\emph{Phys. Rev. D}
  {\bfseries 69} (2004) 105009}
  [\href{https://arxiv.org/abs/hep-th/0312310}{{\ttfamily hep-th/0312310}}].

\bibitem{Super-Kamiokande:1998kpq}
{\scshape Super-Kamiokande} collaboration, \emph{{Evidence for oscillation of
  atmospheric neutrinos}},
  \href{https://doi.org/10.1103/PhysRevLett.81.1562}{\emph{Phys. Rev. Lett.}
  {\bfseries 81} (1998) 1562}
  [\href{https://arxiv.org/abs/hep-ex/9807003}{{\ttfamily hep-ex/9807003}}].

\bibitem{SNO:2001kpb}
{\scshape SNO} collaboration, \emph{{Measurement of the rate of $\nu_e+d \to
  p+p+e^-$ interactions produced by $^8$B solar neutrinos at the Sudbury
  Neutrino Observatory}},
  \href{https://doi.org/10.1103/PhysRevLett.87.071301}{\emph{Phys. Rev. Lett.}
  {\bfseries 87} (2001) 071301}
  [\href{https://arxiv.org/abs/nucl-ex/0106015}{{\ttfamily nucl-ex/0106015}}].

\bibitem{IceCube:2020wum}
{\scshape IceCube} collaboration, \emph{{The IceCube high-energy starting event
  sample: Description and flux characterization with 7.5 years of data}},
  \href{https://doi.org/10.1103/PhysRevD.104.022002}{\emph{Phys. Rev. D}
  {\bfseries 104} (2021) 022002}
  [\href{https://arxiv.org/abs/2011.03545}{{\ttfamily 2011.03545}}].

\bibitem{IC75yrHESEPublicDataRelease}
{IceCube Collaboration}, ``{HESE 7.5 year data release}.''
  \url{https://icecube.wisc.edu/data-releases/2021/12/hese-7-5-year-data/},
  2021.
\newblock 10.21234/4EQJ-BB17.

\bibitem{Telalovic:2023tcb}
B.~Telalovic and M.~Bustamante, \emph{{Flavor Anisotropy in the High-Energy
  Astrophysical Neutrino Sky}},
  \href{https://arxiv.org/abs/2310.15224}{{\ttfamily 2310.15224}}.

\bibitem{ParticleDataGroup:2024cfk}
{\scshape Particle Data Group} collaboration, \emph{{Review of particle
  physics}}, \href{https://doi.org/10.1103/PhysRevD.110.030001}{\emph{Phys.
  Rev. D} {\bfseries 110} (2024) 030001}.

\bibitem{Kostelecky:2002hh}
V.A.~Kosteleck{\'y} and M.~Mewes, \emph{{Signals for Lorentz violation in
  electrodynamics}},
  \href{https://doi.org/10.1103/PhysRevD.66.056005}{\emph{Phys. Rev. D}
  {\bfseries 66} (2002) 056005}
  [\href{https://arxiv.org/abs/hep-ph/0205211}{{\ttfamily hep-ph/0205211}}].

\bibitem{Esteban:2020cvm}
I.~Esteban, M.C.~Gonz\'alez-Garc\'ia, M.~Maltoni, T.~Schwetz and A.~Zhou,
  \emph{{The fate of hints: updated global analysis of three-flavor neutrino
  oscillations}}, \href{https://doi.org/10.1007/JHEP09(2020)178}{\emph{JHEP}
  {\bfseries 09} (2020) 178}
  [\href{https://arxiv.org/abs/2007.14792}{{\ttfamily 2007.14792}}].

\bibitem{Learned:1994wg}
J.G.~Learned and S.~Pakvasa, \emph{{Detecting tau-neutrino oscillations at PeV
  energies}},  1995.
\newblock 10.1016/0927-6505(94)00043-3.

\bibitem{Gorski:2004by}
K.M.~G\'orski, E.~Hivon, A.J.~Banday, B.D.~Wandelt, F.K.~Hansen, M.~Reinecke
  et~al., \emph{{HEALPix - A Framework for high resolution discretization, and
  fast analysis of data distributed on the sphere}},
  \href{https://doi.org/10.1086/427976}{\emph{Astrophys. J.} {\bfseries 622}
  (2005) 759} [\href{https://arxiv.org/abs/astro-ph/0409513}{{\ttfamily
  astro-ph/0409513}}].

\bibitem{healpix_url}
{HEALPix Team}, ``{HEALPix}.'' \url{https://healpix.sourceforge.io/}, 2019.

\bibitem{LSND:2005oop}
{\scshape LSND} collaboration, \emph{{Tests of Lorentz violation in
  $\bar{\nu}_\mu \to \bar{\nu}_e$ oscillations}},
  \href{https://doi.org/10.1103/PhysRevD.72.076004}{\emph{Phys. Rev. D}
  {\bfseries 72} (2005) 076004}
  [\href{https://arxiv.org/abs/hep-ex/0506067}{{\ttfamily hep-ex/0506067}}].

\bibitem{MINOS:2008fnv}
{\scshape MINOS} collaboration, \emph{{Testing Lorentz Invariance and CPT
  Conservation with NuMI Neutrinos in the MINOS Near Detector}},
  \href{https://doi.org/10.1103/PhysRevLett.101.151601}{\emph{Phys. Rev. Lett.}
  {\bfseries 101} (2008) 151601}
  [\href{https://arxiv.org/abs/0806.4945}{{\ttfamily 0806.4945}}].

\bibitem{MINOS:2010kat}
{\scshape MINOS} collaboration, \emph{{A Search for Lorentz Invariance and CPT
  Violation with the MINOS Far Detector}},
  \href{https://doi.org/10.1103/PhysRevLett.105.151601}{\emph{Phys. Rev. Lett.}
  {\bfseries 105} (2010) 151601}
  [\href{https://arxiv.org/abs/1007.2791}{{\ttfamily 1007.2791}}].

\bibitem{MiniBooNE:2011pix}
{\scshape MiniBooNE} collaboration, \emph{{Test of Lorentz and CPT violation
  with Short Baseline Neutrino Oscillation Excesses}},
  \href{https://doi.org/10.1016/j.physletb.2012.12.020}{\emph{Phys. Lett. B}
  {\bfseries 718} (2013) 1303}
  [\href{https://arxiv.org/abs/1109.3480}{{\ttfamily 1109.3480}}].

\bibitem{Kelner:2006tc}
S.R.~Kelner, F.A.~Aharonian and V.V.~Bugayov, \emph{{Energy spectra of
  gamma-rays, electrons and neutrinos produced at proton-proton interactions in
  the very high energy regime}},
  \href{https://doi.org/10.1103/PhysRevD.74.034018}{\emph{Phys. Rev. D}
  {\bfseries 74} (2006) 034018}
  [\href{https://arxiv.org/abs/astro-ph/0606058}{{\ttfamily
  astro-ph/0606058}}].

\bibitem{Kelner:2008ke}
S.R.~Kelner and F.A.~Aharonian, \emph{{Energy spectra of gamma-rays, electrons
  and neutrinos produced at interactions of relativistic protons with low
  energy radiation}},
  \href{https://doi.org/10.1103/PhysRevD.82.099901}{\emph{Phys. Rev. D}
  {\bfseries 78} (2008) 034013}
  [\href{https://arxiv.org/abs/0803.0688}{{\ttfamily 0803.0688}}].

\bibitem{Fiorillo:2022rft}
D.F.G.~Fiorillo and M.~Bustamante, \emph{{Bump hunting in the diffuse flux of
  high-energy cosmic neutrinos}},
  \href{https://doi.org/10.1103/PhysRevD.107.083008}{\emph{Phys. Rev. D}
  {\bfseries 107} (2023) 083008}
  [\href{https://arxiv.org/abs/2301.00024}{{\ttfamily 2301.00024}}].

\bibitem{Naab:2023xcz}
{\scshape IceCube} collaboration, \emph{{Measurement of the astrophysical
  diffuse neutrino flux in a combined fit of IceCube's high energy neutrino
  data}},  in \emph{{38th International Cosmic Ray Conference}}, 7, 2023
  [\href{https://arxiv.org/abs/2308.00191}{{\ttfamily 2308.00191}}].

\bibitem{IceCube:2023qpn}
{\scshape IceCube} collaboration, \emph{{Measurement of the astrophysical
  diffuse neutrino flux in a combined fit of IceCube\textquoteright{}s high
  energy neutrino data}}, \href{https://doi.org/10.22323/1.444.1064}{\emph{PoS}
  {\bfseries ICRC2023} (2023) 1064}.

\bibitem{Bustamante:2016ciw}
M.~Bustamante, J.F.~Beacom and K.~Murase, \emph{{Testing decay of astrophysical
  neutrinos with incomplete information}},
  \href{https://doi.org/10.1103/PhysRevD.95.063013}{\emph{Phys. Rev. D}
  {\bfseries 95} (2017) 063013}
  [\href{https://arxiv.org/abs/1610.02096}{{\ttfamily 1610.02096}}].

\bibitem{Yuksel:2008cu}
H.~Yuksel, M.D.~Kistler, J.F.~Beacom and A.M.~Hopkins, \emph{{Revealing the
  High-Redshift Star Formation Rate with Gamma-Ray Bursts}},
  \href{https://doi.org/10.1086/591449}{\emph{Astrophys. J. Lett.} {\bfseries
  683} (2008) L5} [\href{https://arxiv.org/abs/0804.4008}{{\ttfamily
  0804.4008}}].

\bibitem{Mena:2006eq}
O.~Mena, I.~Mocioiu and S.~Razzaque, \emph{{Oscillation effects on high-energy
  neutrino fluxes from astrophysical hidden sources}},
  \href{https://doi.org/10.1103/PhysRevD.75.063003}{\emph{Phys. Rev. D}
  {\bfseries 75} (2007) 063003}
  [\href{https://arxiv.org/abs/astro-ph/0612325}{{\ttfamily
  astro-ph/0612325}}].

\bibitem{Razzaque:2009kq}
S.~Razzaque and A.Y.~Smirnov, \emph{{Flavor conversion of cosmic neutrinos from
  hidden jets}}, \href{https://doi.org/10.1007/JHEP03(2010)031}{\emph{JHEP}
  {\bfseries 03} (2010) 031} [\href{https://arxiv.org/abs/0912.4028}{{\ttfamily
  0912.4028}}].

\bibitem{Sahu:2010ap}
S.~Sahu and B.~Zhang, \emph{{Effect of Resonant Neutrino Oscillation on TeV
  Neutrino Flavor Ratio from Choked GRBs}},
  \href{https://doi.org/10.1088/1674-4527/10/10/001}{\emph{Res. Astron.
  Astrophys.} {\bfseries 10} (2010) 943}
  [\href{https://arxiv.org/abs/1007.4582}{{\ttfamily 1007.4582}}].

\bibitem{Varela:2014mma}
K.~Varela, S.~Sahu, A.F.~Osorio~Oliveros and J.C.~Sanabria, \emph{{High energy
  neutrinos from choked GRBs and their flavor ratio measurement by the
  IceCube}}, \href{https://doi.org/10.1140/epjc/s10052-015-3524-4}{\emph{Eur.
  Phys. J. C} {\bfseries 75} (2015) 289}
  [\href{https://arxiv.org/abs/1411.7992}{{\ttfamily 1411.7992}}].

\bibitem{Xiao:2015gea}
D.~Xiao and Z.G.~Dai, \emph{{TeV-PeV Neutrino Oscillation of Low-luminosity
  Gamma-ray Bursts}},
  \href{https://doi.org/10.1088/0004-637X/805/2/137}{\emph{Astrophys. J.}
  {\bfseries 805} (2015) 137}
  [\href{https://arxiv.org/abs/1504.01603}{{\ttfamily 1504.01603}}].

\bibitem{Dev:2023znd}
P.S.B.~Dev, S.~Jana and Y.~Porto, \emph{{Flavor Matters, but Matter Flavors:
  Matter Effects on Flavor Composition of Astrophysical Neutrinos}},
  \href{https://arxiv.org/abs/2312.17315}{{\ttfamily 2312.17315}}.

\bibitem{IceCube:2023ame}
{\scshape IceCube} collaboration, \emph{{Observation of high-energy neutrinos
  from the Galactic plane}},
  \href{https://doi.org/10.1126/science.adc9818}{\emph{Science} {\bfseries 380}
  (2023) adc9818} [\href{https://arxiv.org/abs/2307.04427}{{\ttfamily
  2307.04427}}].

\bibitem{Bustamante:2023iyn}
M.~Bustamante, \emph{{The Milky Way shines in high-energy neutrinos}},
  \href{https://doi.org/10.1038/s42254-023-00679-9}{\emph{Nature Rev. Phys.}
  {\bfseries 6} (2024) 8} [\href{https://arxiv.org/abs/2312.08102}{{\ttfamily
  2312.08102}}].

\bibitem{Liu:2023flr}
Q.~Liu, D.F.G.~Fiorillo, C.A.~Arg\"uelles, M.~Bustamante, N.~Song and
  A.C.~Vincent, \emph{{Identifying Energy-Dependent Flavor Transitions in
  High-Energy Astrophysical Neutrino Measurements}},
  \href{https://arxiv.org/abs/2312.07649}{{\ttfamily 2312.07649}}.

\bibitem{Bustamante:2019sdb}
M.~Bustamante and M.~Ahlers, \emph{{Inferring the flavor of high-energy
  astrophysical neutrinos at their sources}},
  \href{https://doi.org/10.1103/PhysRevLett.122.241101}{\emph{Phys. Rev. Lett.}
  {\bfseries 122} (2019) 241101}
  [\href{https://arxiv.org/abs/1901.10087}{{\ttfamily 1901.10087}}].

\bibitem{Song:2020nfh}
N.~Song, S.W.~Li, C.A.~Arg\"uelles, M.~Bustamante and A.C.~Vincent, \emph{{The
  Future of High-Energy Astrophysical Neutrino Flavor Measurements}},
  \href{https://doi.org/10.1088/1475-7516/2021/04/054}{\emph{JCAP} {\bfseries
  04} (2021) 054} [\href{https://arxiv.org/abs/2012.12893}{{\ttfamily
  2012.12893}}].

\bibitem{Pakvasa:2007dc}
S.~Pakvasa, W.~Rodejohann and T.J.~Weiler, \emph{{Flavor Ratios of
  Astrophysical Neutrinos: Implications for Precision Measurements}},
  \href{https://doi.org/10.1088/1126-6708/2008/02/005}{\emph{JHEP} {\bfseries
  02} (2008) 005} [\href{https://arxiv.org/abs/0711.4517}{{\ttfamily
  0711.4517}}].

\bibitem{Lipari:2007su}
P.~Lipari, M.~Lusignoli and D.~Meloni, \emph{{Flavor Composition and Energy
  Spectrum of Astrophysical Neutrinos}},
  \href{https://doi.org/10.1103/PhysRevD.75.123005}{\emph{Phys. Rev. D}
  {\bfseries 75} (2007) 123005}
  [\href{https://arxiv.org/abs/0704.0718}{{\ttfamily 0704.0718}}].

\bibitem{Kashti:2005qa}
T.~Kashti and E.~Waxman, \emph{{Flavoring astrophysical neutrinos: Flavor
  ratios depend on energy}},
  \href{https://doi.org/10.1103/PhysRevLett.95.181101}{\emph{Phys. Rev. Lett.}
  {\bfseries 95} (2005) 181101}
  [\href{https://arxiv.org/abs/astro-ph/0507599}{{\ttfamily
  astro-ph/0507599}}].

\bibitem{Winter:2014tta}
W.~Winter, J.~Becker~Tjus and S.R.~Klein, \emph{{Impact of secondary
  acceleration on the neutrino spectra in gamma-ray bursts}},
  \href{https://doi.org/10.1051/0004-6361/201423745}{\emph{Astron. Astrophys.}
  {\bfseries 569} (2014) A58}
  [\href{https://arxiv.org/abs/1403.0574}{{\ttfamily 1403.0574}}].

\bibitem{Kawanaka:2015qza}
N.~Kawanaka and K.~Ioka, \emph{{Neutrino Flavor Ratios Modified by Cosmic Ray
  Secondary Acceleration}},
  \href{https://doi.org/10.1103/PhysRevD.92.085047}{\emph{Phys. Rev. D}
  {\bfseries 92} (2015) 085047}
  [\href{https://arxiv.org/abs/1504.03417}{{\ttfamily 1504.03417}}].

\bibitem{Farzan:2021gbx}
Y.~Farzan, \emph{{On the \ensuremath{\tau} flavor of the cosmic neutrino
  flux}}, \href{https://doi.org/10.1007/JHEP07(2021)174}{\emph{JHEP} {\bfseries
  07} (2021) 174} [\href{https://arxiv.org/abs/2105.03272}{{\ttfamily
  2105.03272}}].

\bibitem{IceCube:2020fpi}
{\scshape IceCube} collaboration, \emph{{Detection of astrophysical tau
  neutrino candidates in IceCube}},
  \href{https://doi.org/10.1140/epjc/s10052-022-10795-y}{\emph{Eur. Phys. J. C}
  {\bfseries 82} (2022) 1031}
  [\href{https://arxiv.org/abs/2011.03561}{{\ttfamily 2011.03561}}].

\bibitem{Ouellette:2024ggl}
A.~Ouellette and G.~Holder, \emph{{Cross-correlating IceCube neutrinos with a
  large set of galaxy samples around redshift $z \sim 1$}},
  \href{https://doi.org/10.1103/PhysRevD.110.103025}{\emph{Phys. Rev. D}
  {\bfseries 110} (2024) 103025}
  [\href{https://arxiv.org/abs/2405.09633}{{\ttfamily 2405.09633}}].

\bibitem{IceCube:2022ham}
{\scshape IceCube} collaboration, \emph{{Constraints on Populations of Neutrino
  Sources from Searches in the Directions of IceCube Neutrino Alerts}},
  \href{https://doi.org/10.3847/1538-4357/acd2ca}{\emph{Astrophys. J.}
  {\bfseries 951} (2023) 45}
  [\href{https://arxiv.org/abs/2210.04930}{{\ttfamily 2210.04930}}].

\bibitem{Zhou:2024kzp}
Z.~Zhou, J.~Cisewski-Kehe, K.~Fang and A.~Banerjee, \emph{{High-energy Neutrino
  Source Cross-correlations with Nearest-neighbor Distributions}},
  \href{https://doi.org/10.3847/1538-4357/ad924c}{\emph{Astrophys. J.}
  {\bfseries 979} (2025) 194}
  [\href{https://arxiv.org/abs/2406.00796}{{\ttfamily 2406.00796}}].

\bibitem{Schonert:2008is}
S.~Sch{\"o}nert, T.K.~Gaisser, E.~Resconi and O.~Schulz, \emph{{Vetoing
  atmospheric neutrinos in a high energy neutrino telescope}},
  \href{https://doi.org/10.1103/PhysRevD.79.043009}{\emph{Phys. Rev. D}
  {\bfseries 79} (2009) 043009}
  [\href{https://arxiv.org/abs/0812.4308}{{\ttfamily 0812.4308}}].

\bibitem{Gaisser:2014bja}
T.K.~Gaisser, K.~Jero, A.~Karle and J.~van Santen, \emph{{Generalized self-veto
  probability for atmospheric neutrinos}},
  \href{https://doi.org/10.1103/PhysRevD.90.023009}{\emph{Phys. Rev. D}
  {\bfseries 90} (2014) 023009}
  [\href{https://arxiv.org/abs/1405.0525}{{\ttfamily 1405.0525}}].

\bibitem{Beacom:2004jb}
J.F.~Beacom and J.~Candia, \emph{{Shower power: Isolating the prompt
  atmospheric neutrino flux using electron neutrinos}},
  \href{https://doi.org/10.1088/1475-7516/2004/11/009}{\emph{JCAP} {\bfseries
  11} (2004) 009} [\href{https://arxiv.org/abs/hep-ph/0409046}{{\ttfamily
  hep-ph/0409046}}].

\bibitem{IceCube:2013low}
{\scshape IceCube} collaboration, \emph{{Evidence for High-Energy
  Extraterrestrial Neutrinos at the IceCube Detector}},
  \href{https://doi.org/10.1126/science.1242856}{\emph{Science} {\bfseries 342}
  (2013) 1242856} [\href{https://arxiv.org/abs/1311.5238}{{\ttfamily
  1311.5238}}].

\bibitem{Beacom:2003nh}
J.F.~Beacom, N.F.~Bell, D.~Hooper, S.~Pakvasa and T.J.~Weiler, \emph{{Measuring
  Flavor Ratios of High-Energy Astrophysical Neutrinos}},
  \href{https://doi.org/10.1103/PhysRevD.68.093005}{\emph{Phys. Rev. D}
  {\bfseries 68} (2003) 093005}
  [\href{https://arxiv.org/abs/hep-ph/0307025}{{\ttfamily hep-ph/0307025}}].

\bibitem{Bugaev:2003sw}
E.~Bugaev, T.~Montaruli, Y.~Shlepin and I.A.~Sokalski, \emph{{Propagation of
  tau neutrinos and tau leptons through the earth and their detection in
  underwater / ice neutrino telescopes}},
  \href{https://doi.org/10.1016/j.astropartphys.2004.03.002}{\emph{Astropart.
  Phys.} {\bfseries 21} (2004) 491}
  [\href{https://arxiv.org/abs/hep-ph/0312295}{{\ttfamily hep-ph/0312295}}].

\bibitem{IceCube:2013dkx}
{\scshape IceCube} collaboration, \emph{{Energy Reconstruction Methods in the
  IceCube Neutrino Telescope}},
  \href{https://doi.org/10.1088/1748-0221/9/03/P03009}{\emph{JINST} {\bfseries
  9} (2014) P03009} [\href{https://arxiv.org/abs/1311.4767}{{\ttfamily
  1311.4767}}].

\bibitem{IceCube:2015rro}
{\scshape IceCube} collaboration, \emph{{Flavor Ratio of Astrophysical
  Neutrinos above 35 TeV in IceCube}},
  \href{https://doi.org/10.1103/PhysRevLett.114.171102}{\emph{Phys. Rev. Lett.}
  {\bfseries 114} (2015) 171102}
  [\href{https://arxiv.org/abs/1502.03376}{{\ttfamily 1502.03376}}].

\bibitem{IceCube:2015gsk}
{\scshape IceCube} collaboration, \emph{{A combined maximum-likelihood analysis
  of the high-energy astrophysical neutrino flux measured with IceCube}},
  \href{https://doi.org/10.1088/0004-637X/809/1/98}{\emph{Astrophys. J.}
  {\bfseries 809} (2015) 98}
  [\href{https://arxiv.org/abs/1507.03991}{{\ttfamily 1507.03991}}].

\bibitem{Mena:2014sja}
O.~Mena, S.~Palomares-Ruiz and A.C.~Vincent, \emph{{Flavor Composition of the
  High-Energy Neutrino Events in IceCube}},
  \href{https://doi.org/10.1103/PhysRevLett.113.091103}{\emph{Phys. Rev. Lett.}
  {\bfseries 113} (2014) 091103}
  [\href{https://arxiv.org/abs/1404.0017}{{\ttfamily 1404.0017}}].

\bibitem{Palomares-Ruiz:2015mka}
S.~Palomares-Ruiz, A.C.~Vincent and O.~Mena, \emph{{Spectral analysis of the
  high-energy IceCube neutrinos}},
  \href{https://doi.org/10.1103/PhysRevD.91.103008}{\emph{Phys. Rev. D}
  {\bfseries 91} (2015) 103008}
  [\href{https://arxiv.org/abs/1502.02649}{{\ttfamily 1502.02649}}].

\bibitem{Vincent:2016nut}
A.C.~Vincent, S.~Palomares-Ruiz and O.~Mena, \emph{{Analysis of the 4-year
  IceCube high-energy starting events}},
  \href{https://doi.org/10.1103/PhysRevD.94.023009}{\emph{Phys. Rev. D}
  {\bfseries 94} (2016) 023009}
  [\href{https://arxiv.org/abs/1605.01556}{{\ttfamily 1605.01556}}].

\bibitem{Coloma:2023ixt}
P.~Coloma, M.C.~Gonz\'alez-Garc\'ia, M.~Maltoni, J.a.P.~Pinheiro and S.~Urrea,
  \emph{{Global constraints on non-standard neutrino interactions with quarks
  and electrons}}, \href{https://doi.org/10.1007/JHEP08(2023)032}{\emph{JHEP}
  {\bfseries 08} (2023) 032}
  [\href{https://arxiv.org/abs/2305.07698}{{\ttfamily 2305.07698}}].

\bibitem{Hummer:2010vx}
S.~H{\"u}mmer, M.~R{\"u}ger, F.~Spanier and W.~Winter, \emph{{Simplified models
  for photohadronic interactions in cosmic accelerators}},
  \href{https://doi.org/10.1088/0004-637X/721/1/630}{\emph{Astrophys. J.}
  {\bfseries 721} (2010) 630}
  [\href{https://arxiv.org/abs/1002.1310}{{\ttfamily 1002.1310}}].

\bibitem{CTEQ:1993hwr}
{\scshape CTEQ} collaboration, \emph{{Handbook of perturbative QCD: Version
  1.0}}, \href{https://doi.org/10.1103/RevModPhys.67.157}{\emph{Rev. Mod.
  Phys.} {\bfseries 67} (1995) 157}.

\bibitem{Conrad:1997ne}
J.M.~Conrad, M.H.~Shaevitz and T.~Bolton, \emph{{Precision measurements with
  high-energy neutrino beams}},
  \href{https://doi.org/10.1103/RevModPhys.70.1341}{\emph{Rev. Mod. Phys.}
  {\bfseries 70} (1998) 1341}
  [\href{https://arxiv.org/abs/hep-ex/9707015}{{\ttfamily hep-ex/9707015}}].

\bibitem{Formaggio:2012cpf}
J.A.~Formaggio and G.P.~Zeller, \emph{{From eV to EeV: Neutrino Cross Sections
  Across Energy Scales}},
  \href{https://doi.org/10.1103/RevModPhys.84.1307}{\emph{Rev. Mod. Phys.}
  {\bfseries 84} (2012) 1307}
  [\href{https://arxiv.org/abs/1305.7513}{{\ttfamily 1305.7513}}].

\bibitem{Glashow:1960zz}
S.L.~Glashow, \emph{{Resonant Scattering of Antineutrinos}},
  \href{https://doi.org/10.1103/PhysRev.118.316}{\emph{Phys. Rev.} {\bfseries
  118} (1960) 316}.

\bibitem{IceCube:2021rpz}
{\scshape IceCube} collaboration, \emph{{Detection of a particle shower at the
  Glashow resonance with IceCube}},
  \href{https://doi.org/10.1038/s41586-021-03256-1}{\emph{Nature} {\bfseries
  591} (2021) 220} [\href{https://arxiv.org/abs/2110.15051}{{\ttfamily
  2110.15051}}].

\bibitem{Buchner:2021cql}
J.~Buchner, \emph{{UltraNest -- a robust, general purpose Bayesian inference
  engine}},  \href{https://arxiv.org/abs/2101.09604}{{\ttfamily 2101.09604}}.

\bibitem{Buchner:2014}
J.~Buchner, \emph{{A statistical test for Nested Sampling algorithms}},
  \href{https://doi.org/10.1007/s11222-014-9512-y}{\emph{Statistics and
  Computing} {\bfseries 26} (2016) 383}
  [\href{https://arxiv.org/abs/1407.5459}{{\ttfamily 1407.5459}}].

\bibitem{Buchner:2017}
J.~Buchner, \emph{{Collaborative Nested Sampling: Big Data vs. complex physical
  models}}, \href{https://doi.org/10.1088/1538-3873/aae7fc}{\emph{Publications
  of the Astronomical Society of the Pacific} {\bfseries 131} (2019) 108005}
  [\href{https://arxiv.org/abs/1707.04476}{{\ttfamily 1707.04476}}].

\bibitem{Esteban:2024eli}
I.~Esteban, M.C.~Gonz\'alez-Garc\'ia, M.~Maltoni, I.~Mart\'inez-Soler,
  J.a.P.~Pinheiro and T.~Schwetz, \emph{{NuFit-6.0: updated global analysis of
  three-flavor neutrino oscillations}},
  \href{https://doi.org/10.1007/JHEP12(2024)216}{\emph{JHEP} {\bfseries 12}
  (2024) 216} [\href{https://arxiv.org/abs/2410.05380}{{\ttfamily
  2410.05380}}].

\bibitem{KATRIN:2022qou}
{\scshape KATRIN} collaboration, \emph{{Search for Lorentz-invariance violation
  with the first KATRIN data}},
  \href{https://doi.org/10.1103/PhysRevD.107.082005}{\emph{Phys. Rev. D}
  {\bfseries 107} (2023) 082005}
  [\href{https://arxiv.org/abs/2207.06326}{{\ttfamily 2207.06326}}].

\bibitem{KM3Net:2016zxf}
{\scshape KM3Net} collaboration, \emph{{Letter of intent for KM3NeT 2.0}},
  \href{https://doi.org/10.1088/0954-3899/43/8/084001}{\emph{J. Phys. G}
  {\bfseries 43} (2016) 084001}
  [\href{https://arxiv.org/abs/1601.07459}{{\ttfamily 1601.07459}}].

\bibitem{Baikal-GVD:2019kwy}
{\scshape Baikal-GVD} collaboration, \emph{{Neutrino Telescope in Lake Baikal:
  Present and Future}}, \href{https://doi.org/10.22323/1.358.1011}{\emph{PoS}
  {\bfseries ICRC2019} (2020) 1011}
  [\href{https://arxiv.org/abs/1908.05427}{{\ttfamily 1908.05427}}].

\bibitem{Schumacher:2025qca}
L.J.~Schumacher, M.~Bustamante, M.~Agostini, F.~Oikonomou and E.~Resconi,
  \emph{{Beyond first light: global monitoring for high-energy neutrino
  astronomy}},  \href{https://arxiv.org/abs/2503.07549}{{\ttfamily
  2503.07549}}.

\bibitem{Abbasi:2022ypr}
R.~Abbasi et~al., \emph{{Graph Neural Networks for low-energy event
  classification \& reconstruction in IceCube}},
  \href{https://doi.org/10.1088/1748-0221/17/11/P11003}{\emph{JINST} {\bfseries
  17} (2022) P11003} [\href{https://arxiv.org/abs/2209.03042}{{\ttfamily
  2209.03042}}].

\bibitem{Sogaard:2022qgg}
A.~S\o{}gaard et~al., \emph{{GraphNeT: Graph neural networks for neutrino
  telescope event reconstruction}},
  \href{https://doi.org/10.21105/joss.04971}{\emph{J. Open Source Softw.}
  {\bfseries 8} (2023) 4971}
  [\href{https://arxiv.org/abs/2210.12194}{{\ttfamily 2210.12194}}].

\bibitem{Bukhari:2023ezc}
H.~Bukhari, D.~Chakraborty, P.~Eller, T.~Ito, M.V.~Shugaev and R.~\O{}rs\o{}e,
  \emph{{IceCube \textendash{} Neutrinos in Deep Ice: The top 3 solutions from
  the public Kaggle competition}},
  \href{https://doi.org/10.1140/epjc/s10052-024-12977-2}{\emph{Eur. Phys. J. C}
  {\bfseries 84} (2024) 646}
  [\href{https://arxiv.org/abs/2310.15674}{{\ttfamily 2310.15674}}].

\bibitem{IceCube:2023fgt}
{\scshape IceCube} collaboration, \emph{{Summary of IceCube tau neutrino
  searches and flavor composition measurements of the diffuse astrophysical
  neutrino flux}}, \href{https://doi.org/10.22323/1.444.1122}{\emph{PoS}
  {\bfseries ICRC2023} (2023) 1122}
  [\href{https://arxiv.org/abs/2308.15213}{{\ttfamily 2308.15213}}].

\bibitem{Li:2016kra}
S.W.~Li, M.~Bustamante and J.F.~Beacom, \emph{{Echo Technique to Distinguish
  Flavors of Astrophysical Neutrinos}},
  \href{https://doi.org/10.1103/PhysRevLett.122.151101}{\emph{Phys. Rev. Lett.}
  {\bfseries 122} (2019) 151101}
  [\href{https://arxiv.org/abs/1606.06290}{{\ttfamily 1606.06290}}].

\bibitem{Steuer:2017tca}
{\scshape IceCube} collaboration, \emph{{Delayed light emission to distinguish
  astrophysical neutrino flavors in IceCube}},
  \href{https://doi.org/10.22323/1.301.1008}{\emph{PoS} {\bfseries ICRC2017}
  (2018) 1008}.

\bibitem{Farrag:2023jut}
K.~Farrag et~al., \emph{{Distinguishing $\nu_\tau$ neutrinos using the neutron
  echo technique with next generation ice Cherenkov telescopes}},
  \href{https://doi.org/10.22323/1.444.1211}{\emph{PoS} {\bfseries ICRC2023}
  (2023) 1211}.

\bibitem{Valera:2022wmu}
V.B.~Valera, M.~Bustamante and C.~Glaser, \emph{{Near-future discovery of the
  diffuse flux of ultrahigh-energy cosmic neutrinos}},
  \href{https://doi.org/10.1103/PhysRevD.107.043019}{\emph{Phys. Rev. D}
  {\bfseries 107} (2023) 043019}
  [\href{https://arxiv.org/abs/2210.03756}{{\ttfamily 2210.03756}}].

\bibitem{KM3NeT:2025npi}
{\scshape KM3NeT} collaboration, \emph{{Observation of an ultra-high-energy
  cosmic neutrino with KM3NeT}},
  \href{https://doi.org/10.1038/s41586-024-08543-1}{\emph{Nature} {\bfseries
  638} (2025) 376}.

\bibitem{Testagrossa:2023ukh}
F.~Testagrossa, D.F.G.~Fiorillo and M.~Bustamante, \emph{{Two-detector flavor
  sensitivity to ultrahigh-energy cosmic neutrinos}},
  \href{https://doi.org/10.1103/PhysRevD.110.083026}{\emph{Phys. Rev. D}
  {\bfseries 110} (2024) 083026}
  [\href{https://arxiv.org/abs/2310.12215}{{\ttfamily 2310.12215}}].

\bibitem{Coleman:2024scd}
A.~Coleman, O.~Ericsson, C.~Glaser and M.~Bustamante, \emph{{Flavor composition
  of ultrahigh-energy cosmic neutrinos: Measurement forecasts for in-ice
  radio-based EeV neutrino telescopes}},
  \href{https://doi.org/10.1103/PhysRevD.110.023044}{\emph{Phys. Rev. D}
  {\bfseries 110} (2024) 023044}
  [\href{https://arxiv.org/abs/2402.02432}{{\ttfamily 2402.02432}}].

\end{thebibliography}

\providecommand{\href}[2]{#2}\begingroup\raggedright\endgroup


\end{document}